\documentclass[twocolumn]{aastex62}
\usepackage{textcomp}
\usepackage{graphicx}
\usepackage{amsmath}
\usepackage{amssymb}
\usepackage{amsfonts}
\usepackage{epsfig}
\usepackage{lipsum}
\usepackage{lmodern}
\usepackage{hyperref}
\usepackage{natbib}
\usepackage{tabularx}
\usepackage{nameref}
\usepackage{float}

\setcitestyle{authoryear,square,aysep={},yysep={;}}
\newcommand{\ssim}{\mathord{\sim}}

\def\kms{\ifmmode{~{\rm km~s^{-1}}}\else{~km s$^{-1}$}\fi}
\def\cm3{\ifmmode{~{\rm cm^{-3}}}\else{~cm$^{-3}$}\fi}
\def\Ms{\ifmmode{~{\rm M_\odot}}\else{M$_\odot$}\fi}
\def\cm3{cm$^{-3}$}

\def\ltsima{$\; \buildrel < \over \sim \;$}
\def\simlt{\lower.5ex\hbox{\ltsima}} 
\def\gtsima{$\; \buildrel > \over \sim \;$}
\def\simgt{\lower.5ex\hbox{\gtsima}} 

\def\deg{\hbox{$^\circ$}}

\def\gray{$\gamma$-ray}
\def\grays{$\gamma$-rays}

\def\Fermi{\textit{Fermi}}

\newcommand{\fermi}{\textit{Fermi}-LAT}
\newcommand{\swiftXRT}{\textit{Swift}-XRT}
\newcommand{\swiftUVOT}{\textit{Swift}-UVOT}

\newcommand{\gName}[2]{#1\,#2}
\newcommand{\octtwentyninth}{2016 October 29}

\newcommand{\octthirtyfirst}{2016 October 31}
\newcommand{\decthirtieth}{2016 December 30}
\newcommand{\decthirtyfirst}{2016 December 31}
\newcommand{\janfirst}{2017 January 1}
\newcommand{\jansecond}{2017 January 2}
\newcommand{\janthird}{2017 January 3}

\begin{document}

\title{The exceptional 2017 gamma-ray flare of the radio galaxy NGC 1275: VERITAS and Multiwavelength Observations}

\author[0000-0002-2028-9230]{A.~Acharyya}\affiliation{CP3-Origins, University of Southern Denmark, Campusvej 55, 5230 Odense M, Denmark}
\author{A.~Archer}\affiliation{Department of Physics and Astronomy, DePauw University, Greencastle, IN 46135-0037, USA}
\author[0000-0002-3886-3739]{P.~Bangale}\affiliation{Department of Physics, Temple University, Philadelphia, PA 19122, USA}
\author[0000-0002-9675-7328]{J.~T.~Bartkoske}\affiliation{Department of Physics and Astronomy, University of Utah, Salt Lake City, UT 84112, USA}
\author[0000-0003-2098-170X]{W.~Benbow}\affiliation{Center for Astrophysics $|$ Harvard \& Smithsonian, Cambridge, MA 02138, USA}
\author[0009-0001-5719-936X]{Y.~Chen}\affiliation{Department of Physics and Astronomy, University of California, Los Angeles, CA 90095, USA}
\author{J.~L.~Christiansen}\affiliation{Physics Department, California Polytechnic State University, San Luis Obispo, CA 94307, USA}
\author{A.~J.~Chromey}\affiliation{Center for Astrophysics $|$ Harvard \& Smithsonian, Cambridge, MA 02138, USA}
\author[0000-0003-1716-4119]{A.~Duerr}\affiliation{Department of Physics and Astronomy, University of Utah, Salt Lake City, UT 84112, USA}
\author[0000-0002-1853-863X]{M.~Errando}\affiliation{Department of Physics, Washington University, St. Louis, MO 63130, USA}
\author{M.~Escobar~Godoy}\affiliation{Santa Cruz Institute for Particle Physics and Department of Physics, University of California, Santa Cruz, CA 95064, USA}
\author[0000-0002-5068-7344]{A.~Falcone}\affiliation{Department of Astronomy and Astrophysics, 525 Davey Lab, Pennsylvania State University, University Park, PA 16802, USA}
\author{S.~Feldman}\affiliation{Department of Physics and Astronomy, University of California, Los Angeles, CA 90095, USA}
\author[0000-0001-6674-4238]{Q.~Feng}\affiliation{Department of Physics and Astronomy, University of Utah, Salt Lake City, UT 84112, USA}
\author[0000-0002-2636-4756]{S.~Filbert}\affiliation{Department of Physics and Astronomy, University of Utah, Salt Lake City, UT 84112, USA}
\author[0000-0002-1067-8558]{L.~Fortson}\affiliation{School of Physics and Astronomy, University of Minnesota, Minneapolis, MN 55455, USA}
\author[0000-0003-1614-1273]{A.~Furniss}\affiliation{Santa Cruz Institute for Particle Physics and Department of Physics, University of California, Santa Cruz, CA 95064, USA}
\author[0000-0002-0109-4737]{W.~Hanlon}\affiliation{Center for Astrophysics $|$ Harvard \& Smithsonian, Cambridge, MA 02138, USA}
\author[0000-0003-3878-1677]{O.~Hervet}\email{ohervet@ucsc.edu}\affiliation{Santa Cruz Institute for Particle Physics and Department of Physics, University of California, Santa Cruz, CA 95064, USA}
\author[0000-0001-6951-2299]{C.~E.~Hinrichs}\affiliation{Center for Astrophysics $|$ Harvard \& Smithsonian, Cambridge, MA 02138, USA and Department of Physics and Astronomy, Dartmouth College, 6127 Wilder Laboratory, Hanover, NH 03755 USA}
\author[0000-0002-6833-0474]{J.~Holder}\affiliation{Department of Physics and Astronomy and the Bartol Research Institute, University of Delaware, Newark, DE 19716, USA}
\author{Z.~Hughes}\affiliation{Department of Physics, Washington University, St. Louis, MO 63130, USA}
\author{M.~Iskakova}\affiliation{Department of Physics, Washington University, St. Louis, MO 63130, USA}
\author[0000-0002-1089-1754]{W.~Jin}\affiliation{Department of Physics and Astronomy, University of California, Los Angeles, CA 90095, USA}
\author[0009-0008-2688-0815]{M.~N.~Johnson}\affiliation{Santa Cruz Institute for Particle Physics and Department of Physics, University of California, Santa Cruz, CA 95064, USA}
\author[0000-0002-3638-0637]{P.~Kaaret}\affiliation{Department of Physics and Astronomy, University of Iowa, Van Allen Hall, Iowa City, IA 52242, USA}
\author{M.~Kertzman}\affiliation{Department of Physics and Astronomy, DePauw University, Greencastle, IN 46135-0037, USA}
\author{M.~Kherlakian}\affiliation{Fakult\"at f\"ur Physik \& Astronomie, Ruhr-Universit\"at Bochum, D-44780 Bochum, Germany}
\author[0000-0003-4785-0101]{D.~Kieda}\affiliation{Department of Physics and Astronomy, University of Utah, Salt Lake City, UT 84112, USA}
\author[0000-0002-4260-9186]{T.~K.~Kleiner}\affiliation{DESY, Platanenallee 6, 15738 Zeuthen, Germany}
\author[0000-0002-4289-7106]{N.~Korzoun}\affiliation{Department of Physics and Astronomy and the Bartol Research Institute, University of Delaware, Newark, DE 19716, USA}
\author{F.~Krennrich}\affiliation{Department of Physics and Astronomy, Iowa State University, Ames, IA 50011, USA}
\author{S.~Kundu}\affiliation{Department of Physics and Astronomy, University of Alabama, Tuscaloosa, AL 35487, USA}
\author[0000-0003-4641-4201]{M.~J.~Lang}\affiliation{School of Natural Sciences, University of Galway, University Road, Galway, H91 TK33, Ireland}
\author[0000-0003-3802-1619]{M.~Lundy}\affiliation{Physics Department, McGill University, Montreal, QC H3A 2T8, Canada}
\author[0000-0001-9868-4700]{G.~Maier}\affiliation{DESY, Platanenallee 6, 15738 Zeuthen, Germany}
\author{E.~Meyer}\affiliation{Department of Physics, University of Maryland, Baltimore County, Baltimore MD 21250, USA}
\author[0000-0002-1499-2667]{P.~Moriarty}\affiliation{School of Natural Sciences, University of Galway, University Road, Galway, H91 TK33, Ireland}
\author[0000-0002-3223-0754]{R.~Mukherjee}\affiliation{Department of Physics and Astronomy, Barnard College, Columbia University, NY 10027, USA}
\author[0000-0002-6121-3443]{W.~Ning}\affiliation{Department of Physics and Astronomy, University of California, Los Angeles, CA 90095, USA}
\author{M.~Ohishi}\affiliation{Institute for Cosmic Ray Research, University of Tokyo, 5-1-5, Kashiwa-no-ha, Kashiwa, Chiba 277-8582, Japan}
\author[0000-0002-4837-5253]{R.~A.~Ong}\affiliation{Department of Physics and Astronomy, University of California, Los Angeles, CA 90095, USA}
\author[0000-0003-3820-0887]{A.~Pandey}\affiliation{Department of Physics and Astronomy, University of Utah, Salt Lake City, UT 84112, USA}
\author[0000-0002-4131-655X]{J.~Escudero Pedrosa}\affiliation{Center for Astrophysics $|$ Harvard \& Smithsonian, Cambridge, MA 02138, USA}
\author[0000-0001-7861-1707]{M.~Pohl}\affiliation{Institute of Physics and Astronomy, University of Potsdam, 14476 Potsdam-Golm, Germany and DESY, Platanenallee 6, 15738 Zeuthen, Germany}
\author[0000-0002-0529-1973]{E.~Pueschel}\affiliation{Fakult\"at f\"ur Physik \& Astronomie, Ruhr-Universit\"at Bochum, D-44780 Bochum, Germany}
\author[0000-0002-4855-2694]{J.~Quinn}\affiliation{School of Physics, University College Dublin, Belfield, Dublin 4, Ireland}
\author[0000-0002-5104-5263]{P.~L.~Rabinowitz}\affiliation{Department of Physics, Washington University, St. Louis, MO 63130, USA}
\author[0000-0002-5351-3323]{K.~Ragan}\affiliation{Physics Department, McGill University, Montreal, QC H3A 2T8, Canada}
\author{P.~T.~Reynolds}\affiliation{Department of Physical Sciences, Munster Technological University, Bishopstown, Cork, T12 P928, Ireland}
\author[0000-0002-7523-7366]{D.~Ribeiro}\affiliation{School of Physics and Astronomy, University of Minnesota, Minneapolis, MN 55455, USA}
\author{E.~Roache}\affiliation{Center for Astrophysics $|$ Harvard \& Smithsonian, Cambridge, MA 02138, USA}
\author[0000-0001-7483-4348]{C.~Rulten}\email{cameron.b.rulten@durham.ac.uk, cbrulten@umn.edu}\affiliation{School of Physics and Astronomy, University of Minnesota, Minneapolis, MN 55455, USA}\affiliation{Centre for Advanced Instrumentation, Department of Physics, University of Durham, South Road, Durham DH1 3LE, UK}
\author[0000-0003-1387-8915]{I.~Sadeh}\affiliation{DESY, Platanenallee 6, 15738 Zeuthen, Germany}
\author[0000-0002-3171-5039]{L.~Saha}\affiliation{Center for Astrophysics $|$ Harvard \& Smithsonian, Cambridge, MA 02138, USA}
\author{M.~Santander}\affiliation{Department of Physics and Astronomy, University of Alabama, Tuscaloosa, AL 35487, USA}
\author{G.~H.~Sembroski}\affiliation{Department of Physics and Astronomy, Purdue University, West Lafayette, IN 47907, USA}
\author[0000-0002-9856-989X]{R.~Shang}\affiliation{Department of Physics and Astronomy, Barnard College, Columbia University, NY 10027, USA}
\author[0000-0003-3407-9936]{M.~Splettstoesser}\affiliation{Santa Cruz Institute for Particle Physics and Department of Physics, University of California, Santa Cruz, CA 95064, USA}
\author[0000-0002-9852-2469]{D.~Tak}\affiliation{SNU Astronomy Research Center, Seoul National University, Seoul 08826, Republic of Korea.}
\author{A.~K.~Talluri}\affiliation{School of Physics and Astronomy, University of Minnesota, Minneapolis, MN 55455, USA}
\author{J.~V.~Tucci}\affiliation{Department of Physics, Indiana University Indianapolis, Indianapolis, Indiana 46202, USA}
\author[0000-0002-8090-6528]{J.~Valverde}\affiliation{Department of Physics, University of Maryland, Baltimore County, Baltimore MD 21250, USA and NASA GSFC, Greenbelt, MD 20771, USA}
\author{V.~V.~Vassiliev}\affiliation{Department of Physics and Astronomy, University of California, Los Angeles, CA 90095, USA}
\author[0000-0003-2740-9714]{D.~A.~Williams}\affiliation{Santa Cruz Institute for Particle Physics and Department of Physics, University of California, Santa Cruz, CA 95064, USA}
\author[0000-0002-2730-2733]{S.~L.~Wong}\affiliation{Physics Department, McGill University, Montreal, QC H3A 2T8, Canada}
\author{T.~Yoshikoshi}\affiliation{Institute for Cosmic Ray Research, University of Tokyo, 5-1-5, Kashiwa-no-ha, Kashiwa, Chiba 277-8582, Japan}

\collaboration{(VERITAS Collaboration)}

\author{P. S. Smith}
\affiliation{Steward Observatory, University of Arizona, Tucson, AZ 85721 USA}
\author{J. Kataoka}
\affiliation{Research Institute for Science and Engineering, Waseda University, 3-4-1, Okubo, Shinjuku, Tokyo, 169-8555, Japan}


\accepted{for publication in ApJ, May 18, 2026}
\begin{abstract}
The radio galaxy NGC~1275 is the Brightest Cluster 
Galaxy in the Perseus cluster. It is well-studied across all wavebands, including Very High Energy (VHE; $\mathrm{E \geqslant 100GeV})$ $\gamma$-rays, and with radio observations over the last 20 years tracking an unusual radio component, ``C3". NGC~1275 was observed in an exceptional VHE flaring state between 2016 December 31 and 2017 January 3. The flare peak reached $\sim$1.5 Crab units as measured by the MAGIC observatory. We report on the observations of NGC~1275 conducted by VERITAS and multi-wavelength data collected during this flaring state, and for context, data taken between 2009 and 2017 inclusive. VERITAS detected the declining state of the flare on 2017 January 2 (MJD 57755) and 3 (MJD 57756) at an average flux state of 0.5 Crab units. VERITAS spectra show an overall long-term trend of harder-when-brighter. During the flare, the $\gamma$-ray spectrum obtained from the combined \textit{Fermi}-LAT, MAGIC, and VERITAS observations, changes from a power law with an exponential cut-off on January 1 to a log-parabola on January 2. To study the evolution of the flare in more detail, multi-band spectral energy distributions (SEDs) were constructed for the nights of 2017 January 1 and 2 corresponding to the shift from the peak to the decline of the flare. A blob-in-jet modeling of the SEDs results in support for a two-component model with a $\theta = 10^{\circ}$ jet angle to the line of sight and the $\gamma$-ray emission zone located in the vicinity of the C3 radio component.

\end{abstract}

\section{Introduction}
\label{sec-introduction}

The unified model of active galactic nuclei (AGN) \citep{Urry:1995aa} suggests that radio-loud AGN comprise two major categories based on the orientation of jets with respect to the observer's line of sight (LOS). The jets of radio galaxies are oriented at large inclination angles, 
while blazars have jets beamed nearly along the LOS. Unlike the large Doppler factors ($\mathrm{\delta \ssim 10}$) typically seen for blazars \citep{DermerGiebels:2016}, the non-thermal radiation emitted by radio galaxies is therefore only modestly beamed with smaller Doppler factors ($\delta \ssim 1-5$) seen \citep{Ghisellini:1993aa,Urry:1995aa, Giovannini-2001-ApJ,Hovatta:2009aa,Rieger:2017}. 
Thus, despite radio-loud AGN accounting for the majority of observed extragalactic sources at Very High Energies (VHE, $\mathrm{E \geqslant 100 GeV})$, the proportion of those classified as radio galaxies is small ($\mathrm{\leqslant 2\%}$) 
\citep{Wakely:2008}. The five known TeV-emitting radio galaxies (\gName{M}{87} \citep{Acciari-2008-ApJ, Abramowski_2012}, \gName{Cen}{A} \citep{Aharonian_2009}, \gName{NGC}{1275} \citep{Aleksic:2012aa, VERITAS:2017ATEL9931}, \gName{IC}{310} \citep{Aleksic-2010-ApJ} and \gName{3C}{264} \citep{Archer_2020}) are typically only weakly detected at a few percent of the Crab Nebula flux \citep{Rieger:2017}.


The broad band spectral energy distributions (SEDs) of radio galaxies are expected to be similar to the two-peaked distributions for blazars with the radio/optical/soft X-ray data tracing the low-energy synchrotron peak and the hard X-ray/MeV, GeV \& TeV \gray{} data forming the high-energy inverse Compton (IC) peak. In general, the most effective model to describe the high-energy \gray{} emission is the Synchrotron Self-Compton (SSC) model where the low-energy synchrotron peak photons are upscattered via inverse Compton scattering to higher energies. If a population of additional low energy seed photons is available (e.g., from the dusty torus or other components of the AGN or jet system), the Compton peak luminosity can be greater than the synchrotron peak luminosity yielding a Compton Dominance (CD) $> 1$.

Despite their similarities to blazar SEDs, radio galaxy SEDs can be challenging to reproduce with one-zone emission models, which usually work well for blazars. This is due to the expected lower Doppler factors for radio galaxies as well as the relatively large frequency separations between the synchrotron and inverse Compton peaks of their SEDs. In particular, the jet viewing angle is an important constraint on the Doppler factor in SED fitting of radio galaxies.

While extreme VHE flaring states from radio galaxies have rarely been observed, exceptions can provide observational constraints on the location of the VHE emission, the accretion processes, as well as the jet acceleration mechanisms of radio-loud AGN.  For example, \gName{M}{87} exhibited VHE flares with day-scale variability in 2005, 2008 and 2010 \citep{Aharonian-2006-Sci, Acciari-2008-ApJ, VERITAS-M87-2009-Sci, Aliu-2012-ApJ, Abramowski_2012} pointing to the \gray{} emission region being located at the radio galaxy core in two of the three recorded VHE flares and possibly in the HST-1 knot of the jet in the third VHE flare \citep{Cheung-2007-ApJ, Harris-2008-ASPC, Acciari2010M87}.
The great VHE 2012 flare of \gName{IC}{310} showed a 4.8 minute doubling time constraining the VHE emission region to be smaller than the scale of the black hole \citep{Aleksic-2014-Sci}.

\gName{NGC}{1275} (R.A. = 03h19m48.1s, Decl. = +41d30m42s; EquJ2000.0) is the Brightest Cluster Galaxy of the Perseus Cluster and was discovered as a VHE \gray{} emitter in 2010 by MAGIC \citep{Aleksic:2012aa}. With a redshift of $z=0.0176$ it is the third-nearest radio galaxy to be detected in VHE \grays{}. 
Over the Fall of 2016, highly variable VHE \gray{} emission was detected from \gName{NGC}{1275} by both MAGIC and VERITAS, culminating in an extreme flaring state detected by MAGIC equivalent to $\sim$150\% of the Crab Nebula flux on the night of \decthirtyfirst{} to \janfirst{}  \citep{MAGIC-2018-AA}. Upon receiving a direct alert from MAGIC observers, VERITAS carried out follow-up observations over the nights of \jansecond{} and 3. 

The \citet{MAGIC-2018-AA} publication describes the MAGIC analysis of this ``New Year's Flare" with 
the shortest variability timescale measured as $611 \pm  101$ minutes occurring over the peak flux period (\janfirst{}). A power law with exponential cutoff provided the best fit to the VHE data from the peak of the flare with a flux normalization of $(16.1 \pm 2.3) \times 10^{-10} \, \mathrm{TeV^{-1} cm^{-2} s^{-1}}$, a spectral index of $\Gamma = 2.11 \pm 0.14$, and a cutoff energy of $E_{c}=0.56 \pm 0.11 \, \mathrm{TeV}$. A fit combining \fermi{} and MAGIC data confirmed the MAGIC-only result of a best fit by a power-law with exponential cutoff. No further SED analysis was done. 


This paper describes VERITAS observations of \gName{NGC}{1275} spanning an eight-year period from 2009 January 15 (MJD 54846) to 2017 February 26 (MJD 57810) inclusive,
with a focus on the exceptional TeV flare occurring \decthirtyfirst{}, through \janthird{}. We present analysis and discussion of multiwavelength observations recorded during the 2016/17 observing season with SEDs produced for both the nights of \janfirst{} and 2. 

As \gName{NGC}{1275} has been well-studied across multiple wavebands for decades, it is important particularly in the context of our study to review the evolution of the source in radio frequencies. Extensive VLBI observations show a parsec-scale jet and a faint counter jet \citep{Walker:1994ApJ}. From approximately 2005 to 2018 the jet displayed three key radio components: the radio core (hereafter C1), a slow-moving diffuse emission component (hereafter C2), and a hotspot (hereafter C3) associated with an outburst recorded in 2005, and linked to much of the multiwavelength activity seen in \gName{NGC}{1275} during this full time span.
A multi-epoch VLBI study of \gName{NGC}{1275} conducted between 2006 and 2009 confirmed that the 2005 outburst was associated with the inner-jet ($\mathrm{\ssim 1 \;pc}$) and the emergence of a new jet component C3 with a projected speed of 0.23c \citep{Nagai:2010PASJ}. This  was confirmed by \citet{Jorstad:2017ApJ} who reported a speed of 0.2c for the C3 component. 

Most AGN exhibit ridge-brightened jets, while limb-brightened jets -- where there is a high intensity ratio between the jet's sheath and the spine -- are rare and are observed only in a few nearby radio galaxies \citep{Giovannini-2018-NatAs}. A key common feature of the VHE-detected radio galaxies appears to be the presence of a limb-brightened jet \citep{Rulten-2022-Galaxies}.
A 2013 space-VLBI study of \gName{NGC}{1275} provides radio imagery of the central parsec showing a limb-brightened jet connecting the core and the C3 component described above \citep{Giovannini-2018-NatAs}, confirmed by \citet{Jorstad:2017ApJ}. The limb-brightening is also confirmed in a 43 GHz VLBA study \citep{Nagai-2014-ApJ} in which the authors associate the limb-brightening with the 2005 radio outburst, and highlight that prior observations of similar resolution conducted in the 1990s show \gName{NGC}{1275}'s jet to be ridge-brightened.  
Furthermore, \citet{Nagai-2014-ApJ} suggest that the change in jet morphology coincides with an increase in the \gray{} flux, a conclusion obtained by comparing the \fermi{} photon flux with the upper-limit from EGRET, which did not obtain a statistically significant detection from \gName{NGC}{1275} \citep{Lin-EGRET-1993-ApJ}. 
The limb-brightened radio images of the jet provide potential observational evidence of a transverse velocity structure in the jet \citep{Komissarov-1990-SvAL}, and possibly point toward \gray{} emission resulting from a decelerating flow in a structured jet \citep{Georganopoulos-2003-ApJ, Ghisellini:2005aa}. In structured jets, different parts of the jet move at different speeds. For example, a fast-moving spine in the center and a slow-moving sheath near the edge. When particles scatter across these layers of different velocities, they undergo systematic energy gains called shear acceleration. In the low-power jets of FR-I type radio galaxies, particles can gain energy through shear acceleration by repeated crossings of the interface between the faster and slower regions \citep{Stawarz-2002-ApJ, Rieger-2004-ApJ, Tavecchio_2008, Laing-2014-MNRAS}.


Using archival 43 GHz VLBA observations taken in 2015 -- 2016 from the Boston University Blazar Program, \citet{Nagai:2017ApJ} report the detection of significant polarized emission at the C3 hotspot of the inner-jet located approximately 1 pc south of the radio core C1. In addition, \citet{Nagai:2017ApJ} report that the location of the C3 hotspot changed, from its southerly position on the western limb \citep{Nagai-2014-ApJ}, to a new southerly position on the eastern limb suggesting a possible interaction of the jet with the ambient medium. Indeed, using multi-epoch 43 GHz observations from the KVN and VERA array, \citet{Kino:2018ApJ} provide evidence of this interaction by discovering a 0.4 mas (angular) / 0.14 parsec (physical) flip of the C3 component, after which it spent a few months wobbling at the same location before progressing further southward again.  This type of behavior can be characterized as a ``frustration" of the jet \citep{Zensus1997}. In their study, \citet{Kino:2018ApJ} suggested that the directional flip was in good agreement with simulations of jets inside clumpy ambient media. In a follow-up study, \citet{Kino2021ApJ} suggest that the C3 hotspot underwent a year long frustration in 2017, which is indicative of a strong collision between the jet and a dense compact cloud. This frustration of the jet coincides with the VHE flare detected by MAGIC, VERITAS and others, the subject of this paper. Since this period of frustration, the C3 component subsequently continued to move southward and indeed began to break up and distort with its radio flux becoming fainter \citep{Kino2021ApJ}. A recent long-term light curve study of \gName{NGC}{1275} using data from a range of radio instruments and \fermi{} found a positive correlation between the \gray{} and radio data, with a tentative association of \gray{} flares with the ejection of radio features \citep{Paraschos2023}.



Section \ref{sec:observations} of this paper presents the details of the multiwavelength observational data obtained from a number of different instruments across radio, optical, X-ray, GeV and TeV \grays{}. In Section \ref{sec:results} we include the results of variability and spectral analysis for some of these multiwavelength data. The long-term observations provide context for the interpretation of the evolution of near-simultaneous SEDs constructed for observations on \janfirst{} and 2. Finally, in Section \ref{sec:discussion} we discuss the possible physical processes for producing such a flaring outburst of \grays{}, and using these multiwavelength data we attempt to model the process across the broad-band electromagnetic spectrum. We make concluding remarks in Section \ref{sec:conclusion}.
Throughout this paper, a flat $\Lambda$CDM cosmology is used, with $H_0$ = 69.6 km s$^{-1}$ Mpc$^{-1}$ \citep{Bennett_2014}.

\section{Observations}
\label{sec:observations}

\begin{figure*}[hp]
\centering
{\includegraphics[width=0.85\textwidth]{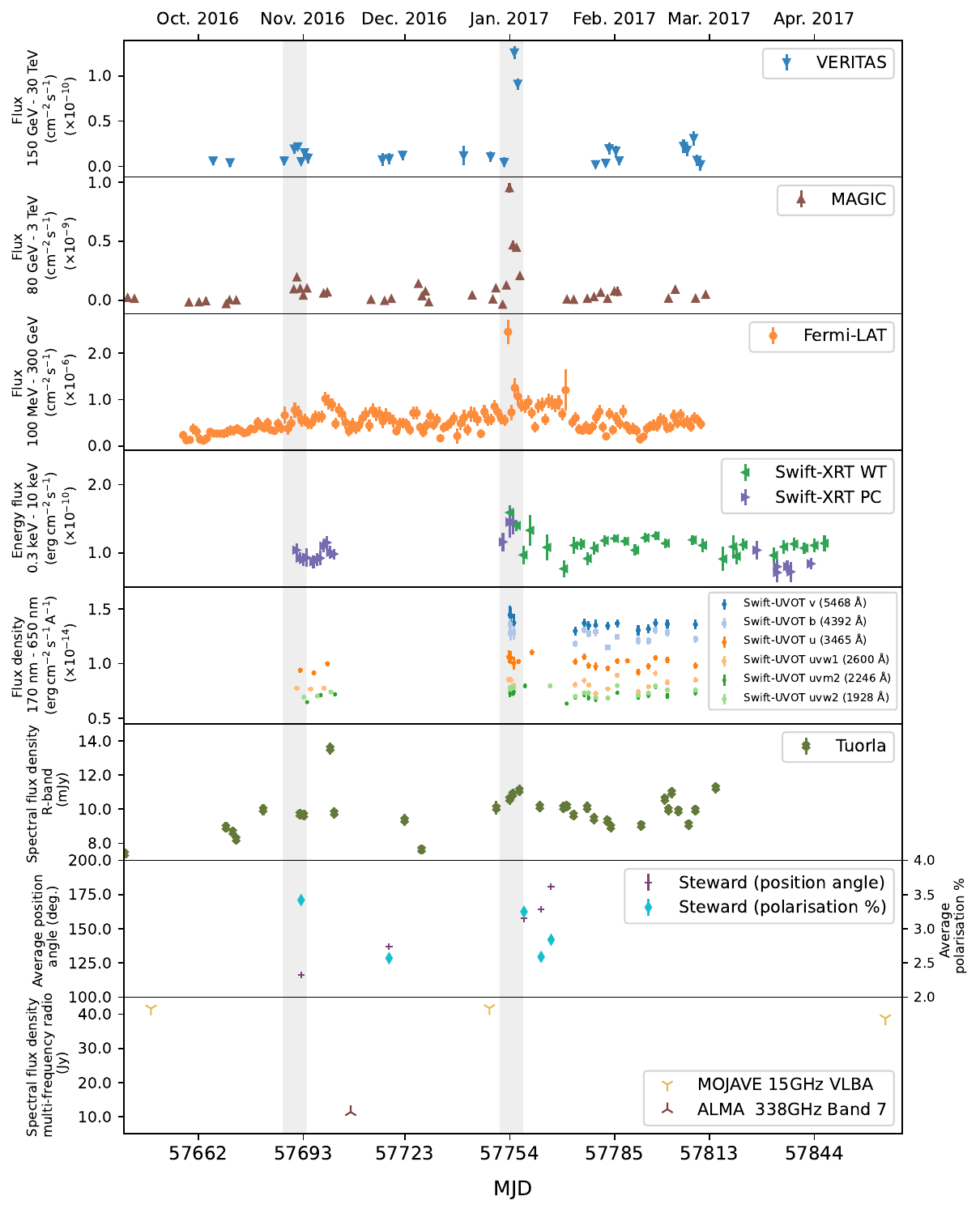}}
\caption{The daily-binned multiwavelength light curve of \gName{NGC}{1275} for the VERITAS observing season 2016/17. The light curves include data recorded with VERITAS (first / top panel), MAGIC \citep{MAGIC-2018-AA} (second panel), \fermi{} (third panel), \swiftXRT{} (fourth panel), \swiftUVOT{} (fifth panel), Tuorla \citep{MAGIC-2018-AA} (sixth panel), Steward Optical Polarization percentage and  Position Angle  (seventh panel), and finally MOJAVE \citep{Lister-2018-ApJS} and ALMA (eighth / bottom panel). The \swiftXRT{} data includes observations for both the Windowed Timing mode (green points) and the Photon Counting mode (purple points). The gray vertical bands each highlight a 7-day period centered on \octtwentyninth{} and \janfirst{} respectively.}
\label{figure:multiwavelength_daily_lightcurve}
\end{figure*}

Figure \ref{figure:multiwavelength_daily_lightcurve} shows the daily-binned light curves for the range of multiwavelength data obtained for the 2016/17 observing season conducted by the Very Energetic Radiation Imaging Telescope Array System (VERITAS). The gray vertical bands each highlight a 7-day period centered on \octtwentyninth{} and \janfirst{} respectively, when \gName{NGC}{1275} was observed to be in a flaring state at TeV energies. The following subsections provide details on the various instruments and their observations used in this work.

\subsection{VERITAS}\label{subsec:veritas}
The VERITAS array of four $\mathrm{12\;m}$ imaging atmospheric Cherenkov telescopes (IACTs) located in southern Arizona at the Fred Lawrence Whipple Observatory \citep{Holder:2011ICRC}, is sensitive to \grays{} in the very high-energy (VHE) range from $\mathrm{100 \,GeV \; to \; 30\, TeV}$, with an energy resolution of $\mathrm{\ssim15 \%}$ and an energy-dependent angular resolution (68\% containment radius) of $\mathrm{\ssim0.1^{\circ}}$ at $\mathrm{1\, TeV}$.

VERITAS observations of \gName{NGC}{1275} taken 2012 December -- 2013 February ($\mathrm{\ssim16 \, hours}$) and 2013 October -- 2013 November ($\mathrm{\ssim16 \, hours}$), subsequent to the MAGIC discovery of the source, yield a detection with a statistical significance of $\mathrm{\ssim7\sigma}$ with an average flux state at the $\ssim1\%$ Crab flux level \citep{Benbow:2015ICRC}. During a  monitoring campaign, a snapshot taken on \octtwentyninth{} (MJD 57690) resulted in a flux estimated 
as $\mathrm{\ssim15\%}$ Crab, a 5-fold increase on previous observations \citep{VERITAS:2016ATEL9690}. On the same day, observations conducted independently by the MAGIC telescopes also yielded a detection of \gName{NGC}{1275} at a similar flux level \citep{MAGIC:2016ATEL9689}.
At the time, this flare was the highest ever flux state seen for \gName{NGC}{1275} and indeed any radio galaxy detected at energies above $\mathrm{\ssim100\, GeV}$.


VERITAS continued to monitor the central region of the Perseus cluster with intermittent $\mathrm{30\, min.}$ snapshot observations. One such snapshot conducted on \decthirtieth{} (MJD 57752) resulted in no statistically significant detection of VHE emission at the location of \gName{NGC}{1275}.
Then VERITAS received a direct communication from the MAGIC observing team of a giant flare from \gName{NGC}{1275} that they had detected during observations between \decthirtyfirst{} and \janfirst{} \citep{MAGIC:2017ATEL9929}. VERITAS was able to conduct follow-up observations of \gName{NGC}{1275} on \jansecond{} (MJD 57755) and 3 (MJD 57756). A total of $\mathrm{2.2\, hours}$ (MJD 57755) and $\mathrm{1.3\, hours}$ (MJD 57756) data were recorded after quality selection, respectively.
\gName{NGC}{1275} was detected with a statistical significance of $\mathrm{\ssim31\sigma}$ (MJD 57755) and $\mathrm{\ssim22\sigma}$ (MJD 57756) recording an average flux state at the $\mathrm{50\%}$ Crab flux level. 

All VERITAS data were analyzed and compared for consistency using the independent VERITAS analysis software packages VEGAS and EventDisplay \citep{Cogan:2008ICRC, Maier2017ICRC}. The events are reconstructed using a standard Hillas-style \citep{Hillas:1985-ICRC} analysis, and statistical significance is calculated using the  \citet{LiMa:1983apj} (Page 320, Equation 17) method generalized for data subsets with different $\mathrm{\alpha}$, the ratio of the on-source time to the off-source time, values \citep{Generalized-LiMa-2004aa} (Page 531). Observations were conducted in ``wobble'' mode, which enables the \gray{} background in the source region to be simultaneously estimated from events in the field of view using the reflected-region method \citep{Berge:2007aa}. The event selection (e.g. see \citet{Acciari:2008aa}) was done using ``soft'' cuts, which yields an energy threshold obtained from simulations of $\mathrm{\ssim110\, GeV}$ for observations before 2012 September, and $\mathrm{\ssim150\, GeV}$ after 2012 September.

The VERITAS observations presented here comprise both \gName{NGC}{1275} and \gName{IC}{310} pointings. While the reconstruction algorithms were not optimized for the multiple pointings for the data set, this did not have a significant effect on the results. 

\subsection{\fermi{}}
\label{subsec:fermi}

To provide context for the VHE observations conducted by VERITAS, data from the Fermi Large Area Telescope \citep[\Fermi-LAT;][]{Atwood:2009apj} were analyzed. In summary, we include all photons with energies between 0.1 and 300 GeV within a $\mathrm{15^{\circ}}$ circular region of interest (ROI) centred on NGC\,1275 (R.A. = 49\fdg95, Decl.~= +45\fdg51; J2000.0). All photons were selected from \Fermi-LAT sky-survey observations following the \texttt{PASS8} data analysis criteria. In addition we used a zenith angle cut of 90\deg\ to reduce contamination from Earth limb \grays{} as well as the additional quality cuts: \texttt{(DATA\_QUAL>0)\&\&(LAT\_CONFIG==1)} and \texttt{abs}(\texttt{rock\_angle})$\mathrm{\,<52}$. Our analysis was conducted using the \Fermi-LAT \texttt{Fermitools} v1.0.1 and the open-source Python package \texttt{Fermipy} v.0.17.4 \citep{Wood-2017-ICRC}. We used a binned maximum-likelihood analysis \citep{Mattox:1996apj} applying the \texttt{P8R3$\_$SOURCE$\_$V2} instrument response functions. For the background, sources in the first data release of the 4FGL catalog \citep{Fermi-4FGL-2020-ApJS} within $\mathrm{20^{\circ}}$ of the ROI center were considered and the Galactic ({\tt gll$\_$iem$\_$v07.fits}) and isotropic diffuse ({\tt iso$\_$P8R3$\_$SOURCE$\_$V2$\_$v1.txt}) templates provided with \texttt{Fermitools} were used.

Our \fermi{} analysis spans the following periods: for the daily-binned light curve (Figure \ref{figure:multiwavelength_daily_lightcurve}) we consider MJD 57640 -- MJD 57810 where \gName{NGC}{1275} is detected with a test statistic (TS)\footnote{The test statistic is defined as twice the difference between the log-likelihoods of two different models, $2 (\log \mathcal{L}_{0} - \log \mathcal{L}_{1})$, where $\mathcal{L}_{0}$ and $\mathcal{L}_{1}$ are the likelihoods of individual model fits \citep{Mattox:1996apj}.} value of TS = 3869 (equivalent to a statistical significance of $\ssim$62$\sigma$), and for flux variability studies (Figure \ref{figure:fermiLAT_variability}) we consider MJD 57751 -- MJD 57760. For the combined GeV and TeV spectral analysis (see Section \ref{subsec:combined_GeV_TeV_spectral_analysis}) we consider \fermi{} mission elapsed time (MET) 504900004 - 505008004 seconds for \decthirtyfirst{}/\janfirst{}, and MET 505008005 - 505094404 seconds for \jansecond{}. For the SED (Figure \ref{figure:sed_modelling}) the data spans a period determined by the Bayesian Block binning scheme highlighted in Section \ref{subsec:fermi_variability}. For all \fermi{} data analyzed we use the \texttt{Fermipy} \texttt{find\_sources} algorithm to search for additional sources within the ROI where $\mathrm{TS\geqslant}25$ even though we only consider periods shorter than that covered by the 4FGL catalog.

\subsection{\swiftXRT{} and \swiftUVOT{}}
\label{subsec:swift}

Contemporaneous observations of \gName{NGC}{1275} between MJD 57691 and MJD 57809 with the \textit{Swift} X-ray Telescope (XRT) and the UltraViolet/Optical Telescope (UVOT) were also analyzed.

The Swift-XRT archival data between MJD 57691 and MJD 57844 
included observations in both windowed timing (WT) and photon counting (PC) modes \citep{Burrows-2005-SSR}. Data analysis was performed with tools provided by \texttt{HEASOFT} v6.32.1, and spectral fitting with the \texttt{Sherpa} package 
provided by \texttt{ciao} v4.16. 
Pileup was found in the central four pixels ($\ssim$9") of the PC mode observations. Therefore, an annular region (9"-25") centered on \gName{NGC}{1275} was used as the source region. 
Since no pileup was found in the WT mode observations, a circular source extraction region of radius $\ssim$25" was used. To account for the hot thermal cluster emission, annular background extraction regions (47"-71") were utilized for both PC and WT mode observations. 
The XRT spectra over 0.3-10.0 keV were fitted to the model phabs*(apec+zphabs*pl) to account for the Galactic absorption, thermal emission from the surrounding hot cluster gas, the local nH absorption, and the AGN emission. The Galactic hydrogen column density was frozen to $\mathrm{1.35 \times 10^{21} \, cm^{-2}}$ and the local absorption was frozen to $\mathrm{1.69 \times 10^{21} \, cm^{-2}}$. 
After a global fit with the metallicity abundance left free to vary, we obtained a value of 0.8 Solar. For subsequent fits the abundance was frozen at this value. A joint fitting to obtain a single background fit for all epochs in the PC and WT modes resulted in an `apec' temperature and a normalization constant for each mode.  With these values held frozen, all spectra were fitted to the model with the power law parameters set free. The best-fitting power-law index and normalization constant, as well as the integral flux over 0.3-10 keV, are listed in Table \ref{table:xrt_obervations} located in the Appendix \ref{Appendix::spec}, and shown in Figure \ref{figure:multiwavelength_daily_lightcurve} for the \swiftXRT{} light curve.

The \swiftUVOT{} data were analyzed with software provided by the \textit{Swift} Science Center\footnote{https://swift.gsfc.nasa.gov}. Specifically, the \texttt{uvotsource} tool was used to perform aperture photometry on individual bands (uvw1, uvm2, uvw2, u, b, v) and measure the source flux density in each exposure. A circular region with 5$''$ radius centered on \gName{NGC}{1275} coordinates was used for the source region, while a background region was defined within a 20$''$-radius circle located in a nearby source-free area. Foreground Galactic extinction correction was applied with E(B-V) = 0.163~\citep{Schlafly:2011apj} and extinction coefficients were derived for individual bands using the~\citet{Fitzpatrick:1999PASP} reddening law. The effective wavelength for each band was taken from~\citet{Breeveld:2011aipc}.
To generate the UVOT SED points (included in Figure~\ref{figure:sed_modelling}), observations between MJD 57752 and MJD 57756 were first combined before following the above procedure. The light curves shown in Figure~\ref{figure:multiwavelength_daily_lightcurve} were generated by applying the same procedure on individual observations.

\subsection{Tuorla}
\label{subsec:tuorla}
\gName{NGC}{1275} R-band optical data is obtained from observations with the KVA $\mathrm{35\; cm}$ telescope located in La Palma as part of the Tuorla blazar monitoring program. A description of the data reduction methodology is given in \cite{MAGIC-2018-AA}. In summary, all the optical fluxes shown in Figure \ref{figure:multiwavelength_daily_lightcurve} are host galaxy and line contamination subtracted (12 mJy, see details in \cite{Aleksic:2014ab}) and the fluxes were also corrected for Galactic extinction with $\mathrm{A_R = 0.354}$ \citep{Schlafly:2011apj}.

\subsection{Steward Observatory}
\label{subsec:steward}
Spectral polarimetry data for  \gName{NGC}{1275} was obtained using the Steward Observatory 1.54 m Kuiper Telescope with the SPOL instrument \citep{SPOL1992}. 
The source was observed on 5 nights from \octthirtyfirst{} to 2017 January 13. All observations consisted of 2 complete polarization measurements averaged together. Each measurement was obtained by combining 45 second exposures at 16 positions of a half-waveplate and derived by binning data within 5000-7000 Angstroms on each night.
The individual measurements are generally consistent with each other, but there is some evidence for a small amount of polarization variability seen in \gName{NGC}{1275} between the two nightly observations made on 2017 January 10 and 13 (see Figure \ref{figure:multiwavelength_daily_lightcurve}).  Generally, large changes in optical polarization are observed between nights.

The turbulent nature of jets typically translates to ``jitter" or high variability in the polarization angle over time. However, the optical linear polarization position angle (\textit{t}) shows a systematic increase during the period of observation (also shown on Figure \ref{figure:multiwavelength_daily_lightcurve}).  While the sampling is very sparse,  \textit{t} rotates from $116^{\circ}$ to $181^ {\circ}$ from \octthirtyfirst{} to 2017 January 13.  Such a consistent change in \textit{t} over 5 observations is actually hard to come by for blazars regardless of the timescale involved.  The rate of the rotation of \textit{t} varies between epochs, so it is likely that \gName{NGC}{1275} would show much more complex polarization behavior if it were monitored more intensively.


\subsection{ALMA}
\label{subsec:alma}
\gName{NGC}{1275} has been observed several times in different bands by the Atacama Large Millimeter/sub-millimeter Array (ALMA), both as a target and as a bright source suitable for gain and bandpass calibration. We have reduced a single observation in each of bands 3, 5, 6, 7, and 9, choosing the observation nearest in time to the VERITAS observations when there was more than one archival dataset at the same frequency. The details of the ALMA observations (shown in Figure \ref{figure:multiwavelength_daily_lightcurve}, bottom panel) are summarized in Table~\ref{table:alma_obervations} where we list the date, band, central frequency, observation ID, synthesized beam size in arcseconds, the final image RMS in Jy, and total flux of the source. All calibration and imaging were conducted using the Common Astronomy Software Applications (CASA) package \citep{casa}. In all cases, the data were initially calibrated using the pipeline script included with the uncalibrated download from the ALMA archives. Initial imaging deconvolution was performed with the \texttt{clean} algorithm in `mfs' mode and nterms=2 (this allows us to account for spectral curvature over the wide bandpass). We used a Briggs weighting with a robust parameter of 0.5 which roughly balances resolution and sensitivity in the resulting image. We used several rounds of self-calibration to improve the imaging (reducing the RMS) and to look for any extended structure in the images. No extended structures were found and in all images the central point source is unresolved. 

\begin{deluxetable}{ccccccc}[]
\caption{ALMA observations of \gName{NGC}{1275} reduced for this project. Only data from Band 7 are shown in Figure~\ref{figure:multiwavelength_daily_lightcurve}; data from all five bands were used in the SED construction shown in Figure~\ref{figure:sed_modelling}.}
\label{table:alma_obervations}
\tabletypesize{\scriptsize}
    \tablehead{
        \colhead{Obs. Date} & \colhead{Band} & \colhead{Freq.} & \colhead{Obs. ID} & \colhead{Beam} & \colhead{RMS} & \colhead{Flux} \\
        \colhead{} & \colhead{} & \colhead{[GHz]} & \colhead{} & \colhead{[arsecs]} & \colhead{[Jy]} & \colhead{[Jy]}
    }
    \startdata
    2014-01-01  & 3   & 93.6    & 2012.1.00394  & 20.9$\times$15.00  &  3.2e-2  & 17.2 \\
    2019-08-14  & 5   & 178.2   & 2018.1.01438  & 0.30$\times$0.14   &  6.8e-3  & 11.0 \\
    2017-11-27  & 6   & 232.1   & 2017.1.01257  & 0.13$\times$0.07   &  3.2e-3  & 7.48 \\
    2016-11-15  & 7   & 338.2   & 2016.1.01305  & 9.29$\times$4.42   &  3.3e-5  & 11.4 \\
    2019-08-13  & 9   & 693.5   & 2018.1.01438  & 3.02$\times$2.23   &  4.1e-2  & 4.27 \\
    \hline
    \enddata
\end{deluxetable}

\subsection{MOJAVE / VLBA}
\label{subsec:mojave}
Additional radio data used for the multiwavelength light curve (Figure \ref{figure:multiwavelength_daily_lightcurve}, bottom panel) and broad-band SED (Figure \ref{figure:sed_modelling}) are taken from the MOJAVE 2\,cm Survey Data Archive \citep{Lister-2018-ApJS}. This is a collection of radio data largely obtained with the Very Long Baseline Array (VLBA) at 15 GHz, but also includes some data from the National Radio Astronomy Observatory (NRAO) data archives. Details of the MOJAVE survey observations and data reduction can be found in \cite{Lister-2018-ApJS}. Unfortunately, none of the radio data is coincident with the January flare. Instead the radio observations tend to be taken at epochs typically 6 months apart. As a result we only have 3 data points for the time period we are considering, one of which is contemporaneous with the MJD 57755 flare and used in the broadband SED modeling.

\subsection{Extraction of the host galaxy emission}
\label{subsec:host_extraction}

The contribution of the host galaxy emission as well as emission line contribution in the R band of the KVA telescope of the Tuorla observatory has been removed. The host galaxy flux within the $5''$ aperture radius of the KVA was estimated at $11.08 \pm 0.55$ mJy (($5.05 \pm 0.25) \times10^{-11}$ erg cm$^{-2}$ s$^{-1}$) \citep{Aleksic:2014ab}.
We detail here the process to remove the host contribution in the UVOT filters, in order to build a consistent MWL SED.

The host galaxy has a stellar surface brightness profile which follows a de Vaucouleurs' $R^{1/4}$ law to at least 150 kpc from the center \citep{Prestwich_1997}. 
The effective radius of the host was measured at $16.9^{+2.5}_{-2.2}$ arcsec from the third reference catalog of bright galaxies \citep[RC3,][]{deVaucouleurs_1991}.
From the de Vaucouleurs' law, we estimate a luminosity fraction of $21.0^{+2.7}_{-2.4} \%$ of the host within a $5''$ aperture used by Tuorla and UVOT.

As \gName{NGC}{1275} is significantly bluer than a normal elliptical (E) galaxy, it was assumed by \cite{Mathews_2006} that the stellar mass of \gName{NGC}{1275} is dominated by an old stellar population, as in normal E galaxies, intermixed with an additional population of young, luminous stars that does not contribute significantly to the total mass.
In the same paper, the total stellar mass of the host was estimated at $M_{host} = 2.43 \times10^{11} M_{\odot}$.

Using an elliptical galaxy template from PEGASE.2 \citep{Fioc_1999} scaled by the measured host mass, and with an age of 13 GYr, we estimate a flux in the R band within an aperture of $5''$ of $5.26 \pm 0.63 \times10^{-11}$ erg cm$^{-2}$ s$^{-1}$, consistent with the Tuorla estimation of $5.05 \pm 0.25 \times10^{-11}$ erg cm$^{-2}$ s$^{-1}$.

The host estimation in the UVOT bands within a $5''$ aperture is estimated at V: $(4.31\pm0.51) \times10^{-11}$, B: $(2.69\pm0.33) \times10^{-11}$, U: $(8.13\pm0.97)\times10^{-12}$, UVW1: $(3.01\pm0.36)\times10^{-12}$, UVM2: $(2.60\pm0.32)\times10^{-12}$, UVW2: $(2.88\pm0.35)\times10^{-12}$ [erg cm$^{-2}$ s${-1}$].
Given the fact that \gName{NGC}{1275} is bluer than normal E galaxies, a likely caveat from this calculation is that these values are underestimated in most of the UVOT bands.


\section{Results: Gamma-Ray Variability and Spectral Analysis}\label{sec:results}
To provide context for the analysis of the flaring state, we first investigate the variability of VERITAS long-term light curves to determine different states, including the two extreme-flare dates. We then carry out a spectral analysis of each of the resulting four states. Turning to an analysis of the flaring state, we first look for short-term variability in the VERITAS highest-flux state (the night following the MAGIC highest-flux state). We then use the short-timescale \Fermi-LAT light curve covering both the MAGIC and VERITAS highest-flux states of the flare to determine the change in flare state as it decayed. The resulting two combined GeV and TeV \gray\ SEDs are constructed and analyzed, with one comprising \Fermi-LAT plus MAGIC data covering \decthirtyfirst{}/\janfirst{} and the second comprising \Fermi-LAT plus VERITAS data covering \jansecond{}. 



\subsection{VERITAS variability analysis} \label{subsec:variability_analysis}
\begin{figure}[htb]
\centering
{\includegraphics[width=\columnwidth]{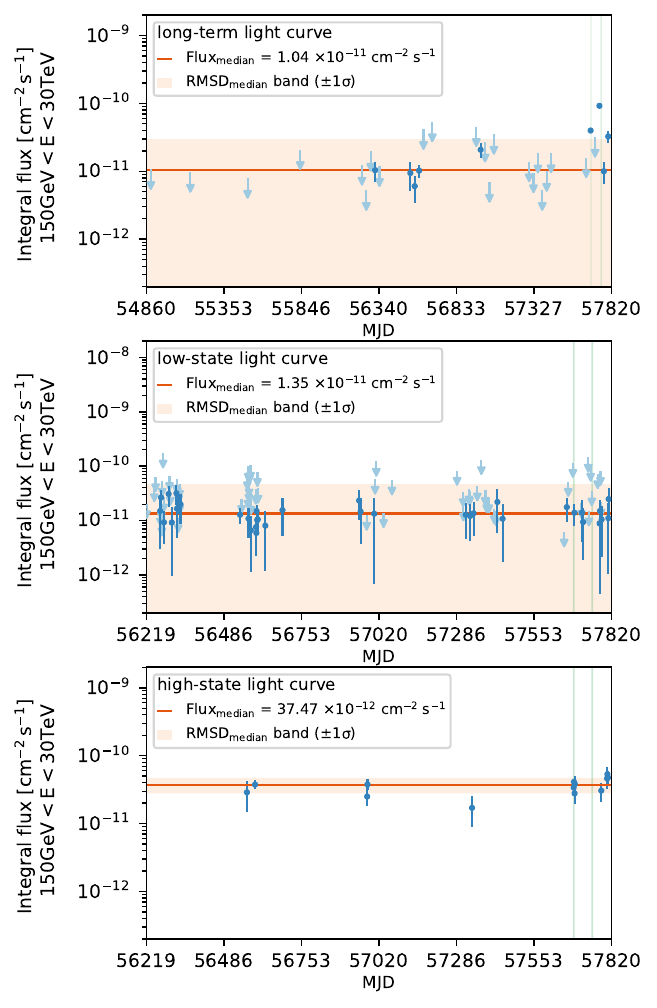}}
\caption{Top panel: long-term VERITAS TeV light curve for all 4-telescope observations of radio galaxy \gName{NGC}{1275} for the energy range $\mathrm{0.15\,TeV \leqslant E \leqslant 30\,TeV}$, spanning more than 8-years (2009-2017) and binned in 28-day intervals. The median flux (solid orange line) and $\mathrm{1\sigma}$ root mean squared deviation (RMSD; orange band) are shown, with 95\% confidence-level upper limits plotted for flux points $\mathrm{<2\sigma}$.
Center panel: average low-state light curve calculated for \gName{NGC}{1275} during 2012 October to 2017 June, with median flux and $\mathrm{1\sigma}$ RMSD as above; here 95\% confidence-level upper limits are shown for flux points $\mathrm{<1\sigma}$.
Bottom panel: average high-state light curve calculated for \gName{NGC}{1275} during 2012 September to 2017 June, with median flux and $\mathrm{1\sigma}$ RMSD.
The two vertical green lines mark the \octthirtyfirst{} and \janfirst{} flares repectively.}
\label{figure:veritas_lightcurves}
\end{figure}

The top panel of Figure \ref{figure:veritas_lightcurves} shows the long-term VERITAS light curve for all observations of \gName{NGC}{1275} from 2009 January 15 to 2017 February 26 for the energy range $\mathrm{0.15\, TeV \leqslant E \leqslant 30\, TeV}$ and binned in 28-day intervals. The top panel also shows the median flux (solid orange line) and the orange band highlights the $\mathrm{1\sigma}$ root mean squared deviation (RMSD). The RMSD is computed following the central value of each data point. However, flux points detected with $\mathrm{<2\sigma}$  are shown in Figure~\ref{figure:veritas_lightcurves} as 95\% confidence level upper limits. 
The RMSD provides a variability diagnostic given the limited detections and uneven sampling. Notably, the 28-day flux bin incorporating the 2017 January flare lies above the $\mathrm{1\sigma}$ RMSD band in the top panel of Figure \ref{figure:veritas_lightcurves}, suggestive of long-term variability relative to the apparent steadier low- and high-states (center and bottom panels).

When considering the daily-binned light curve recorded for \gName{NGC}{1275} with the VERITAS upgrade array (2012 October 19 to 2017 February 26) it is possible to separate out an average low- and high-state as well as an extreme-high-state. Flux states are defined such that the high-state corresponds to fluxes $\mathrm{\geqslant 1\sigma}$ above the mean, and the extreme-high state to fluxes $\mathrm{\geqslant 3\sigma}$ above the mean. The center panel of Figure \ref{figure:veritas_lightcurves} shows the average low-state of \gName{NGC}{1275} during the period 2012 October to 2017 February. As in the upper panel, we plot the median flux (solid orange line) and the $\mathrm{1\sigma}$ RMSD (orange band), computed including censored data with 50\% confidence-level upper limits, while the plotted upper limits are at the 95\% level; in this case, upper limits are shown for flux points $\mathrm{<1\sigma}$ in order to increase the available statistics.

The bottom panel of Figure \ref{figure:veritas_lightcurves} shows the average high-state of \gName{NGC}{1275} during the period 2012 October to 2017 February. As in the upper and center panels, we plot the median flux (solid orange line) and the $\mathrm{1\sigma}$ RMSD (orange band). We define the extreme-high-state of \gName{NGC}{1275} to be the observations recorded on MJD57755 and MJD57756 respectively, and do not include these extreme-high-state observations in the calculation of the high-state light curve. The analysis results for each of the four states are summarized in Table \ref{table:vegas_results}.


\begin{deluxetable}{lrrrrrr}[]
\caption{Summary of VERITAS analysis results.}
\label{table:vegas_results}
\tabletypesize{\scriptsize}
    \tablehead{
        \colhead{Details} & \colhead{Exp.} & \colhead{On} & \colhead{Off} & \colhead{Alpha} & \colhead{Sig.} & \colhead{Rate}
    }
    \startdata
    Average low state & 3377 & 9468 & 54349 & 0.5 & $\mathrm{9\sigma}$ & 0.3 \\
    Average high state & 833 & 2869 & 9493 & 0.5 & $\mathrm{24\sigma}$ & 1.5 \\
    Flare MJD57755 & 119 & 971 & 978 & 0.33 & $\mathrm{31\sigma}$ & 6.3 \\
    Flare MJD57756 & 81 & 648 & 671 & 0.33 & $\mathrm{22\sigma}$ & 5.9 \\
    \hline
    \enddata
    \tablecomments{ Exp. is the total exposure time in minutes, On is the number of on-region counts, Off is the number of off-region counts, Alpha is the off-region source normalization, Sig. is the statistical significance calculated using a generalized form of the \citet{LiMa:1983apj} method (see Section \ref{subsec:veritas}), and Rate is the number of \grays{}/minute. The definition of high-state and low-state is detailed in the main text.}
\end{deluxetable}

\subsection{VERITAS spectral analysis} \label{subsec:spectral_analysis}

\begin{figure}[t]
\centering
{\includegraphics[width=1.0\columnwidth]{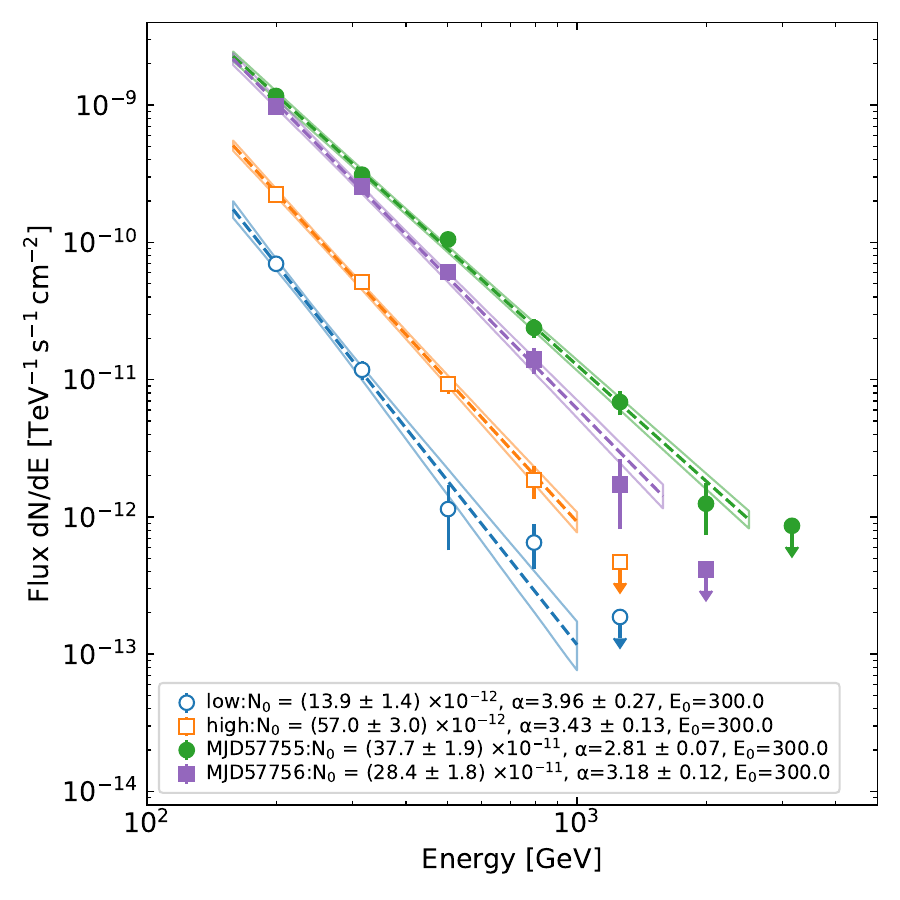}}
\caption{The spectra calculated for the average low-state (open blue circles), the average high-state (open orange squares) and for the extreme-high-state flares that occurred on MJD57755 (green-filled circles) and MJD57756 (purple-filled squares). For each of these states, the VHE \gray{} emission falls according to a power law spectrum. As \gName{NGC}{1275} increases in flux brightness the spectral indices get harder and detected spectra extend to higher energies.}
\label{figure:veritas_spectrum_all_states}
\end{figure}

Figure \ref{figure:veritas_spectrum_all_states} shows the VERITAS spectral energy distributions (SEDs) calculated for the average low-state (open blue circles), the average high-state (open orange squares) and for the extreme-high-state flares that occurred on \jansecond{} MJD57755 (green-filled circles) and \janthird{} MJD57756 (purple-filled squares). In each of these cases the \gName{NGC}{1275} VHE spectrum decays according to a power law function as shown in Equation \ref{equation:powerlaw}:

\begin{equation}
    dN/dE = f_{0}(E/E_{0})^{-\alpha}
    \label{equation:powerlaw}
\end{equation}

\noindent where $f_{0}$ is the normalization, $E_{0}$ the scale-energy and $\alpha$ the spectral index. In addition, we also fitted a log parabola function, like that shown in Equation \ref{equation:logparabola}, to the VERITAS spectra and perform statistical tests to compare which model best represents the data.

\begin{equation}
    dN/dE = f_{0}(E/E_{0})^{-\alpha - \beta log(-E/E_{0})}
    \label{equation:logparabola}
\end{equation}
Equation \ref{equation:logparabola} shows the log parabola function where $f_{0}$ is the normalization, $E_{0}$ the scale energy, $\alpha$ the spectral index and $\beta$ the spectral curvature. Table \ref{table:veritas_spectral_parameters} highlights the spectral parameters obtained for the VERITAS observations of \gName{NGC}{1275}.

\begin{figure}[htb]
\centering
{\includegraphics[width=1.0\columnwidth]{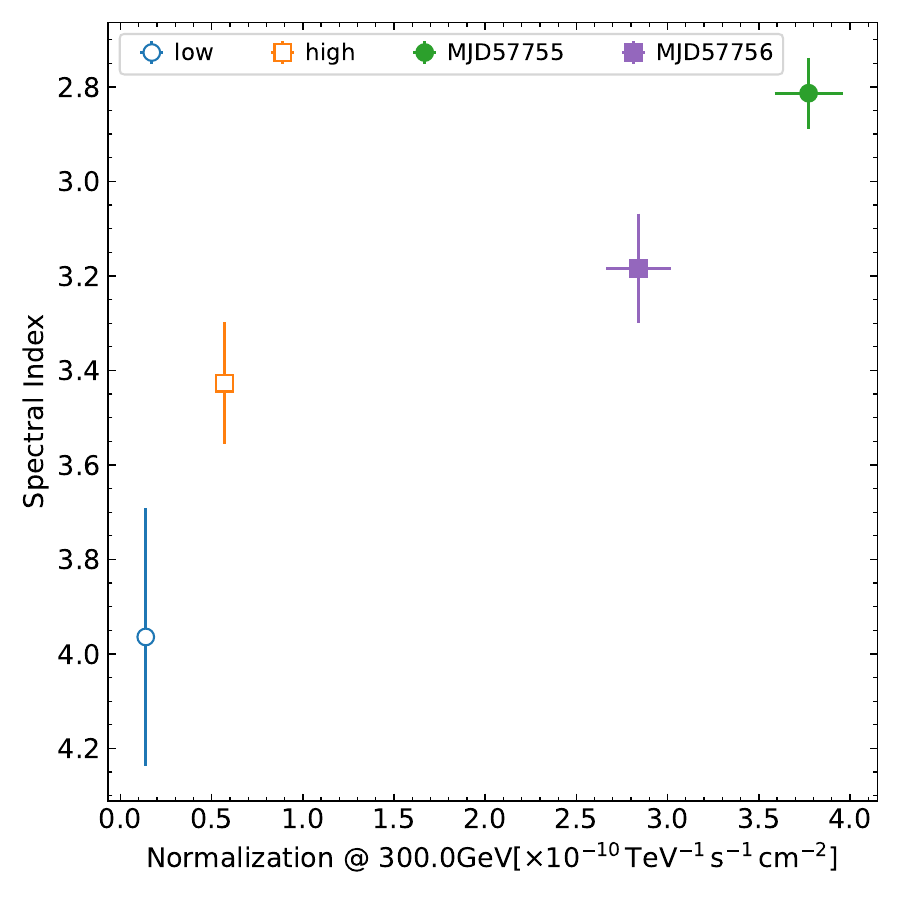}}
\caption{The spectral index of the fitted power laws for each state hardens as the recorded flux brightness increases from low-state (open blue circle) to high-state (open orange square) and even during the two flares of extreme-high-state: MJD57755 (green-filled circle) and MJD57756 (purple-filled square). Note that the direction of the y-axis is inverted.}
\label{figure:veritas_spectral_index_evolution}
\end{figure}

Figure \ref{figure:veritas_spectral_index_evolution} illustrates how the spectral index of the fitted power laws for each state decreases, leading to a harder spectrum as the recorded flux brightness increases from low- to high-state and even during the two flares of extreme-high-state. It is important to note that the VERITAS extreme flare state discussed here occurred the day after the MAGIC detection of the highest flux state of the flare on MJD57754; see Section \ref{subsec:combined_GeV_TeV_spectral_analysis} for spectral analysis including the MAGIC data.

\begin{deluxetable}{lrrrr}[b]
\tablecaption{Power law spectral parameters obtained for the different nominal states identified for \gName{NGC}{1275} with a scale energy $E_{0}=300$ $\mathrm{GeV}$ for all four states. See text for explanation of the p-value in the last column. 
\label{table:veritas_spectral_parameters}}
\tabletypesize{\scriptsize}
    \tablehead{
    \colhead{Details} & \colhead{Index $\alpha$} & \colhead{Norm.} & \colhead{$\mathrm{\chi^{2} / NDF}$} & \colhead{p} \\
    & & [$\mathrm{TeV^{-1} s^{-1} cm^{-2}}$] & &
    }
    \startdata
    Average low state  &  3.96 $\pm$ 0.27  &  (13.9 $\pm$ 1.4) $\times 10^{-12}$  &  $3.78 / 2$ & 1\\
    Average high state &  3.43 $\pm$ 0.13  &  (57.0 $\pm$ 3.0) $\times 10^{-12}$   &  $1.15 / 2$ & 0.37\\
    Flare MJD57755     &  2.81 $\pm$ 0.07  &  (37.7 $\pm$ 1.9) $\times 10^{-11}$   &  $4.05 / 4$ & 0.22\\
    Flare MJD57756     &  3.18 $\pm$ 0.12  &  (28.4 $\pm$ 1.8) $\times 10^{-11}$   &  $3.00 / 3$ & 0.12\\
    \hline
\enddata
\end{deluxetable}

To compare which functional model best represents the VERITAS data, we refitted the spectral energy distribution (SED) data using a likelihood-based approach implemented with \texttt{scipy.optimize.minimize}, employing the L-BFGS-B optimization algorithm to minimize the negative log-likelihood. This method assumes Gaussian-distributed observational uncertainties and returns the maximum likelihood estimates for each model's parameters. We then used the resulting maximum log-likelihood values to compare the nested models with the Likelihood Ratio Test \citep{Wilks-1938}, whose test statistic is defined as $\mathrm{\Lambda = 2(\log{\mathcal{L}}_{2} - \log{\mathcal{L}}_{1})}$, where $\mathrm{\mathcal{L}_{1}}$ and $\mathrm{\mathcal{L}_{2}}$ are the maximum likelihood values of the simpler (power law) and more complex (log parabola) models respectively. Under the null hypothesis that the simpler model is sufficient, the test statistic asymptotically follows a $\mathrm{\chi^{2}}$ distribution with degrees of freedom equal to the difference in the number of free parameters $\mathrm{\Delta k}$ between each model. The p-value is then calculated as $\mathrm{p = 1 - CDF_{\chi^{2}}(\Lambda, \Delta k)}$ and represents the probability of obtaining a test statistic at least as extreme as $\Lambda$, under the assumption that the simpler model is true. A threshold of $\mathrm{p_{\alpha} = 0.05}$ is used to determine statistical significance. Table \ref{table:veritas_spectral_parameters} includes the p-values obtained for the model comparison of each SED. In each case the $\mathrm{p \geqslant 0.05}$ meaning we cannot reject the simpler power law model at the $95\%$ confidence level.


Our low-state parameters are comparable within errors to those published by MAGIC  for data taken between 2009 and 2014 \citep{Ahnen2016-lowstate} which finds for a simple power-law fit a photon index $\alpha = 3.6 \pm0.2_\mathrm{stat} \pm0.2_\mathrm{syst}$ and a normalization constant at 200 GeV of $f_0 = (2.1 \pm0.2_\mathrm{stat} \pm0.3_{\mathrm{syst}}) \times
10^{-11} \mathrm{cm^{-2} s^{-1} TeV^{-1}}. $
We note that our result of an average high state is the first such published result.

\subsection{VERITAS spectral variability} \label{subsec:variability-analysis}


Using data recorded on \jansecond{} (MJD57755), the most extreme state so far detected by VERITAS, we produced a number of light curves with different time binning schemes (5, 10, 18, 32 and 40 minutes respectively) to look for any hints of short-period variability during the night of the flare. By fitting a constant function to all the sampled light curves we were able to determine that all binning schemes tested, apart from the 18 and 32 minute bins, are strongly consistent with a constant function. The 18 minute-binned light curve yielded a reduced $\mathrm{\chi^{2}}$ value of 2.85 (1\% confidence level)
while the
32 minute-binned light curve yielded a reduced $\mathrm{\chi^{2}}$ value of 2.26 (7\% confidence level).

\subsection{Fermi-LAT flux variability}\label{subsec:fermi_variability}
\begin{figure}[htb]
\centering
{\includegraphics[width=\columnwidth]{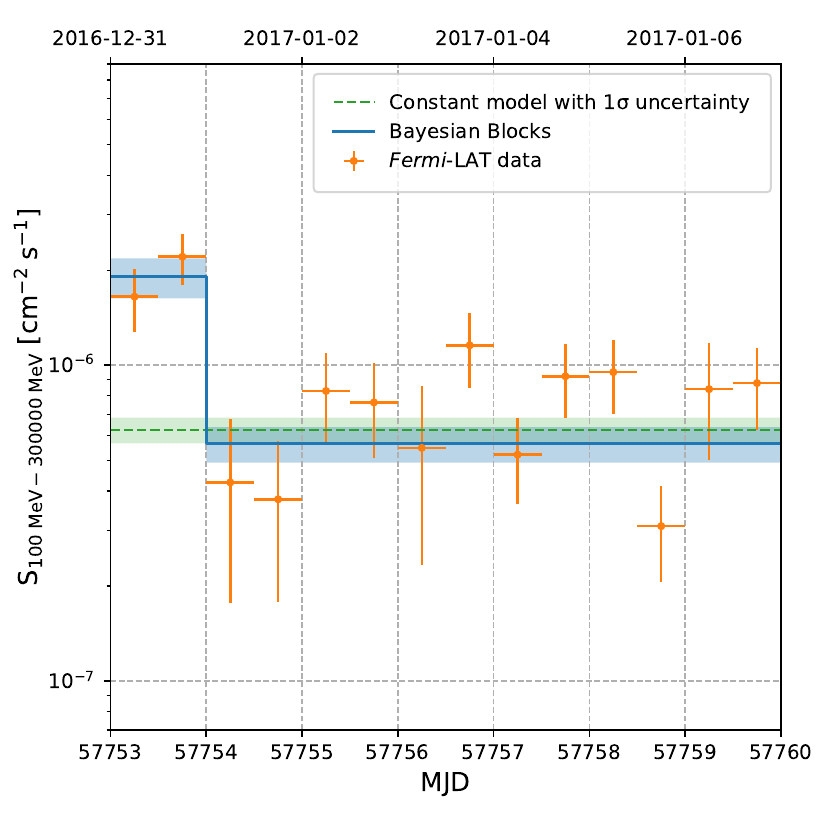}}
\caption{The \Fermi-LAT NGC\,1275 flux versus time for the period MJD 57753 -- MJD 57760 spanning the 2017 flare detected at VHE energies. The orange points show the 12-hour binned light curve data and the solid blue line the mean flux for optimal Bayesian block binning including uncertainty (blue shaded band). For reference we also show a constant model (dashed green line) fitted to the data including the $\mathrm{68\%}$ confidence bound of this fit (green shaded band). The reduced chi-squared statistic for this fit was 3.32.}
\label{figure:fermiLAT_variability}
\end{figure}

To inspect the short-timescale variability around the 2017 January VHE-flaring interval, we produced a short-timescale \Fermi-LAT light curve with 12-hour binning for MJD 57753 -- MJD 57760 (see Figure \ref{figure:fermiLAT_variability}). The 12-hour binning is the minimum time scale achievable with the \fermi{} data to avoid including upper limits in the light curve. We then computed the optimal segmentation of the light curve data using Bayesian blocks as defined by \cite{Scargle-2013-ApJ}. We did this with the \texttt{Astropy} v2.0.16 Bayesian blocks implementation using a false alarm probability of $\mathrm{p0 = 0.05}$. This process yielded two distinct blocks where the first Bayesian block bin, when \fermi{} sees \gName{NGC}{1275} in an elevated state, approximately spans MJDs 57753 - 57754.
The second Bayesian block bin, when \fermi{} sees \gName{NGC}{1275} return to a relatively lower and apparently steady state, approximately spans MJDs 57754 - 57760.


\fermi{} detects \gName{NGC}{1275} with a statistical significance of $\ssim$18.7$\sigma$ (TS=350) and $\ssim$28.3$\sigma$ (TS=802)
respectively in these two bins. Two \fermi{} SEDs were produced using data corresponding to the two periods defined by the Bayesian block analysis. Using the Likelihood Ratio test mentioned above, we compared the log parabola and power law nested models fitted to the $\mathrm{dN/dE}$ SED for each Bayesian Block bin. The resulting p-values ($\mathrm{p_{block 1}=0.26}$, $\mathrm{p_{block 2}=0.06}$)
of each Bayesian Block bin was $\geqslant0.05$. Therefore, we cannot reject the simpler power law model at the 95\% confidence level. For each of these Bayesian block bins, their respective spectral index $\alpha$ agrees within error, $\mathrm{\alpha_{block 1} = 1.97^{+0.08}_{-0.14}}$ versus $\mathrm{\alpha_{block 2} = 1.92^{+0.05}_{-0.05}}$,
meaning we see no significant change with \fermi{} during the flare despite the flux normalization halving at the common scale-energy $\mathrm{E_{0} = 500\,MeV}$, from  $\mathrm{N^{block 1}_{500} = 80^{+11}_{-10}}$ to $\mathrm{N^{block 2}_{500} = 34.2^{+2.6}_{-2.7}}$,
in units of $\mathrm{10^{-8}\, GeV^{-1}\,s^{-1}\,cm^{-2}}$.


We also use the two Bayesian blocks defined above to calculate two corresponding Compton-peak SEDs (combined GeV and TeV data, see Section \ref{subsec:combined_GeV_TeV_spectral_analysis}) as well as two broadband multiwavelength SEDs discussed in Section \ref{sec:discussion}. 
The date at the separation between the two \fermi{} Bayesian blocks is \decthirtyfirst{}, 17$\pm$12h UTC, while the time of the VHE peak can be estimated from MAGIC data to be \janfirst{}, 0$\pm$12h UTC. From these times, we see ambiguity as to whether the VHE flare peaked during the high or the low \fermi{} flux state represented in these Bayesian-defined blocks.
In this work, we assume that the VHE flare is associated with the first Bayesian block, representing the high \fermi{} state (\decthirtyfirst{}/\janfirst{} broadband SED), while Bayesian block 2, representing the low \fermi{} state, corresponds to the \jansecond{} broadband SED. This is the most likely scenario given the expected TeV-GeV correlations arising from blazar standard synchrotron-self-Compton models.

\subsection{Combined GeV and TeV spectral analysis} \label{subsec:combined_GeV_TeV_spectral_analysis}

\begin{figure}[t!]
\centering
\includegraphics[width=1.0\columnwidth]{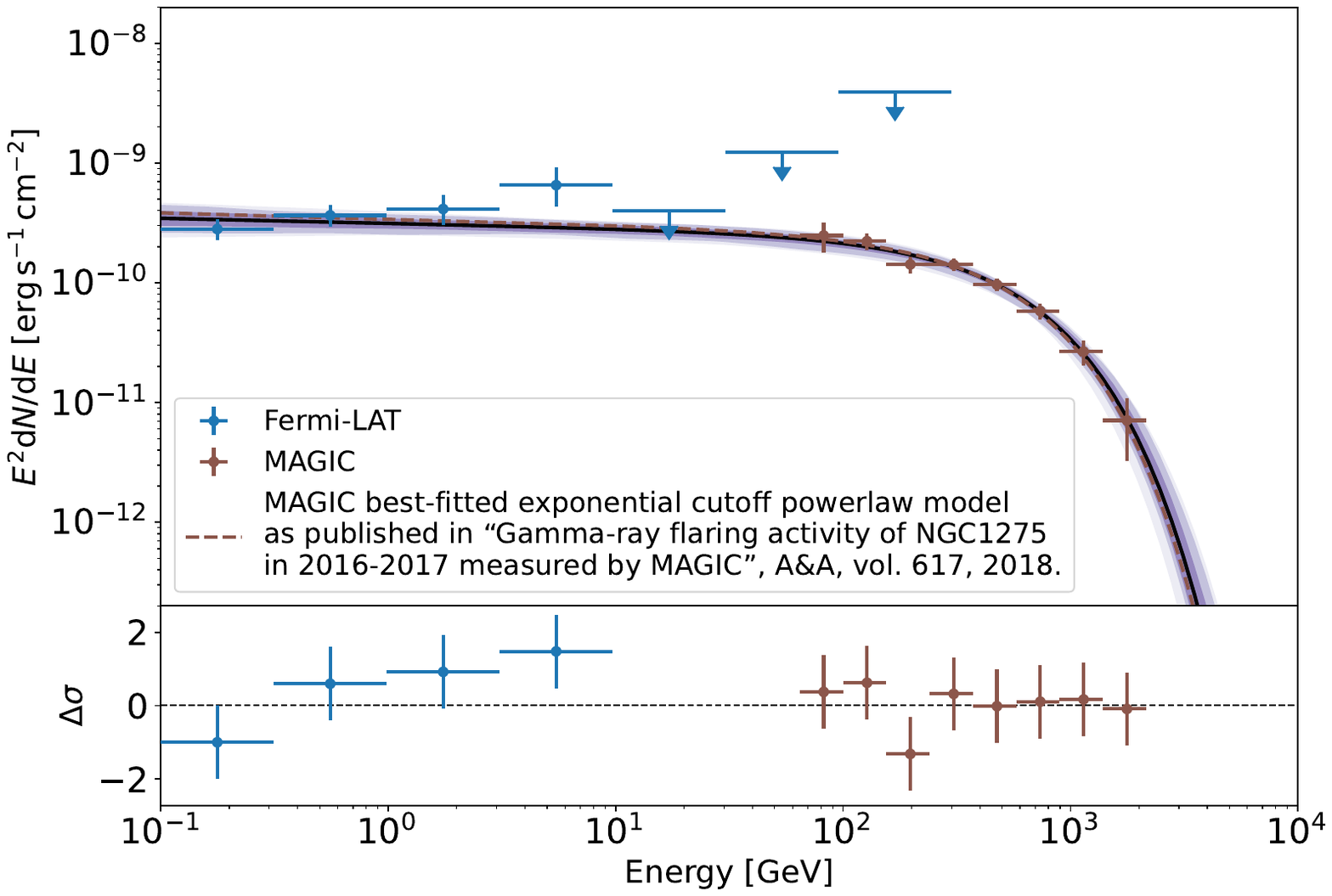}
\put(-210,140){\colorbox{white}{\footnotesize \bf \decthirtyfirst{}/\janfirst{}}}\newline
\includegraphics[width=1.0\columnwidth]{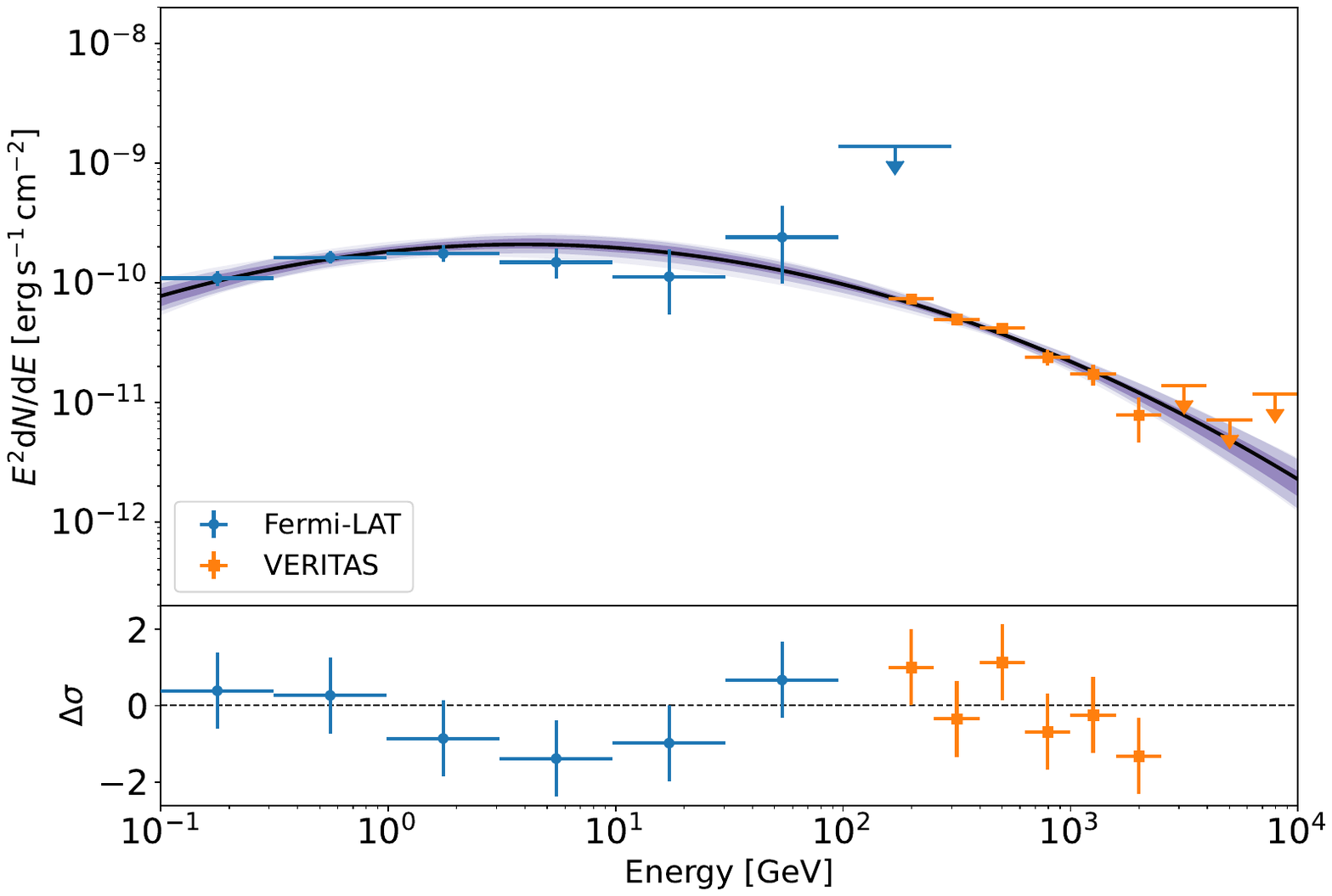}
\put(-210,140){\colorbox{white}{\footnotesize \bf \jansecond{}}}
\caption{Shown here are the best fitted spectral models (including residuals) to the combined Fermi-LAT (blue filled circles) and MAGIC (brown filled circles) data for \decthirtyfirst{}/\janfirst{} (top panel), and the combined Fermi-LAT (blue filled circles) and VERITAS (orange filled squares) data for \jansecond{} (bottom panel). Also shown in the top panel is the best-fitted spectral model (brown dashed line) published by MAGIC \citep{MAGIC-2018-AA}. For \decthirtyfirst{}/\janfirst{} the combined data is best described by a power law with exponential cutoff model (solid black line), whereas on \jansecond{} the combined data is best described by a log parabola model (solid black line). The purple bands highlight the $1\sigma$, $2\sigma$ and $3\sigma$ confidence intervals, respectively. These best-fitted spectral models highlight the stark evolution 
of the Compton peak shape between \janfirst{} and \jansecond{}. Details of the fits can be found in Table \ref{table:fermi_veritas_combined_fit_parameters}.}
\label{figure:fermi_veritas_combined_spectral_analysis}
\end{figure}

We construct the Compton-peak SEDs for the two Bayesian blocks defined above with the 
MAGIC data \citep{MAGIC-2018-AA} combined with our \fermi{} analysis for the block corresponding
to \decthirtyfirst{}/\janfirst{}, and the combined \fermi{} and VERITAS data for the block corresponding to \jansecond{}. 

We then tested two spectral models against the combined GeV and TeV data for each SED; a power law with exponential cut off (ECPL):
\begin{equation}
    dN/dE = f_{0}(E/E_{0})^{-\alpha}exp(-E/E_{c})
    \label{equation:ECPL}
\end{equation}
where $f_{0}$ is the normalization, $E_{0}$ the scale energy, $\alpha$ the spectral index and $E_{c}$ the cutoff energy; and a log parabola (LP) shown in Equation \ref{equation:logparabola}. 

\begin{deluxetable*}{cccccc}[t]
\tablecaption{Parameters of the spectral models fitted to the combined data.\label{table:fermi_veritas_combined_fit_parameters}}
\tabletypesize{\scriptsize}
    \tablehead{
    \colhead{} & \multicolumn{3}{c}{\fermi{} and MAGIC: \decthirtyfirst{}/\janfirst{}} & \multicolumn{2}{c}{\fermi{} and VERITAS: \jansecond{}}\\[-0.1cm]\hline 
    \colhead{Fit} & \colhead{ECPL\dag} & \colhead{ECPL} & \colhead{LP} &  \colhead{ECPL} & \colhead{LP}\\[-0.4cm]
    }
    \startdata
    $f_{0}$ & (26.24 $\pm$ 1.38) $\times 10^{-11}$ & $(31.2^{+3.4}_{-3.2}) \times 10^{-11}$ & $(50.3^{+5.9}_{-5.5}) \times 10^{-11}$ & $(126.6^{+8.3}_{-8.4}) \times 10^{-12}$ & $(18.1^{+1.4}_{-1.3}) \times 10^{-11}$\\
    $\alpha$ & $-0.05 \pm 0.03$ & $0.04^{+0.02}_{-0.03}$ & $-0.35^{+0.06}_{-0.07}$ & $0.07^{+0.02}_{-0.02}$ & $-0.20^{+0.04}_{-0.05}$\\ 
    $\beta$ & $-$ & $-$ & $0.11^{+0.01}_{-0.01}$  & $-$ & $0.07^{+0.01}_{-0.01}$ \\
    $E_{c}$ & $492 \pm 35$ & $516^{+65}_{-56}$ & $-$ & $716^{+118}_{-100}$ & $-$\\
    $E_{0}$ & $198.21$ & $1$ & $1$ & $1$  & $1$ \\
    $\mathrm{\log(\mathcal{L})}$ & $-$ & -3.70 & -13.41 & -9.51 & -4.31\\
    BIC & $-$ & 15.5 & 34.9 & 27.3 & 16.9\\
    AICc & $-$ & 13.4 & 32.8 & 25.0 & 14.6\\
   $\chi^{2}/d.o.f$ & $19.18/11$ & $-$ & $-$ & $-$ & $-$ \\ [0.1cm]
   \hline
\enddata
\tablecomments{where ECPL is the power law with exponential cut off, LP is the log parabola, $f_{0}$ is the flux normalization in units $\mathrm{erg\,s^{-1}\,cm^{-2}}$, $\alpha$ the spectral index, $\beta$ the power law curvature, $E_{c}$ the cut off energy in units GeV, the scale-energy $E_{0}$ in units GeV, $\mathrm{\log(\mathcal{L})}$ the maximum log likelihood values, the Bayesian Information Criterion (BIC) and the corrected Akaike Information Criterion (AICc) used to determine the favoured model amongst competing models. \dag Parameters of MAGIC's published model (see Table 2 \citep{MAGIC-2018-AA}) converted to units used here.}
\end{deluxetable*}

Figure \ref{figure:fermi_veritas_combined_spectral_analysis} shows the best fitted spectral models to the SED for \decthirtyfirst{}/\janfirst{} (top panel) and the SED for \jansecond{} (bottom panel). For comparison, the best-fitted exponential-cutoff power law model (brown dashed line) published by MAGIC \citep{MAGIC-2018-AA} is also shown in the top panel.\footnote{Note this published model was fit to \fermi{} analysis carried out in conjunction with the MAGIC publication and not corresponding to the analysis carried out for this work.} We fitted these combined SEDs using a Bayesian approach implemented in the Python package \texttt{naima} \citep{Zabalza-2015-ICRC}, employing Markov Chain Monte Carlo (MCMC) sampling of their likelihood distributions with \texttt{emcee} \citep{Foreman-Mackey-2013-PASP}. Table \ref{table:fermi_veritas_combined_fit_parameters} highlights the parameter estimates as the median of the posterior distribution, along with 68\% credible intervals. The best-fitted models (solid black lines) were computed using the maximum-likelihood parameters. Since the ECPL and LP models have an equal number of free parameters and are non-nested models -- models that cannot be expressed as special cases of each by constraining certain parameters -- the Likelihood Ratio Test to determine which spectral model from the competing models best fits the data is no longer applicable \citep{Protassov-2002-ApJ}. Instead, we calculated the Bayesian Information Criterion (BIC) in Equation \ref{equation:BIC} and the Akaike Information Criterion (AIC). Given the small number of flux bins ($\mathrm{n < 40}$) in the SEDs we used the corrected Akaike Information Criterion (AICc) shown in Equation \ref{equation:AICc} as recommended by \citet{Burnham-2011} and \citet{Wagenmakers-2004} for example. The AICc includes a bias-corrected estimate of model quality for small sample sizes, where the standard (non-corrected) AIC can overfit.

\begin{equation}
    \mathrm{BIC} = k\log(n) - 2\log(\mathcal{L})
    \label{equation:BIC}
\end{equation}
\begin{equation}
    \mathrm{AICc} = 2k - 2\log(\mathcal{L}) + \frac{2k(k+1)}{(n-k-1)}
    \label{equation:AICc}
\end{equation}

The BIC and corrected AIC are given in Equations \ref{equation:BIC} and \ref{equation:AICc} where $\mathcal{L}$ is the maximized likelihood function, $n$ the number of data points, and $k$ the number of free parameters. These criteria estimate model quality by balancing goodness-of-fit and model complexity. Using this approach, the model with the lowest BIC or AIC value is the favored model (see Table \ref{table:fermi_veritas_combined_fit_parameters}). However, to quantify the relative support for each model under the AIC framework, Equation \ref{equation:AIC-weights} defines the Akaike weights for $i$ models, which we have also computed:

\begin{equation}
    w_{i} = \dfrac{\exp{(-\Delta_{i} / 2)}}{\sum_{j=1}^J \exp({-\Delta_{j} / 2})}
    \label{equation:AIC-weights}
\end{equation}

\noindent where $\mathrm{\Delta_{i}(AIC) = AIC_{i} - min\, AIC}$. The Akaike weights allow us to directly compare probabilities for each model tested. For the combined \fermi{} and MAGIC SED $\mathrm{\Delta_{AICc}}$ = 19.4 and the ECPL model is decisively preferred by AIC with a 99.99\% probability of being the better model. For the combined \fermi{} and VERITAS SED $\mathrm{\Delta_{AICc}}$ = 10.4 and the log parabola model is decisively preferred by AIC with a 99.45\% probability of being the better model. While it may be tempting to conclude from Figure~\ref{figure:fermi_veritas_combined_spectral_analysis} that the source is evolving to a harder-when-dimmer state, we can only conclude from the above statistical analysis that the best-fitted 
spectral model for the Compton peak evolves from an ECPL to a LP from \decthirtyfirst{}/\janfirst{} to \jansecond{}. 

\section{Discussion}\label{sec:discussion}

\begin{figure*}[t!]
\centering
\includegraphics[width=9cm]{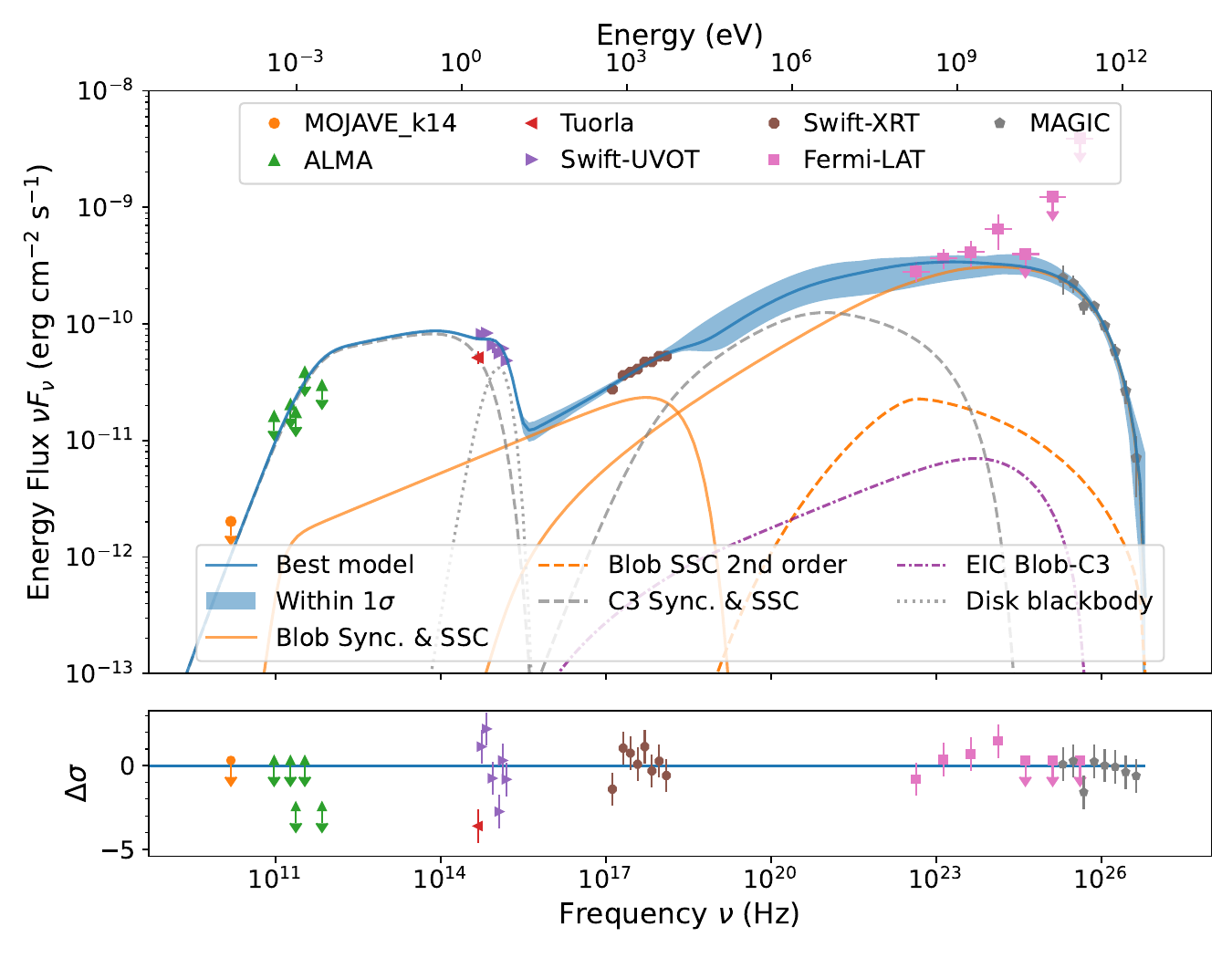}
\put(-210,150){\footnotesize{\bf 2016 Dec. 31 / 2017 Jan. 1}}
\includegraphics[width=9cm]{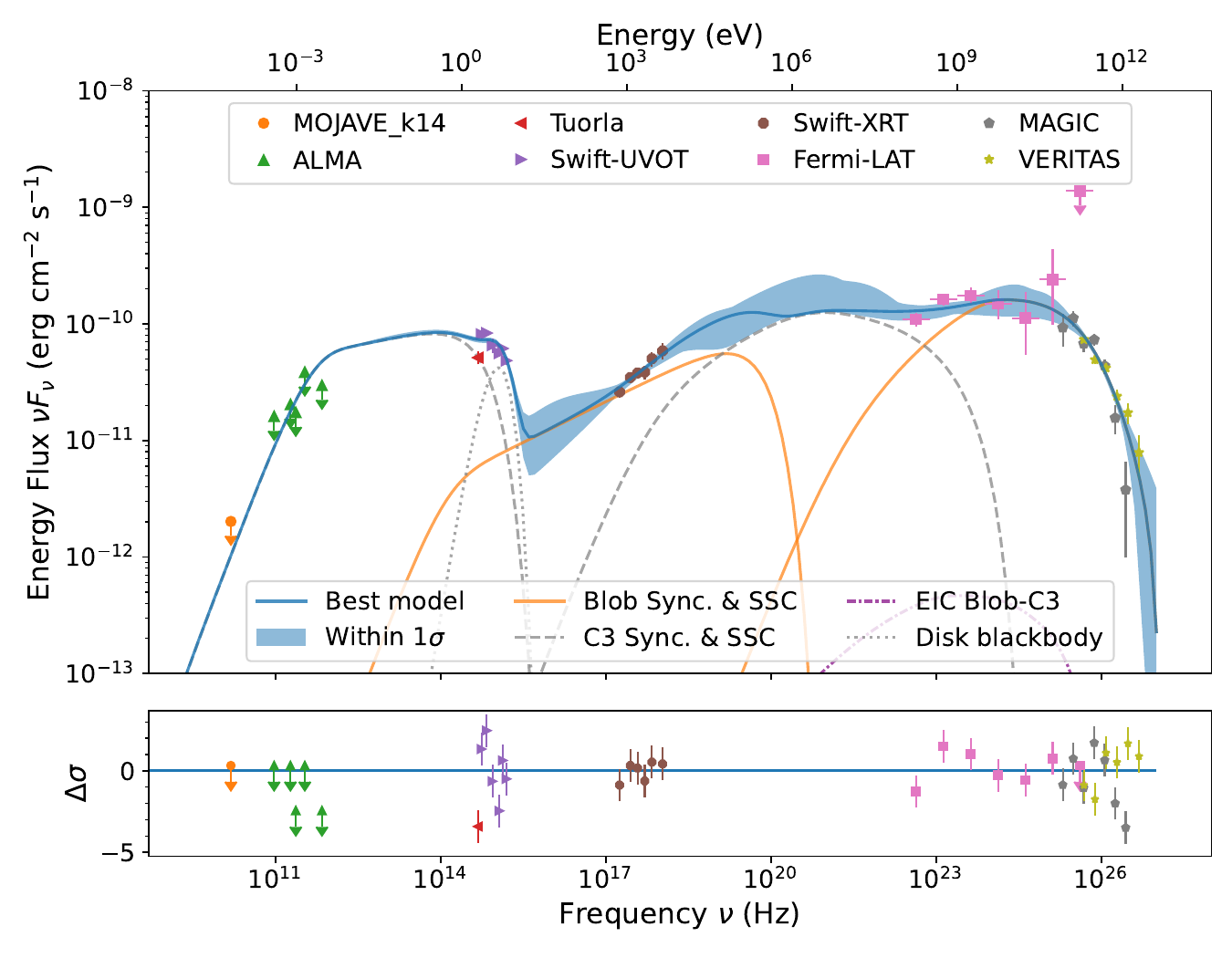}
\put(-210,150){\footnotesize{\bf 2017 Jan. 2}}
\caption{Multiwavelength SEDs with models and residuals of \gName{NGC}{1275} during the 2017 VHE flare (left) and one day after (right). Gray lines are for components considered steady over the two days: the C3 synchrotron and SSC emission (dashed), and the thermal emission from the accretion disk (dotted). These two components are fitted by eye and constrained from optical and radio data. Colored lines are linked to the blob emission that varies between the two states, fitted to the data with \texttt{Bjet\_MCMC}: the blob synchrotron and SSC contribution (plain orange), the second-order SSC emission (dashed orange), and the combined EIC emission of both the blob's particles on C3 photons and C3 particles on blob's photons (dotted-dashed purple). The sum of all components and associated $1 \sigma$ contours are shown in blue. 
The \gray{} emission is absorbed by the EBL following the model of \cite{Franceschini_2017}.} 
\label{figure:sed_modelling}
\end{figure*}

To investigate the possible evolution of the source during the flare, we construct two broadband SEDs covering the periods \decthirtyfirst{}/\janfirst{} and \jansecond{}. The period corresponds with the TeV observations conducted by both MAGIC and VERITAS. 
In addition to the VHE components, our SEDs comprise contemporaneous data in radio (MOJAVE, ALMA), and simultaneous data acquired from optical (Tuorla, \swiftUVOT{}), \swiftXRT{}, and \fermi{}. A geometric scheme of the radiative model is displayed in Figure \ref{figure:ngc1275_jet_components}. The broadband SEDs with modeled emission are shown in Figure \ref{figure:sed_modelling}, and their associated physical parameters in Table \ref{table:sed_model_params}.

\subsection{Multiwavelength approach}
Multiwavelength data sets provide specific constraints for our modeling process. 
During the night between \decthirtyfirst{} and \janfirst{}, the peak of the VHE flare was observed by MAGIC, reaching $\ssim 150\%$ Crab above 100 GeV \citep{MAGIC:2017ATEL9929}. From our Bayesian block \fermi{} analysis (see Figure \ref{figure:fermiLAT_variability}) we can see a simultaneous elevated state of the \fermi{} flux, which quickly died off after \janfirst{}.

This elevated \gray{} state does not show any simultaneous counterpart in other wavelengths (see Figure~\ref{figure:multiwavelength_daily_lightcurve}). The \janfirst{} hard X-ray spectral index of $\alpha = 1.81 \pm 0.06$ 
links its emission to the inverse-Compton process. 
However, the lack of a simultaneous joint-X-ray flare with \grays{}, which are also associated with the inverse-Compton process, provides strong evidence of multiple non-thermal radiation fields, possibly originating from different emission zones.

In this context, one would associate the \gray{} emission with a compact emission zone of fast variability of $\ssim 10$h \citep{MAGIC-2018-AA}. The X-ray emission displays slower variability of at least 10 days (only minor X-ray flux change is detected from the light curve shown in Figure \ref{figure:multiwavelength_daily_lightcurve}). This could originate from a more extended/slower emission zone, or from the same compact zone but with a much longer cooling time and/or some local equilibrium between injection and cooling at this energy range.

As discussed in the Introduction (Section~\ref{sec-introduction}), due to excellent radio-VLBI imaging, we have a good representation of the jet structure at \emph{mas} scale around the time of the flare. It appears that the brightest component emitting in radio at the closest epoch to the \gray{} flare is not the core, but the radio knot C3 \citep{Nagai-2014-ApJ} / k14 (MOJAVE). This component was measured by MOJAVE shortly before the \gray{} flare (2016 December 26) at a flux of 3.5 times the core at 15.3 GHz, at a distance of 2.61 \emph{mas} from the core (0.940 pc, sky-projected), and fitted by a symmetrical Gaussian with a size (FWHM) of 1.08 \emph{mas} (0.389 pc, sky-projected) \citep{Lister-2019-ApJ}.

\begin{figure}[h!]
\centering
\includegraphics[width=8.5cm]{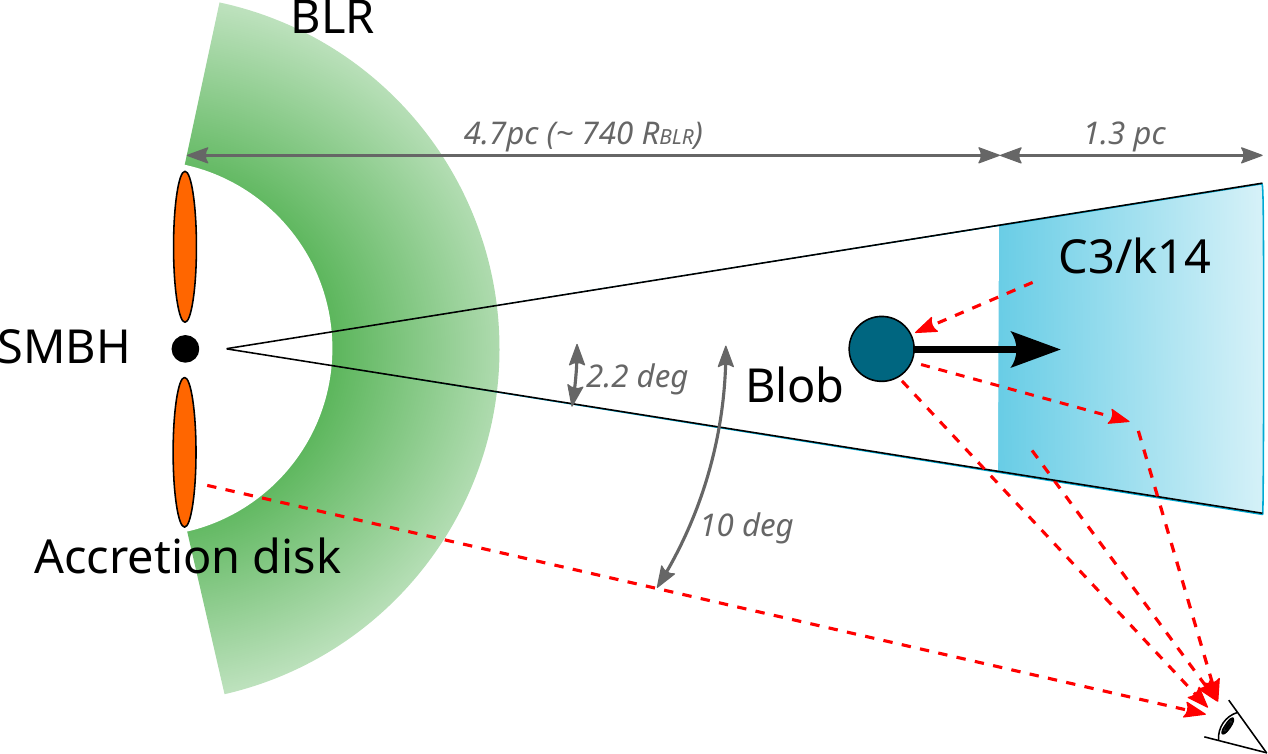}
\caption{Geometrical scheme of radiative components considered for the broadband SED modeling of \gName{NGC}{1275} (not to scale). The red-dashed lines represent the multiple radiative transfers taken into account.
In our code, the accretion disk is considered as a point-like source. With 5.4pc as the deprojected distance to the middle of C3, we take 4.7pc as the deprojected distance to the edge of C3.}
\label{figure:ngc1275_jet_components}
\end{figure}

In the following, we propose a multi-zone scenario where the compact emission zone producing the \gray{} flare interacts with C3/k14 (hereafter just referenced as C3). We use the leptonic SSC multi-zone package \texttt{Bjet\_MCMC} to model the multiwavelength SEDs of \janfirst{} and \jansecond{} \citep{Hervet_2015, Hervet_2024}.
The compact zone is approximated by a spherical blob filled with an isotropic magnetic field and a homogeneous population of electrons. The particle spectrum follows a broken power-law shape. The external inverse-Compton and \gray{} absorption induced by the blob interaction with the broad line region (BLR) is considered when the blob is close enough to the Supermassive Black Hole (SMBH). In this model, the C3 region consists of an SSC conical jet section. This section is further discretized into cylindrical slices with particle density and magnetic field decreasing according to the relative radius of each slice from the black hole. The radiative transfer of the blob emission through the jet is calculated, as well as the inverse-Compton interactions between the two components.
Our model's geometrical scheme (not to scale) is presented in Figure \ref{figure:ngc1275_jet_components}.

\subsection{Radiative structure of NGC 1275}

We consider an angle $\theta$ of $10^{\circ}$ between the pc-scale jet direction and the line of sight. 
As discussed further in the paper,  the value of $\theta$ in \gName{NGC}{1275} is still controversial and can significantly impact the validity of the proposed emission scenario. 
We consider the accretion disk spectrum as a monothermal blackbody that we adjust to the minimum observed optical flux of Tuorla corrected by the host galaxy emission (set as Tuorla\_min in Figure \ref{figure:sed_modelling}). From our best model, we favor an accretion disk with a temperature of $1.4 \times 10^4$ K and a luminosity of $5.0 \times 10^{43}$ erg s$^{-1}$.

At $\theta = 10^{\circ}$, the projected distance of C3 from the core translates to $\sim 5.4$pc ($D = D_{proj}/\sin \theta$). We estimate its length ($L$) and width ($W$) from the fitted Gaussian FWHM by MOJAVE as $L=1.3$ pc and $W=0.4$ pc.
We set the distance to the core, length, and average width of our conical jet section to the ones of C3. This leads to a half-opening angle of the jet of $2.2 ^{\circ}$, which is within the standard range for AGN jets \citep[e.g.,][]{Hervet_2016}.

The Doppler factor of C3 in our model is set at $\delta = 2.0$, with an associated apparent speed of $\beta_{\mathrm{app}} = 0.27$c. This is in good accordance with previous radio velocity measurements of C3, such as 0.23 c \citep{Nagai:2010PASJ}, 0.2 c \citep{Jorstad:2017ApJ}, or 0.33 c \citep{Kino2021ApJ}.

C3 appears to be fully outside the influence of the BLR radiation field. Indeed, the BLR distance at maximum density (or BLR radius, $r_{BLR}$) is usually deduced from the bolometric disk luminosity ($L_d$) as $r_{BLR} = 0.1 \sqrt{L_d / 1 \times 10^{46}~ \mathrm{erg~s}^{-1}}$ pc \citep{Sikora-2009-ApJ, Ghisellini-2009-MNRAS}.
In our case, $r_{BLR} = 6.3 \times 10^{-3}$ pc. Assuming a compact emission zone in the vicinity of C3, it would lie at $D \gtrsim 700~r_{BLR}$, making any absorption by pair creation or external inverse-Compton emission on the BLR photon field insignificant.

Applying the above geometric constraints, we first do a fit-by-eye of the C3 component to obtain the C3 parameters of our model. While the C3 parameters are then held fixed, we set most of the blob's parameters free to keep an agnostic view on how parameters change from \janfirst{} to \jansecond{}.
The constraints on the blob are:
\begin{itemize}
    \item The angle with the line of sight, set at $10^\circ$, which effectively constrains the Doppler factor to $\delta \leq 5.76$ ($\delta \leq 1/\sin(\theta)$).
    \item The blob's fastest variability, set at $t_{var} = 13.6$ h, corresponding to a $\sim 2$ sigma upper limit from the fastest doubling time scale deduced by MAGIC of $10.2 \pm 1.7$h. It constrains the Doppler factor and radius of the blob ($R/\delta \leq t_{var} c/(1+z)$).
    \item The blob distance to the SMBH is set at 4.63 pc, which is as close as possible to C3, but without having C3 obscuring the blob along the line of sight. Indeed, we observe that C3 is highly VHE opaque in our model from the pair production process.
\end{itemize}


In this specific model, we opt for what we term as a ``radio galaxy setup" linking the Lorentz factor $\Gamma$ to the Doppler factor $\delta$ and angle $\theta$.
The standard Lorenz factor equation 
is of quadratic form and consequently always admits two solutions of $\Gamma$, such as
\begin{equation}
    \Gamma = \frac{1}{\sqrt{1-\beta^2}},
    \label{Eq::Lorentz}
\end{equation} 

\noindent with
\begin{equation}
    \beta = \frac{v}{c} = \frac{\delta^2 \cos\theta  \boldsymbol{\pm} \sqrt{1-\delta^2 \sin^2\theta}}{1+\delta^2 \cos^2\theta}.
    \label{Eq::Speed}
\end{equation}

Considering example values for \gName{NGC}{1275} with $\delta = 3$ and  $\theta = 10^\circ$, we get the two possible solutions $\Gamma_\mathrm{min} = 1.8$ and $\Gamma_\mathrm{max} = 20$ for the blob. 
By convention, $\Gamma_\mathrm{min}$ is always favored for blazars. This choice is rooted in the general principle of stationary action, where natural processes tend to use the least amount of energy. 

In the case of a radio galaxy such as \gName{NGC}{1275}, with an expected low Doppler factor due to its large angle with the line of sight, the principle of stationary action loses its relevance. From the general AGN unification scheme, we actually expect to have similar jet Lorentz factors between radio galaxies and blazars ($\Gamma \sim 5-50$). Radio galaxies are, per definition, AGN with a misaligned jet with the line of sight. From this consideration and for the remainder of this study, we use the solution of Equation \ref{Eq::Lorentz} that maximizes the Lorentz factor, $\Gamma = \Gamma_\mathrm{max} = 20$, which is within standard boundaries of blazar values. 

\subsection{Blob-C3 radiative interactions}

Our model considers radiative interactions between the blob and the conical jet section C3.  The blob-jet interactions are based on \cite{Hervet_2015}, which considers a framework relatively similar to the ``spine-layer" (or ``spine-sheath") model \citep[e.g.][]{Ghisellini_2005, Tavecchio_2008, Sikora_2016}. 
In our case, the blob acts as the fast inner spine, while C3 acts as a slower layer. In the following, we reference C3 as the ``jet" component.

The relative Lorentz factor $\Gamma'$ between the blob and the jet is given by the equation 
\begin{equation}
\Gamma' =  \Gamma_b \Gamma_{j}  (1 - \beta_b \beta_{j}).
\end{equation}

Considering isotropic radiation fields in the blob (jet) frame, the specific intensity of jet (blob) $I_{\nu}$ is boosted by $\Gamma'$ with its radiation frequencies also boosted by $\Gamma'$, such as $I'_{\nu'} =\Gamma' I_{\nu \Gamma'}$. So, the external radiation field density in the blob (jet) frame $U'_\mathrm{ext}$ can be expressed as
\begin{equation}
U'_\mathrm{ext} = \frac{4 \pi}{c} \int_\nu I'_{\nu'} d\nu' = \Gamma'^2 U_\mathrm{ext}.
\end{equation}

In the \cite{Hervet_2015} study, only the EIC in the blob's frame was considered. In this study, we now include the EIC in the larger jet component frame between its particles and the blob synchrotron emission.
The jet being larger than the blob, we consider a dilution of the blob emission over the jet particles, in each slice of the jet as
\begin{equation}
I_{b,i}' = \Gamma' I_{b,i} R_b^2/R_{j,i}^2,
\end{equation}
with $I_{b,i}$ the intensity of the blob in each slice of the jet after a radiation transfer, including synchrotron self-absorption and photon-photon annihilation by pair creation.

Although we included this additional physical process in our model, it does not appear to be a significant player in the emission scenario displayed in Figure \ref{figure:sed_modelling}. In fact, we see that the combined external inverse-Compton emissions of blob-C3 (both the blob's particles scattering off C3 radiation and C3 particles scattering off the blob's radiation) at its peak energy during the flare, accounts for only up to $\sim 2\%$ of the total \gray ~emission.

\begin{deluxetable}{cccc}[h]
\tablecaption{{ Model parameters associated with the SEDs shown in Figure \ref{figure:sed_modelling}. Parameters without uncertainties and dashes for 2017 Jan 2 are manually fixed.}\label{table:sed_model_params}}
\tabletypesize{\scriptsize}
    \tablehead{
    \colhead{Parameter} & \colhead{2016 Dec 31/2017 Jan 1} & \colhead{2017 Jan 2} & \colhead{Unit}\\ \hline
    \colhead{$\theta$} & \colhead{$10.0$} & $-$ & \colhead{deg}\\ \hline
    \colhead{Blob} &  & &
    }
    \startdata
    $\delta$        & $5.55_{-1.07}^{+0.15}$ & $4.94_{-1.68}^{+0.76}$ & $-$ \\ 
    $N_{e}^{(1)}$   & $7.29_{-2.84}^{+0.46}$ & $4.06_{-2.73}^{+3.65}$ & $\log_{10}$ cm$^{-3}$\\ 
    $n_1$           & $2.43_{-1.26}^{+0.19}$ & $1.40_{-0.40}^{+1.05}$ & $-$\\
    $n_2$           & $2.61_{-0.09}^{+0.06}$ & $2.53_{-0.06}^{+0.12}$ & $-$\\
    $\gamma_{\mathrm{min}}$ & $1.83_{-1.80}^{+0.52}$ & $3.33_{-3.27}^{+0.38}$ & $\log_{10}$ $-$\\
    $\gamma_{\mathrm{max}}$ & $6.01_{-0.08}^{+0.11}$ & $6.29_{-0.10}^{+0.42}$ & $\log_{10}$ $-$\\
    $\gamma_{\mathrm{brk}}$ & $2.38_{-2.02}^{+3.38}$ & $3.61_{-0.15}^{+0.74}$ & $\log_{10}$ $-$\\
    $B$             & $-0.93_{-0.21}^{+0.16}$ & $-0.05_{-0.55}^{+0.05}$ & $\log_{10}$ G\\
    $R$ & $15.62_{-0.16}^{+0.22}$ & $14.89_{-0.11}^{+0.78}$ & $\log_{10}$ cm\\
    $D_{BH}$\tablenotemark{*} & $4.63$ & $-$ & pc\\
    \hline
    \colhead{Nucleus} &  &  & \\
    \hline
    $L_{disk}$ & $5.0\times 10^{43}$ & $-$ & erg s$^{-1}$\\ 
    $T_{disk}$ & $1.4\times 10^{4}$  & $-$ & K\\
    \hline
    \colhead{C3} &  &  & \\
    \hline
    $\delta$ & $2.0$ & $-$ & $-$\\ 
    $N_{e}^{(1)}$ & $3.5\times 10^{4}$ & $-$ & cm$^{-3}$\\ 
    $n$ & $2.80$ & $-$ & $-$\\
    $\gamma_{\mathrm{min}}$ & $7.0\times 10^{2}$ & $-$ & $-$\\
    $\gamma_{\mathrm{max}}$ & $2.8\times 10^{4}$ & $-$ & $-$\\
    $B_1$ & $1.0\times 10^{-1}$ & $-$ & G\\
    $R_1$ & $5.6\times 10^{17}$ & $-$ & cm\\
    $L$\tablenotemark{*} & $1.3$ & $-$ & pc\\
    $\alpha/2$\tablenotemark{*} & $2.2$ & $-$ & deg
\enddata
\tablenotetext{}{$\theta$ is the angle between the jet direction and the
line of sight. The electron energy distribution between Lorentz factors $\gamma_{\mathrm{min}}$ and
$\gamma_{\mathrm{max}}$ is given by a broken power law with indices $n_1$ and $n_1$ below and above
$\gamma_{\mathrm{brk}}$ , with $N_{e}^{(1)}$ the normalization factor at $\gamma = 1$. The blob Doppler factor,
magnetic field, radius, and distance to the black hole are given by $\delta$, $B$, $R$, and
$D_{BH}$ , respectively. The disk luminosity and temperature are given by $L_{disk}$ and
$T_{disk}$. The jet is characterized by a length of $L$, and an opening angle of $\alpha$. Its radius and magnetic field strength are set for the first slice as $R_1$ and $B_1$, respectively.}
\tablenotetext{*}{\textit{ Host galaxy frame.}}
\vspace{-6ex}
\end{deluxetable}

\subsection{Modeling results}

As mentioned above, we consider that only the blob varies between the two activity states reported. 
The main differences in the broadband SED between \janfirst{} and \jansecond{} are between the flux and shape of the inverse-Compton emission peak, as reported in Figure \ref{figure:fermi_veritas_combined_spectral_analysis}.
Overall, from \janfirst{} to \jansecond{}, the IC peak gets dimmer and broader.

Both states display overall good SED fits within our multi-zone scheme with a reduced chi-squared $\chi^2/\mathrm{dof} = 52.97 / 27 = 1.96$ for \janfirst{} and $\chi^2/\mathrm{dof} = 78.03 / 30 = 2.60$ for \jansecond{}. We observe in Fig. \ref{figure:sed_modelling} that the $\chi^2$ are mainly degraded by some discrepancies between datasets from different instruments, such as in optical between \textit{Swift}-UVOT and Tuorla, and in VHE between MAGIC and VERITAS. We note that neither the VERITAS nor MAGIC systematic uncertainties are included in their SED points, which amplifies their discrepancy.

The fit convergence is achieved through a newly-implemented bootstrapping method in \texttt{Bjet\_MCMC} where the parameters' profile likelihoods resulting from the result of a MCMC run are used to generate the prior distribution of a following MCMC run. MCMC runs are iterated in this manner until the MCMC posterior distributions of parameters are roughly centered on their profile likelihood maximums. This new method ensures a higher confidence in the final estimated parameter uncertainties. For both SED fits, we use 6000 steps, 45 walkers, and a burn-in phase of 300 steps. The convergence is achieved after three successive MCMC runs for \janfirst{}, and two MCMC runs for \jansecond{}. Corner plots of the final MCMC posterior distributions of parameters are given in Appendix \ref{Appendix::corner_plots}.

As shown in Table \ref{table:sed_model_params}, all fitted parameters of the blob are consistent with each other at a $2$ sigma level. This indicates that we cannot highlight any specific physical process that would be responsible for the observed dimming of the source after the flare. It subsequently means that, given the spectral point error bars and the parameter degeneracy of our model, the amplitude of the \gray{} dimming observed from \janfirst{} to \jansecond{} is not strong enough to significantly impact the parameters deduced from the fit.
Our initial tests with a fit-by-eye approach showed that changing only the two blob parameters $R$ and $N_{e}^{(1)}$ (an expanding blob with density decreasing in an adiabatic way) can provide a good fit for both SEDs. However, once freeing all the blob's parameters, this scenario is no longer favored.

\subsection{Modeling limitations and caveats}

An unconstrained multi-zone model fit with all parameters free would likely yield meaningless results. While there are good physical arguments for the constraints applied in our model, nonetheless by manually fixing all parameters of the C3 emission zone, and the location of the blob in the jet, we have to remind the reader that the full parameter space of our multi-zone model is not fully explored, and that the parameter uncertainties noted in Table \ref{table:sed_model_params} need to be treated with caution.

As seen in Figure \ref{figure:sed_modelling}, the blob is found 
to be heavily Compton-dominated. Hence, inverse-Compton (IC) is the main process that cools particles at the highest energies. The IC cooling timescale in the observer frame at the maximum particle energy for \janfirst{} is $T_\mathrm{cool}(\gamma_\mathrm{max}) \sim 2$ min.  
If the particle acceleration processes were to stop at the flare maximum, the source would effectively go silent in 
VHE at about a blob light-crossing timescale $\tau_\mathrm{min,obs} = 7$ h. In order to maintain the flux at the level observed on \jansecond{}, one would need a powerful 
continuous particle acceleration process in action.
This cooling time issue could be mitigated if we consider a larger Doppler factor (smaller jet angle) which would effectively reduce the energetics in the blob's frame. Other possible scenarios unexplored in this study could involve another emission zone significantly contributing to the \gray{} emission, or a hadronic origin of \grays{}. Indeed, protons have significantly longer cooling times than electrons when in a similar environment.
We leave these doors open for further studies.

\subsection{Previous models of NGC 1275 and the issue of its jet angle with the line of sight}

Radio galaxies, by definition, display jets with large angles to the line of sight. This feature is highlighted by extended radio jets ($\gtrsim  100$ kpc) and/or the observation of counter jets. 
There is usually no clear definition of what exactly large angles are, but the general understanding is that they are large enough to have relatively small effects from Doppler boosting and relativistic aberration compared to blazars. For example, at $\theta \geq 20^{\circ}$, the Doppler factor must be less than 3 ($\delta \leq 1/\sin(\theta)$), which would be incompatible with most blazars, where $\delta \gtrsim 10$.

TeV-detected radio galaxies have challenged the large-angle assumption. Some displayed fast flares or SEDs strongly suggesting a pc-scale ``blazar core" with jets bending out of the LOS at a larger scale, making them appear as radio galaxies, such as IC 310 \citep{Kadler_2012} or  PKS 0625-354 \citep{HESS_2018, HESS_2024}. Some others, such as 3C 264 and M87 display highly superluminal apparent motions in their jets, putting strong constraints for viewing angles $<20^{\circ}$ \citep{Biretta_1999, Meyer_2015}.

NGC 1275 stands out by the inconsistency of the various estimates of its jet viewing angle. From the jet/counter jet flux ratio and an apparent speed of $B_\mathrm{app} \sim [0.3-0.5]$ c in 8.4 GHz VLBA observations, \cite{Walker:1994ApJ} estimated a jet angle within the $30^{\circ}-55^{\circ}$ range. This was further extended to $65^{\circ} \pm 16$ by \cite{Fujita:2017mnras}, using a similar method but with $B_\mathrm{app} \sim 0.23$ c from 43 GHz VLBA observations.
On the other hand, \cite{Lister_2009} noted that the two lobe motions are consistent with a viewing angle of about $11^{\circ}$ and a weak Lorentz factor of 0.6. Using constraints from the observed jet opening angle, \cite{Jorstad:2017ApJ} deduced an even smaller viewing angle of about $4^{\circ}$. However, \citet{Jorstad:2017ApJ} note that this small viewing angle for \gName{NGC}{1275} corresponds to a Doppler factor of $\delta \ssim 11$, implying proper motions of $\ssim0.35$ \textit{mas} per month should be seen, which they and others have not detected.

We highlight that the blob in our model is radio-quiet, as seen in the SEDs in Figure \ref{figure:sed_modelling}. Such a fast-moving blob should be undetectable in radio-VLBI. Previous radio studies constraining the jet Doppler factor from radio-VLBI motion did not consider two-flow jets, and thus do not preclude the existence of a faster, radio-quiet inner jet.

The radio component C3 used in our study, with a measured apparent radius ($R_\mathrm{C3} = 0.2$ pc) and core distance ($D_\mathrm{proj,C3} = 0.94$ pc), also brings constraints on the maximum jet angle of \gName{NGC}{1275}. Indeed, from large blazar samples observed in radio VLBI, we can consider a maximum intrinsic jet opening angle of $\alpha_\mathrm{max} = 10^\circ$ \citep{Hervet_2016, Jorstad:2017ApJ}. Hence, we deduce a maximum jet angle of  
\begin{equation}
\theta_\mathrm{max} = \arcsin \left[ \tan\left(\frac{\alpha_\mathrm{max}}{2}\right)  \frac{D_\mathrm{proj,C3}}{R_\mathrm{C3}} \right] = 24^\circ.
\end{equation}

Prior SED modeling papers have used NGC 1275 to test scenarios that would allow \grays{} to be detected at large viewing angles.
The most notable is likely the ``spine-layer" model set up by \cite{Ghisellini_2005} and applied to NGC~1275 by \cite{Tavecchio-2014-MNRAS}, which is relatively close in its radiative structure to the multizone model \texttt{Bjet} used in this study. In contrast to our assumptions, they suggested that the layer was responsible for the VHE emission, not the blob. They were able to successfully fit an SED considering a viewing angle of $18^{\circ}$. However, since the emission comes from a large emission zone with weak Doppler boosting, they could not address variability timescales less than a week, such as the one studied in this paper. Their angle assumption was used in other studies, such as \cite{Giovannini-2018-NatAs}.
In their 2018 detection paper, the MAGIC Collaboration proposed an innovative scenario to explain NGC1275 VHE flares at large angles, with \grays{} emitted directly from the SMBH magnetosphere \citep{MAGIC-2018-AA}. However, they did not attempt a MWL SED modeling or address how these \grays{} would escape the opacity of the broad-line region.




To date, no model besides the one presented in this study has successfully fitted the SED from radio to \grays{} of \gName{NGC}{1275} while being consistent with a daily flare timescale. 
Our model considers a blob Doppler factor of $\delta \in [\sim3-6]$  
for an angle with the line of sight of $\theta = 10^\circ$. 
This model should remain valid with a Doppler factor of 3 up to an angle with the line of sight of $\sim 20^\circ$, which would only affect the geometry of the jet. No acceptable broadband SED fits were found above this limit, where the Klein-Nishina regime and pair production opacity prevent reaching energies above $\sim 100$ GeV while being consistent with daily variability.

\section{Conclusion}\label{sec:conclusion}

This paper describes the long-term VERITAS VHE \gray{} observations of the radio galaxy \gName{NGC}{1275} with a specific focus on the spectacular \gray{} flare of \decthirtyfirst{} - \janthird{}. The Compton peak SEDs are produced with \fermi{}, MAGIC, and VERITAS data showing a strong evolution in the spectral shape over the flare period.
Multiwavelength lightcurves are presented for the period 2009 September to 2017 June from VERITAS, MAGIC, \fermi{}, \swiftXRT{}, \swiftUVOT{}, as well as optical and radio observatories. For the first time, full multiwavelength SEDs for the nights of \janfirst{} and 2 are constructed. A blob-in-jet multi-zone SED model is fit to both SEDs where the specific conditions of the C3 radio component are used to constrain the model.

Our findings reveal several key insights. First, the \gray{} spectral parameters observed during the low state are consistent with those reported by MAGIC, falling within the margin of error. Notably, we report for the first time the detection of a more general ``high'' flux state that does not correspond to extreme flaring activity. Additionally, model comparison testing of the VHE SEDs derived using VERITAS data indicates that a simple power-law model is statistically favored over more complex alternatives. Through our joint \fermi{}-VHE analysis, we also identify a clear evolution in the shape of the Compton peak during the flare and its subsequent decline. In examining the Compton peak SEDs, we find that the spectral index of the fitted power law models systematically hardens as the flux brightness increases across observed states. 


The broadband SED of \gName{NGC}{1275} displayed a complex shape with a relatively narrow synchrotron peak (from FIR to optical) compared to its high energy peak spanning from X-ray to VHE. Also, the \janfirst{} flare appears to contribute significantly only to \gray{} emission. These observations point to a more complex model than a simple one-zone SSC scenario for the \gName{NGC}{1275} flare.  We instead consider a multi-zone leptonic scenario, also motivated by observations showing that the C3 radio component was observed to be brighter than the radio core during the months around the flaring period.  Thus, in our model, the synchrotron emission is dominated by the radio component C3, while most of the energy output from X-rays and above comes from a compact relativistic blob in the jet. 
The hard spectral index observed in X-rays is interpreted as coming from the combined contributions of the blob synchrotron and the C3 SSC emissions.

The exact location of the blob in the jet remains uncertain, but we exclude it from being within C3 or too close to the nucleus, as their opacities will prevent any detection up to the TeV energies, as observed during the flare. After fitting the two SEDs with our model, the observed flux decrease in the \gray{} band does not yield any significant changes in the blob parameters.  Hence, the main phenomenon responsible for the flare cooling phase remains uncertain.

Our model is consistent with all measured VLBI sizes, distances, and apparent speeds when considering a jet angle with the line of sight of $10^\circ$. We consider that our proposed emission scenario will lose its relevance only for angles above $20^\circ$.

Having this extraordinary VHE flare happening within the same period of the detected acceleration of the C3 component in the jet indicates some links between these two events, especially since the coincidence of \gray{} flares with radio knot ejections has now been reported in multiple blazars \citep[e.g.,][]{Marscher_2010, Abeysekara_2018, Lico_2022}.
One can interpret this flare as the outcome of a strong relativistic jet perturbation interacting with a slower component in the vicinity of C3, leading to diffuse shock acceleration or magnetic reconnection. This relativistic flow would eventually drag and accelerate the C3 zone along the jet as seen in radio VLBI observations.

Due to its proximity to us and relatively high jet angle with the line of sight, \gName{NGC}{1275} is one of the few TeV radio galaxies that allows us to probe its jet features at multiple scales, which is especially relevant in studying the links between fast high-energy flares and its large-scale jet structure, as done in this paper. Further deep multiwavelength campaigns in the low and outburst states of the source, as well as more advanced modeling tools using time evolution,  promise to shed new light on AGN jet emission mechanisms from the ideal cosmic laboratory that is \gName{NGC}{1275}.\\
\\

\section*{Acknowledgments}\label{sec:acknowledgements}

VERITAS is supported by grants from the U.S. Department of Energy Office of Science, the U.S. National Science Foundation and the Smithsonian Institution, and by NSERC in Canada. We acknowledge the excellent work of the technical support staff at the Fred Lawrence Whipple Observatory and at the collaborating institutions in the construction and operation of the instrument.

The \textit{Fermi} LAT Collaboration acknowledges generous ongoing support from a number of agencies and institutes that have supported both the
development and the operation of the LAT as well as scientific data analysis. These include the National Aeronautics and Space Administration and the
Department of Energy in the United States, the Commissariat \`a l'Energie Atomique and the Centre National de la Recherche Scientifique / Institut National de Physique
Nucl\'eaire et de Physique des Particules in France, the Agenzia Spaziale Italiana and the Istituto Nazionale di Fisica Nucleare in Italy, the Ministry of Education,
Culture, Sports, Science and Technology (MEXT), High Energy Accelerator Research Organization (KEK) and Japan Aerospace Exploration Agency (JAXA) in Japan, and
the K.~A.~Wallenberg Foundation, the Swedish Research Council and the Swedish National Space Board in Sweden.
 
Additional support for science analysis during the operations phase is gratefully acknowledged from the Istituto Nazionale di Astrofisica in Italy and the Centre
National d'\'Etudes Spatiales in France. This work performed in part under DOE Contract DE-AC02-76SF00515.

This research has made use of the NASA/IPAC Extragalactic Database (NED) which is operated by the Jet Propulsion Laboratory, California Institute of Technology, under contract with the National Aeronautics and Space Administration.

The National Radio Astronomy Observatory is a facility of the National Science Foundation operated under cooperative agreement by Associated Universities, Inc.  This paper makes use of the following ALMA data sets: 2012.1.00394, 2016.1.01305, 2017.1.01257, and 2018.1.01438. ALMA is a partnership of ESO (representing its member states), NSF (USA) and NINS (Japan), together with NRC (Canada), MOST and ASIAA (Taiwan), and KASI (Republic of Korea), in cooperation with the Republic of Chile. The Joint ALMA Observatory is operated by ESO, AUI/NRAO and NAOJ.

O.H. thanks NSF for support under grants PHY-2011420 and PHY-2310002. L.F.F. acknowledges partial support under NSF grants PHY-2110737 and PHY-2513660.
C.R. acknowledges the financial support of the UK Science and Technology Facilities Council consolidated grant ST/X001075/1.


\bibliographystyle{aasjournal}
\bibliography{references-NGC1275-flare-2016}

@article{Laing-2014-MNRAS,
	adsurl = {https://ui.adsabs.harvard.edu/abs/2014MNRAS.437.3405L},
	archiveprefix = {arXiv},
	author = {{Laing}, R.~A. and {Bridle}, A.~H.},
	doi = {10.1093/mnras/stt2138},
	eprint = {1311.1015},
	journal = {\mnras},
	month = feb,
	number = {4},
	pages = {3405-3441},
	primaryclass = {astro-ph.CO},
	title = {{Systematic properties of decelerating relativistic jets in low-luminosity radio galaxies}},
	volume = {437},
	year = 2014}

@article{Stawarz-2002-ApJ,
	adsurl = {https://ui.adsabs.harvard.edu/abs/2002ApJ...578..763S},
	archiveprefix = {arXiv},
	author = {{Stawarz}, {\L}. and {Ostrowski}, M.},
	doi = {10.1086/342649},
	eprint = {astro-ph/0203040},
	journal = {\apj},
	month = oct,
	number = {2},
	pages = {763-774},
	primaryclass = {astro-ph},
	title = {{Radiation from the Relativistic Jet: A Role of the Shear Boundary Layer}},
	volume = {578},
	year = 2002}

@article{Rieger-2004-ApJ,
	adsurl = {https://ui.adsabs.harvard.edu/abs/2004ApJ...617..155R},
	archiveprefix = {arXiv},
	author = {{Rieger}, Frank M. and {Duffy}, Peter},
	doi = {10.1086/425167},
	eprint = {astro-ph/0410269},
	journal = {\apj},
	month = dec,
	number = {1},
	pages = {155-161},
	primaryclass = {astro-ph},
	title = {{Shear Acceleration in Relativistic Astrophysical Jets}},
	volume = {617},
	year = 2004}

@article{Hovatta:2009aa,
	adsurl = {https://ui.adsabs.harvard.edu/abs/2009A&A...494..527H},
	archiveprefix = {arXiv},
	author = {{Hovatta}, T. and {Valtaoja}, E. and {Tornikoski}, M. and {L{\"a}hteenm{\"a}ki}, A.},
	doi = {10.1051/0004-6361:200811150},
	eprint = {0811.4278},
	journal = {\aap},
	month = feb,
	number = {2},
	pages = {527-537},
	primaryclass = {astro-ph},
	title = {{Doppler factors, Lorentz factors and viewing angles for quasars, BL Lacertae objects and radio galaxies}},
	volume = {494},
	year = 2009}

@article{Giovannini-2001-ApJ,
	adsurl = {https://ui.adsabs.harvard.edu/abs/2001ApJ...552..508G},
	archiveprefix = {arXiv},
	author = {{Giovannini}, G. and {Cotton}, W.~D. and {Feretti}, L. and {Lara}, L. and {Venturi}, T.},
	doi = {10.1086/320581},
	eprint = {astro-ph/0101096},
	journal = {\apj},
	month = may,
	number = {2},
	pages = {508-526},
	primaryclass = {astro-ph},
	title = {{VLBI Observations of a Complete Sample of Radio Galaxies: 10 Years Later}},
	volume = {552},
	year = 2001}

@article{Generalized-LiMa-2004aa,
	adsurl = {https://ui.adsabs.harvard.edu/abs/2004A&A...421..529A},
	archiveprefix = {arXiv},
	author = {{Aharonian}, F. and {Akhperjanian}, A. and {Beilicke}, M. and {Bernl{\"o}hr}, K. and {B{\"o}rst}, H. -G. and {Bojahr}, H. and {Bolz}, O. and {Coarasa}, T. and {Contreras}, J.~L. and {Cortina}, J. and {Denninghoff}, S. and {Fonseca}, V. and {Girma}, M. and {G{\"o}tting}, N. and {Heinzelmann}, G. and {Hermann}, G. and {Heusler}, A. and {Hofmann}, W. and {Horns}, D. and {Jung}, I. and {Kankanyan}, R. and {Kestel}, M. and {Konopelko}, A. and {Kornmeyer}, H. and {Kranich}, D. and {Lampeitl}, H. and {Lopez}, M. and {Lorenz}, E. and {Lucarelli}, F. and {Mang}, O. and {Mazin}, D. and {Meyer}, H. and {Mirzoyan}, R. and {Moralejo}, A. and {Ona-Wilhelmi}, E. and {Panter}, M. and {Plyasheshnikov}, A. and {P{\"u}hlhofer}, G. and {de los Reyes}, R. and {Rhode}, W. and {Ripken}, J. and {Rowell}, G. and {Sahakian}, V. and {Samorski}, M. and {Schilling}, M. and {Siems}, M. and {Sobzynska}, D. and {Stamm}, W. and {Tluczykont}, M. and {Vitale}, V. and {V{\"o}lk}, H.~J. and {Wiedner}, C.~A. and {Wittek}, W.},
	doi = {10.1051/0004-6361:20035764},
	eprint = {astro-ph/0401301},
	journal = {\aap},
	month = jul,
	pages = {529-537},
	primaryclass = {astro-ph},
	title = {{Observations of 54 Active Galactic Nuclei with the HEGRA system of Cherenkov telescopes}},
	volume = {421},
	year = 2004}

@article{Marscher_2010,
	adsurl = {https://ui.adsabs.harvard.edu/abs/2010ApJ...710L.126M},
	archiveprefix = {arXiv},
	author = {{Marscher}, Alan P. and {Jorstad}, Svetlana G. and {Larionov}, Valeri M. and {Aller}, Margo F. and {Aller}, Hugh D. and {L{\"a}hteenm{\"a}ki}, Anne and {Agudo}, Iv{\'a}n and {Smith}, Paul S. and {Gurwell}, Mark and {Hagen-Thorn}, Vladimir A. and {Konstantinova}, Tatiana S. and {Larionova}, Elena G. and {Larionova}, Liudmila V. and {Melnichuk}, Daria A. and {Blinov}, Dmitry A. and {Kopatskaya}, Evgenia N. and {Troitsky}, Ivan S. and {Tornikoski}, Merja and {Hovatta}, Talvikki and {Schmidt}, Gary D. and {D'Arcangelo}, Francesca D. and {Bhattarai}, Dipesh and {Taylor}, Brian and {Olmstead}, Alice R. and {Manne-Nicholas}, Emily and {Roca-Sogorb}, Mar and {G{\'o}mez}, Jos{\'e} L. and {McHardy}, Ian M. and {Kurtanidze}, Omar and {Nikolashvili}, Maria G. and {Kimeridze}, Givi N. and {Sigua}, Lorand A.},
	doi = {10.1088/2041-8205/710/2/L126},
	eprint = {1001.2574},
	journal = {\apjl},
	month = feb,
	number = {2},
	pages = {L126-L131},
	primaryclass = {astro-ph.CO},
	title = {{Probing the Inner Jet of the Quasar PKS 1510-089 with Multi-Waveband Monitoring During Strong Gamma-Ray Activity}},
	volume = {710},
	year = 2010}

@article{Abeysekara_2018,
	adsurl = {https://ui.adsabs.harvard.edu/abs/2018ApJ...856...95A},
	archiveprefix = {arXiv},
	author = {{Abeysekara}, A.~U. and {Benbow}, W. and {Bird}, R. and {Brantseg}, T. and {Brose}, R. and {Buchovecky}, M. and {Buckley}, J.~H. and {Bugaev}, V. and {Connolly}, M.~P. and {Cui}, W. and {Daniel}, M.~K. and {Falcone}, A. and {Feng}, Q. and {Finley}, J.~P. and {Fortson}, L. and {Furniss}, A. and {Gillanders}, G.~H. and {Gunawardhana}, I. and {H{\"u}tten}, M. and {Hanna}, D. and {Hervet}, O. and {Holder}, J. and {Hughes}, G. and {Humensky}, T.~B. and {Johnson}, C.~A. and {Kaaret}, P. and {Kar}, P. and {Kertzman}, M. and {Krennrich}, F. and {Lang}, M.~J. and {Lin}, T.~T.~Y. and {McArthur}, S. and {Moriarty}, P. and {Mukherjee}, R. and {O'Brien}, S. and {Ong}, R.~A. and {Otte}, A.~N. and {Park}, N. and {Petrashyk}, A. and {Pohl}, M. and {Pueschel}, E. and {Quinn}, J. and {Ragan}, K. and {Reynolds}, P.~T. and {Richards}, G.~T. and {Roache}, E. and {Rulten}, C. and {Sadeh}, I. and {Santander}, M. and {Sembroski}, G.~H. and {Shahinyan}, K. and {Wakely}, S.~P. and {Weinstein}, A. and {Wells}, R.~M. and {Wilcox}, P. and {Williams}, D.~A. and {Zitzer}, B. and {VERITAS Collaboration} and {Jorstad}, S.~G. and {Marscher}, A.~P. and {Lister}, M.~L. and {Kovalev}, Y.~Y. and {Pushkarev}, A.~B. and {Savolainen}, T. and {Agudo}, I. and {Molina}, S.~N. and {G{\'o}mez}, J.~L. and {Larionov}, V.~M. and {Borman}, G.~A. and {Mokrushina}, A.~A. and {Tornikoski}, M. and {L{\"a}hteenm{\"a}ki}, A. and {Chamani}, W. and {Enestam}, S. and {Kiehlmann}, S. and {Hovatta}, T. and {Smith}, P.~S. and {Pontrelli}, P.},
	doi = {10.3847/1538-4357/aab35c},
	eid = {95},
	eprint = {1802.10113},
	journal = {\apj},
	month = apr,
	number = {2},
	pages = {95},
	primaryclass = {astro-ph.HE},
	title = {{Multiwavelength Observations of the Blazar BL Lacertae: A New Fast TeV Gamma-Ray Flare}},
	volume = {856},
	year = 2018}

@article{Lico_2022,
	adsurl = {https://ui.adsabs.harvard.edu/abs/2022A&A...658L..10L},
	archiveprefix = {arXiv},
	author = {{Lico}, R. and {Casadio}, C. and {Jorstad}, S.~G. and {G{\'o}mez}, J.~L. and {Marscher}, A.~P. and {Traianou}, E. and {Kim}, J. -Y. and {Zhao}, G. -Y. and {Fuentes}, A. and {Cho}, I. and {Krichbaum}, T.~P. and {Hervet}, O. and {O'Brien}, S. and {Boccardi}, B. and {Myserlis}, I. and {Agudo}, I. and {Alberdi}, A. and {Weaver}, Z.~R. and {Zensus}, J.~A.},
	doi = {10.1051/0004-6361/202142948},
	eid = {L10},
	eprint = {2202.02523},
	journal = {\aap},
	month = feb,
	pages = {L10},
	primaryclass = {astro-ph.HE},
	title = {{New jet feature in the parsec-scale jet of the blazar OJ 287 connected to the 2017 teraelectronvolt flaring activity}},
	volume = {658},
	year = 2022}

@article{Lister_2009,
	adsurl = {https://ui.adsabs.harvard.edu/abs/2009AJ....138.1874L},
	archiveprefix = {arXiv},
	author = {{Lister}, M.~L. and {Cohen}, M.~H. and {Homan}, D.~C. and {Kadler}, M. and {Kellermann}, K.~I. and {Kovalev}, Y.~Y. and {Ros}, E. and {Savolainen}, T. and {Zensus}, J.~A.},
	doi = {10.1088/0004-6256/138/6/1874},
	eprint = {0909.5100},
	journal = {\aj},
	month = dec,
	number = {6},
	pages = {1874-1892},
	primaryclass = {astro-ph.CO},
	title = {{MOJAVE: Monitoring of Jets in Active Galactic Nuclei with VLBA Experiments. VI. Kinematics Analysis of a Complete Sample of Blazar Jets}},
	volume = {138},
	year = 2009}

@article{Wagenmakers-2004,
	author = {Wagenmakers, Eric-Jan and Farrell, Simon},
	date = {2004/02/01},
	doi = {10.3758/BF03206482},
	id = {Wagenmakers2004},
	isbn = {1531-5320},
	journal = {Psychonomic Bulletin \& Review},
	number = {1},
	pages = {192--196},
	title = {AIC model selection using Akaike weights},
	url = {https://doi.org/10.3758/BF03206482},
	volume = {11},
	year = {2004}}

@article{Burnham-2011,
	author = {Burnham, Kenneth P. and Anderson, David R. and Huyvaert, Kathryn P.},
	date = {2011/01/01},
	doi = {10.1007/s00265-010-1029-6},
	id = {Burnham2011},
	isbn = {1432-0762},
	journal = {Behavioral Ecology and Sociobiology},
	number = {1},
	pages = {23--35},
	title = {AIC model selection and multimodel inference in behavioral ecology: some background, observations, and comparisons},
	url = {https://doi.org/10.1007/s00265-010-1029-6},
	volume = {65},
	year = {2011}}

@article{Protassov-2002-ApJ,
	adsurl = {https://ui.adsabs.harvard.edu/abs/2002ApJ...571..545P},
	archiveprefix = {arXiv},
	author = {{Protassov}, Rostislav and {van Dyk}, David A. and {Connors}, Alanna and {Kashyap}, Vinay L. and {Siemiginowska}, Aneta},
	doi = {10.1086/339856},
	eprint = {astro-ph/0201547},
	journal = {\apj},
	month = may,
	number = {1},
	pages = {545-559},
	primaryclass = {astro-ph},
	title = {{Statistics, Handle with Care: Detecting Multiple Model Components with the Likelihood Ratio Test}},
	volume = {571},
	year = 2002}

@article{Wilks-1938,
	author = {S. S. Wilks},
	issn = {00034851, 21688990},
	journal = {The Annals of Mathematical Statistics},
	number = {1},
	pages = {60--62},
	publisher = {Institute of Mathematical Statistics},
	title = {The Large-Sample Distribution of the Likelihood Ratio for Testing Composite Hypotheses},
	url = {http://www.jstor.org/stable/2957648},
	urldate = {2025-04-17},
	volume = {9},
	year = {1938}}

@article{Biretta_1999,
	adsurl = {https://ui.adsabs.harvard.edu/abs/1999ApJ...520..621B},
	author = {{Biretta}, J.~A. and {Sparks}, W.~B. and {Macchetto}, F.},
	doi = {10.1086/307499},
	journal = {\apj},
	month = aug,
	number = {2},
	pages = {621-626},
	title = {{Hubble Space Telescope Observations of Superluminal Motion in the M87 Jet}},
	volume = {520},
	year = 1999}

@article{Meyer_2015,
	adsurl = {https://ui.adsabs.harvard.edu/abs/2015Natur.521..495M},
	author = {{Meyer}, Eileen T. and {Georganopoulos}, Markos and {Sparks}, William B. and {Perlman}, Eric and {van der Marel}, Roeland P. and {Anderson}, Jay and {Sohn}, Sangmo Tony and {Biretta}, John and {Norman}, Colin and {Chiaberge}, Marco},
	doi = {10.1038/nature14481},
	journal = {\nat},
	month = may,
	number = {7553},
	pages = {495-497},
	title = {{A kiloparsec-scale internal shock collision in the jet of a nearby radio galaxy}},
	volume = {521},
	year = 2015}

@article{HESS_2024,
	adsurl = {https://ui.adsabs.harvard.edu/abs/2024A&A...683A..70H},
	archiveprefix = {arXiv},
	author = {{H.~E.~S.~S. Collaboration} and {Aharonian}, F. and {Ait Benkhali}, F. and {Aschersleben}, J. and {Ashkar}, H. and {Backes}, M. and {Baktash}, A. and {Barbosa Martins}, V. and {Barnard}, J. and {Batzofin}, R. and {Becherini}, Y. and {Berge}, D. and {Bernl{\"o}hr}, K. and {Bi}, B. and {B{\"o}ttcher}, M. and {Boisson}, C. and {Bolmont}, J. and {de Bony de Lavergne}, M. and {Borowska}, J. and {Bradascio}, F. and {Breuhaus}, M. and {Brose}, R. and {Brown}, A. and {Brun}, F. and {Bruno}, B. and {Bulik}, T. and {Burger-Scheidlin}, C. and {Bylund}, T. and {Caroff}, S. and {Casanova}, S. and {Cecil}, R. and {Celic}, J. and {Cerruti}, M. and {Chand}, T. and {Chandra}, S. and {Chen}, A. and {Chibueze}, J. and {Chibueze}, O. and {Cotter}, G. and {Damascene Mbarubucyeye}, J. and {Davids}, I.~D. and {Djuvsland}, J. and {Dmytriiev}, A. and {Doroshenko}, V. and {Egberts}, K. and {Einecke}, S. and {Ernenwein}, J. -P. and {Fontaine}, G. and {F{\"u}{\ss}ling}, M. and {Funk}, S. and {Gabici}, S. and {Ghafourizadeh}, S. and {Giavitto}, G. and {Glawion}, D. and {Glicenstein}, J.~F. and {Glombitza}, J. and {Goswami}, P. and {Grolleron}, G. and {Haerer}, L. and {Hinton}, J.~A. and {Holch}, T.~L. and {Holler}, M. and {Horns}, D. and {Jamrozy}, M. and {Jankowsky}, F. and {Joshi}, V. and {Jung-Richardt}, I. and {Kasai}, E. and {Katarzy{\'n}ski}, K. and {Khatoon}, R. and {Kh{\'e}lifi}, B. and {Klu{\'z}niak}, W. and {Komin}, Nu. and {Kosack}, K. and {Kostunin}, D. and {Lang}, R.~G. and {Le Stum}, S. and {Leitl}, F. and {Lemi{\`e}re}, A. and {Lenain}, J. -P. and {Leuschner}, F. and {Luashvili}, A. and {Mackey}, J. and {Marx}, R. and {Mehta}, A. and {Meyer}, M. and {Mitchell}, A. and {Moderski}, R. and {Montanari}, A. and {Moulin}, E. and {de Naurois}, M. and {Niemiec}, J. and {O'Brien}, P. and {Ohm}, S. and {Olivera-Nieto}, L. and {de Ona Wilhelmi}, E. and {Ostrowski}, M. and {Panny}, S. and {Parsons}, R.~D. and {Pita}, S. and {Prokhorov}, D.~A. and {P{\"u}hlhofer}, G. and {Punch}, M. and {Quirrenbach}, A. and {Reichherzer}, P. and {Reimer}, A. and {Reimer}, O. and {Ren}, H. and {Rieger}, F. and {Rudak}, B. and {Sahakian}, V. and {Salzmann}, H. and {Sanchez}, D.~A. and {Sasaki}, M. and {Sch{\"u}ssler}, F. and {Schutte}, H.~M. and {Shapopi}, J.~N.~S. and {Sol}, H. and {Specovius}, A. and {Spencer}, S. and {Stawarz}, {\L}. and {Steenkamp}, R. and {Steinmassl}, S. and {Streil}, K. and {Sushch}, I. and {Suzuki}, H. and {Takahashi}, T. and {Tanaka}, T. and {van Eldik}, C. and {Vecchi}, M. and {Veh}, J. and {Venter}, C. and {Wagner}, S.~J. and {Wierzcholska}, A. and {Zacharias}, M. and {Zargaryan}, D. and {Zdziarski}, A.~A. and {Zech}, A. and {Zouari}, S. and {{\.Z}ywucka}, N.},
	doi = {10.1051/0004-6361/202348063},
	eid = {A70},
	eprint = {2401.07071},
	journal = {\aap},
	month = mar,
	pages = {A70},
	primaryclass = {astro-ph.HE},
	title = {{TeV flaring activity of the AGN PKS 0625-354 in November 2018}},
	volume = {683},
	year = 2024}

@article{HESS_2018,
	adsurl = {https://ui.adsabs.harvard.edu/abs/2018MNRAS.476.4187H},
	archiveprefix = {arXiv},
	author = {{HESS Collaboration} and {Abdalla}, H. and {Abramowski}, A. and {Aharonian}, F. and {Ait Benkhali}, F. and {Akhperjanian}, A.~G. and {Andersson}, T. and {Ang{\"u}ner}, E.~O. and {Arrieta}, M. and {Aubert}, P. and {Backes}, M. and {Balzer}, A. and {Barnard}, M. and {Becherini}, Y. and {Becker Tjus}, J. and {Berge}, D. and {Bernhard}, S. and {Bernl{\"o}hr}, K. and {Blackwell}, R. and {B{\"o}ttcher}, M. and {Boisson}, C. and {Bolmont}, J. and {Bordas}, P. and {Bregeon}, J. and {Brun}, F. and {Brun}, P. and {Bryan}, M. and {Bulik}, T. and {Capasso}, M. and {Carr}, J. and {Casanova}, S. and {Cerruti}, M. and {Chakraborty}, N. and {Chalme-Calvet}, R. and {Chaves}, R.~C.~G. and {Chen}, A. and {Chevalier}, J. and {Chr{\'e}tien}, M. and {Colafrancesco}, S. and {Cologna}, G. and {Condon}, B. and {Conrad}, J. and {Cui}, Y. and {Davids}, I.~D. and {Decock}, J. and {Degrange}, B. and {Deil}, C. and {Devin}, J. and {deWilt}, P. and {Dirson}, L. and {Djannati-Ata{\"\i}}, A. and {Domainko}, W. and {Donath}, A. and {Drury}, L. O'C. and {Dubus}, G. and {Dutson}, K. and {Dyks}, J. and {Dyrda}, M. and {Edwards}, T. and {Egberts}, K. and {Eger}, P. and {Ernenwein}, J. -P. and {Eschbach}, S. and {Farnier}, C. and {Fegan}, S. and {Fernandes}, M.~V. and {Fiasson}, A. and {Fontaine}, G. and {F{\"o}rster}, A. and {Funk}, S. and {F{\"u}{\ss}ling}, M. and {Gabici}, S. and {Gajdus}, M. and {Gallant}, Y.~A. and {Garrigoux}, T. and {Giavitto}, G. and {Giebels}, B. and {Glicenstein}, J.~F. and {Gottschall}, D. and {Goyal}, A. and {Grondin}, M. -H. and {Hadasch}, D. and {Hahn}, J. and {Haupt}, M. and {Hawkes}, J. and {Heinzelmann}, G. and {Henri}, G. and {Hermann}, G. and {Hervet}, O. and {Hinton}, J.~A. and {Hofmann}, W. and {Hoischen}, C. and {Holler}, M. and {Horns}, D. and {Ivascenko}, A. and {Jacholkowska}, A. and {Jamrozy}, M. and {Janiak}, M. and {Jankowsky}, D. and {Jankowsky}, F. and {Jingo}, M. and {Jogler}, T. and {Jouvin}, L. and {Jung-Richardt}, I. and {Kastendieck}, M.~A. and {Katarzy{\'n}ski}, K. and {Katz}, U. and {Kerszberg}, D. and {Kh{\'e}lifi}, B. and {Kieffer}, M. and {King}, J. and {Klepser}, S. and {Klochkov}, D. and {Klu{\'z}niak}, W. and {Kolitzus}, D. and {Komin}, Nu and {Kosack}, K. and {Krakau}, S. and {Kraus}, M. and {Krayzel}, F. and {Kr{\"u}ger}, P.~P. and {Laffon}, H. and {Lamanna}, G. and {Lau}, J. and {Lees}, J. -P. and {Lefaucheur}, J. and {Lefranc}, V. and {Lemi{\`e}re}, A. and {Lemoine-Goumard}, M. and {Lenain}, J. -P. and {Leser}, E. and {Lohse}, T. and {Lorentz}, M. and {Liu}, R. and {L{\'o}pez-Coto}, R. and {Lypova}, I. and {Marandon}, V. and {Marcowith}, A. and {Mariaud}, C. and {Marx}, R. and {Maurin}, G. and {Maxted}, N. and {Mayer}, M. and {Meintjes}, P.~J. and {Meyer}, M. and {Mitchell}, A.~M.~W. and {Moderski}, R. and {Mohamed}, M. and {Mohrmann}, L. and {Mor{\^a}}, K. and {Moulin}, E. and {Murach}, T. and {de Naurois}, M. and {Niederwanger}, F. and {Niemiec}, J. and {Oakes}, L. and {O'Brien}, P. and {Odaka}, H. and {{\"O}ttl}, S. and {Ohm}, S. and {Ostrowski}, M. and {Oya}, I. and {Padovani}, M. and {Panter}, M. and {Parsons}, R.~D. and {Pekeur}, N.~W. and {Pelletier}, G. and {Perennes}, C. and {Petrucci}, P. -O. and {Peyaud}, B. and {Piel}, Q. and {Pita}, S. and {Poon}, H. and {Prokhorov}, D. and {Prokoph}, H. and {P{\"u}hlhofer}, G. and {Punch}, M. and {Quirrenbach}, A. and {Raab}, S. and {Reimer}, A. and {Reimer}, O. and {Renaud}, M. and {de los Reyes}, R. and {Rieger}, F. and {Romoli}, C. and {Rosier-Lees}, S. and {Rowell}, G. and {Rudak}, B. and {Rulten}, C.~B. and {Sahakian}, V. and {Salek}, D. and {Sanchez}, D.~A. and {Santangelo}, A. and {Sasaki}, M. and {Schlickeiser}, R. and {Sch{\"u}ssler}, F. and {Schulz}, A. and {Schwanke}, U. and {Schwemmer}, S. and {Settimo}, M.},
	doi = {10.1093/mnras/sty439},
	eprint = {1802.07611},
	journal = {\mnras},
	month = may,
	number = {3},
	pages = {4187-4198},
	primaryclass = {astro-ph.HE},
	title = {{H.E.S.S. discovery of very high energy {\ensuremath{\gamma}}-ray emission from PKS 0625-354}},
	volume = {476},
	year = 2018}

@article{Kadler_2012,
	adsurl = {https://ui.adsabs.harvard.edu/abs/2012A&A...538L...1K},
	archiveprefix = {arXiv},
	author = {{Kadler}, M. and {Eisenacher}, D. and {Ros}, E. and {Mannheim}, K. and {Els{\"a}sser}, D. and {Bach}, U.},
	doi = {10.1051/0004-6361/201118212},
	eid = {L1},
	eprint = {1201.2870},
	journal = {\aap},
	month = feb,
	pages = {L1},
	primaryclass = {astro-ph.HE},
	title = {{The blazar-like radio structure of the TeV source IC 310}},
	volume = {538},
	year = 2012}

@article{Hervet_2016,
	adsurl = {https://ui.adsabs.harvard.edu/abs/2016A&A...592A..22H},
	archiveprefix = {arXiv},
	author = {{Hervet}, O. and {Boisson}, C. and {Sol}, H.},
	doi = {10.1051/0004-6361/201628117},
	eid = {A22},
	eprint = {1605.02272},
	journal = {\aap},
	month = jul,
	pages = {A22},
	primaryclass = {astro-ph.HE},
	title = {{An innovative blazar classification based on radio jet kinematics}},
	volume = {592},
	year = 2016}

@article{Sikora_2016,
	adsurl = {https://ui.adsabs.harvard.edu/abs/2016MNRAS.457.1352S},
	archiveprefix = {arXiv},
	author = {{Sikora}, Marek and {Rutkowski}, Mieszko and {Begelman}, Mitchell C.},
	doi = {10.1093/mnras/stw107},
	eprint = {1511.08924},
	journal = {\mnras},
	month = apr,
	number = {2},
	pages = {1352-1358},
	primaryclass = {astro-ph.HE},
	title = {{A spine-sheath model for strong-line blazars}},
	volume = {457},
	year = 2016}

@article{Tavecchio_2008,
	adsurl = {https://ui.adsabs.harvard.edu/abs/2008MNRAS.385L..98T},
	archiveprefix = {arXiv},
	author = {{Tavecchio}, Fabrizio and {Ghisellini}, Gabriele},
	doi = {10.1111/j.1745-3933.2008.00441.x},
	eprint = {0801.0593},
	journal = {\mnras},
	month = mar,
	number = {1},
	pages = {L98-L102},
	primaryclass = {astro-ph},
	title = {{Spine-sheath layer radiative interplay in subparsec-scale jets and the TeV emission from M87}},
	volume = {385},
	year = 2008}

@article{Ghisellini_2005,
	adsurl = {https://ui.adsabs.harvard.edu/abs/2005A&A...432..401G},
	archiveprefix = {arXiv},
	author = {{Ghisellini}, G. and {Tavecchio}, F. and {Chiaberge}, M.},
	doi = {10.1051/0004-6361:20041404},
	eprint = {astro-ph/0406093},
	journal = {\aap},
	month = mar,
	number = {2},
	pages = {401-410},
	primaryclass = {astro-ph},
	title = {{Structured jets in TeV BL Lac objects and radiogalaxies. Implications for the observed properties}},
	volume = {432},
	year = 2005}

@article{Burrows-2005-SSR,
	adsurl = {https://ui.adsabs.harvard.edu/abs/2005SSRv..120..165B},
	archiveprefix = {arXiv},
	author = {{Burrows}, David N. and {Hill}, J.~E. and {Nousek}, J.~A. and {Kennea}, J.~A. and {Wells}, A. and {Osborne}, J.~P. and {Abbey}, A.~F. and {Beardmore}, A. and {Mukerjee}, K. and {Short}, A.~D.~T. and {Chincarini}, G. and {Campana}, S. and {Citterio}, O. and {Moretti}, A. and {Pagani}, C. and {Tagliaferri}, G. and {Giommi}, P. and {Capalbi}, M. and {Tamburelli}, F. and {Angelini}, L. and {Cusumano}, G. and {Br{\"a}uninger}, H.~W. and {Burkert}, W. and {Hartner}, G.~D.},
	doi = {10.1007/s11214-005-5097-2},
	eprint = {astro-ph/0508071},
	journal = {\ssr},
	month = oct,
	number = {3-4},
	pages = {165-195},
	primaryclass = {astro-ph},
	title = {{The Swift X-Ray Telescope}},
	volume = {120},
	year = 2005}

@article{Aharonian-2006-Sci,
	adsurl = {https://ui.adsabs.harvard.edu/abs/2006Sci...314.1424A},
	archiveprefix = {arXiv},
	author = {{Aharonian}, F. and {Akhperjanian}, A.~G. and {Bazer-Bachi}, A.~R. and {Beilicke}, M. and {Benbow}, W. and {Berge}, D. and {Bernl{\"o}hr}, K. and {Boisson}, C. and {Bolz}, O. and {Borrel}, V. and {Braun}, I. and {Brown}, A.~M. and {B{\"u}hler}, R. and {B{\"u}sching}, I. and {Carrigan}, S. and {Chadwick}, P.~M. and {Chounet}, L. -M. and {Coignet}, G. and {Cornils}, R. and {Costamante}, L. and {Degrange}, B. and {Dickinson}, H.~J. and {Djannati-Ata{\"\i}}, A. and {Drury}, L. O'C. and {Dubus}, G. and {Egberts}, K. and {Emmanoulopoulos}, D. and {Espigat}, P. and {Feinstein}, F. and {Ferrero}, E. and {Fiasson}, A. and {Fontaine}, G. and {Funk}, Seb. and {Funk}, S. and {F{\"u}{\ss}ling}, M. and {Gallant}, Y.~A. and {Giebels}, B. and {Glicenstein}, J.~F. and {Goret}, P. and {Hadjichristidis}, C. and {Hauser}, D. and {Hauser}, M. and {Heinzelmann}, G. and {Henri}, G. and {Hermann}, G. and {Hinton}, J.~A. and {Hoffmann}, A. and {Hofmann}, W. and {Holleran}, M. and {Hoppe}, S. and {Horns}, D. and {Jacholkowska}, A. and {de Jager}, O.~C. and {Kendziorra}, E. and {Kerschhaggl}, M. and {Kh{\'e}lifi}, B. and {Komin}, Nu. and {Konopelko}, A. and {Kosack}, K. and {Lamanna}, G. and {Latham}, I.~J. and {Le Gallou}, R. and {Lemi{\`e}re}, A. and {Lemoine-Goumard}, M. and {Lenain}, J. -P. and {Lohse}, T. and {Martin}, J.~M. and {Martineau-Huynh}, O. and {Marcowith}, A. and {Masterson}, C. and {Maurin}, G. and {McComb}, T.~J.~L. and {Moulin}, E. and {de Naurois}, M. and {Nedbal}, D. and {Nolan}, S.~J. and {Noutsos}, A. and {Orford}, K.~J. and {Osborne}, J.~L. and {Ouchrif}, M. and {Panter}, M. and {Pelletier}, G. and {Pita}, S. and {P{\"u}hlhofer}, G. and {Punch}, M. and {Ranchon}, S. and {Raubenheimer}, B.~C. and {Raue}, M. and {Rayner}, S.~M. and {Reimer}, A. and {Ripken}, J. and {Rob}, L. and {Rolland}, L. and {Rosier-Lees}, S. and {Rowell}, G. and {Sahakian}, V. and {Santangelo}, A. and {Saug{\'e}}, L. and {Schlenker}, S. and {Schlickeiser}, R. and {Schr{\"o}der}, R. and {Schwanke}, U. and {Schwarzburg}, S. and {Schwemmer}, S. and {Shalchi}, A. and {Sol}, H. and {Spangler}, D. and {Spanier}, F. and {Steenkamp}, R. and {Stegmann}, C. and {Superina}, G. and {Tam}, P.~H. and {Tavernet}, J. -P. and {Terrier}, R. and {Tluczykont}, M. and {van Eldik}, C. and {Vasileiadis}, G. and {Venter}, C. and {Vialle}, J.~P. and {Vincent}, P. and {V{\"o}lk}, H.~J. and {Wagner}, S.~J. and {Ward}, M.},
	doi = {10.1126/science.1134408},
	eprint = {astro-ph/0612016},
	journal = {Science},
	month = dec,
	number = {5804},
	pages = {1424-1427},
	primaryclass = {astro-ph},
	title = {{Fast Variability of Tera-Electron Volt {\ensuremath{\gamma}} Rays from the Radio Galaxy M87}},
	volume = {314},
	year = 2006}

@article{Cheung-2007-ApJ,
	adsurl = {https://ui.adsabs.harvard.edu/abs/2007ApJ...663L..65C},
	archiveprefix = {arXiv},
	author = {{Cheung}, C.~C. and {Harris}, D.~E. and {Stawarz}, {\L}.},
	doi = {10.1086/520510},
	eprint = {0705.2448},
	journal = {\apjl},
	month = jul,
	number = {2},
	pages = {L65-L68},
	primaryclass = {astro-ph},
	title = {{Superluminal Radio Features in the M87 Jet and the Site of Flaring TeV Gamma-Ray Emission}},
	volume = {663},
	year = 2007}

@inproceedings{Harris-2008-ASPC,
	adsurl = {https://ui.adsabs.harvard.edu/abs/2008ASPC..386...80H},
	archiveprefix = {arXiv},
	author = {{Harris}, D.~E. and {Cheung}, C.~C. and {Stawarz}, L. and {Biretta}, J.~A. and {Sparks}, W. and {Perlman}, E.~S. and {Wilson}, A.~S.},
	booktitle = {Extragalactic Jets: Theory and Observation from Radio to Gamma Ray},
	doi = {10.48550/arXiv.0707.3124},
	editor = {{Rector}, T.~A. and {De Young}, D.~S.},
	eprint = {0707.3124},
	month = jun,
	pages = {80},
	primaryclass = {astro-ph},
	series = {Astronomical Society of the Pacific Conference Series},
	title = {{The Continuing Saga of the Explosive Event(s) in the M87 Jet: Is M87 a Blazar?}},
	volume = {386},
	year = 2008}

@article{VERITAS-M87-2009-Sci,
	author = {{The VERITAS Collaboration} and the VLBA 43 GHz M87 Monitoring Team and the H.E.S.S. Collaboration and the MAGIC Collaboration and V. A. Acciari and E. Aliu and T. Arlen and M. Bautista and M. Beilicke and W. Benbow and S. M. Bradbury and J. H. Buckley and V. Bugaev and Y. Butt and K. Byrum and A. Cannon and O. Celik and A. Cesarini and Y. C. Chow and L. Ciupik and P. Cogan and W. Cui and R. Dickherber and S. J. Fegan and J. P. Finley and P. Fortin and L. Fortson and A. Furniss and D. Gall and G. H. Gillanders and J. Grube and R. Guenette and G. Gyuk and D. Hanna and J. Holder and D. Horan and C. M. Hui and T. B. Humensky and A. Imran and P. Kaaret and N. Karlsson and D. Kieda and J. Kildea and A. Konopelko and H. Krawczynski and F. Krennrich and M. J. Lang and S. LeBohec and G. Maier and A. McCann and M. McCutcheon and J. Millis and P. Moriarty and R. A. Ong and A. N. Otte and D. Pandel and J. S. Perkins and D. Petry and M. Pohl and J. Quinn and K. Ragan and L. C. Reyes and P. T. Reynolds and E. Roache and E. Roache and H. J. Rose and M. Schroedter and G. H. Sembroski and A. W. Smith and S. P. Swordy and M. Theiling and J. A. Toner and A. Varlotta and S. Vincent and S. P. Wakely and J. E. Ward and T. C. Weekes and A. Weinstein and D. A. Williams and S. Wissel and M. Wood and R. C. Walker and F. Davies and P. E. Hardee and W. Junor and C. Ly and F. Aharonian and A. G. Akhperjanian and G. Anton and U. Barres de Almeida and A. R. Bazer-Bachi and Y. Becherini and B. Behera and K. Bernl{\"o}hr and A. Bochow and C. Boisson and J. Bolmont and V. Borrel and J. Brucker and F. Brun and P. Brun and R. B{\"u}hler and T. Bulik and I. B{\"u}sching and T. Boutelier and P. M. Chadwick and A. Charbonnier and R. C. G. Chaves and A. Cheesebrough and L.-M. Chounet and A. C. Clapson and G. Coignet and M. Dalton and M. K. Daniel and I. D. Davids and B. Degrange and C. Deil and H. J. Dickinson and A. Djannati-Ata{\"\i} and W. Domainko and L. O'C. Drury and F. Dubois and G. Dubus and J. Dyks and M. Dyrda and K. Egberts and D. Emmanoulopoulos and P. Espigat and C. Farnier and F. Feinstein and A. Fiasson and A. F{\"o}rster and G. Fontaine and M. F{\"u}{\ss}ling and S. Gabici and Y. A. Gallant and L. G{\'e}rard and D. Gerbig and B. Giebels and J. F. Glicenstein and B. Gl{\"u}ck and P. Goret and D. G{\"o}hring and D. Hauser and M. Hauser and S. Heinz and G. Heinzelmann and G. Henri and G. Hermann and J. A. Hinton and A. Hoffmann and W. Hofmann and M. Holleran and S. Hoppe and D. Horns and A. Jacholkowska and O. C. de Jager and C. Jahn and I. Jung and K. Katarzy{\'n}ski and U. Katz and S. Kaufmann and E. Kendziorra and M. Kerschhaggl and D. Khangulyan and B. Kh{\'e}lifi and D. Keogh and W. Klu{\'z}niak and T. Kneiske and Nu. Komin and K. Kosack and G. Lamanna and J.-P. Lenain and T. Lohse and V. Marandon and J. M. Martin and O. Martineau-Huynh and A. Marcowith and D. Maurin and T. J. L. McComb and M. C. Medina and R. Moderski and E. Moulin and M. Naumann-Godo and M. de Naurois and D. Nedbal and D. Nekrassov and B. Nicholas and J. Niemiec and S. J. Nolan and S. Ohm and J.-F. Olive and E. de O{\~n}a Wilhelmi and K. J. Orford and M. Ostrowski and M. Panter and M. Paz Arribas and G. Pedaletti and G. Pelletier and P.-O. Petrucci and S. Pita and G. P{\"u}hlhofer and M. Punch and A. Quirrenbach and B. C. Raubenheimer and M. Raue and S. M. Rayner and M. Renaud and F. Rieger and J. Ripken and L. Rob and S. Rosier-Lees and G. Rowell and B. Rudak and C. B. Rulten and J. Ruppel and V. Sahakian and A. Santangelo and R. Schlickeiser and F. M. Sch{\"o}ck and R. Schr{\"o}der and U. Schwanke and S. Schwarzburg and S. Schwemmer and A. Shalchi and M. Sikora and J. L. Skilton and H. Sol and D. Spangler and {\L}. Stawarz and R. Steenkamp and C. Stegmann and F. Stinzing and G. Superina and A. Szostek and P. H. Tam and J.-P. Tavernet and R. Terrier and O. Tibolla and M. Tluczykont and C. van Eldik and G. Vasileiadis and C. Venter and L. Venter and J. P. Vialle and P. Vincent and M. Vivier and H. J. V{\"o}lk and F. Volpe and S. J. Wagner and M. Ward and A. A. Zdziarski and A. Zech and H. Anderhub and L. A. Antonelli and P. Antoranz and M. Backes and C. Baixeras and S. Balestra and J. A. Barrio and D. Bastieri and J. Becerra Gonz{\'a}lez and J. K. Becker and W. Bednarek and K. Berger and E. Bernardini and A. Biland and R. K. Bock and G. Bonnoli and P. Bordas and D. Borla Tridon and V. Bosch-Ramon and D. Bose and I. Braun and T. Bretz and I. Britvitch and M. Camara and E. Carmona and S. Commichau and J. L. Contreras and J. Cortina and M. T. Costado and S. Covino and V. Curtef and F. Dazzi and A. De Angelis and E. De Cea del Pozo and C. Delgado Mendez and R. De los Reyes and B. De Lotto and M. De Maria and F. De Sabata and A. Dominguez and D. Dorner and M. Doro and D. Elsaesser and M. Errando and D. Ferenc and E. Fern{\'a}ndez and R. Firpo and M. V. Fonseca and L. Font and N. Galante and R. J. Garc{\'\i}a L{\'o}pez and M. Garczarczyk and M. Gaug and F. Goebel and D. Hadasch and M. Hayashida and A. Herrero and D. Hildebrand and D. H{\"o}hne-M{\"o}nch and J. Hose and C. C. Hsu and T. Jogler and D. Kranich and A. La Barbera and A. Laille and E. Leonardo and E. Lindfors and S. Lombardi and F. Longo and M. L{\'o}pez and E. Lorenz and P. Majumdar and G. Maneva and N. Mankuzhiyil and K. Mannheim and L. Maraschi and M. Mariotti and M. Mart{\'\i}nez and D. Mazin and M. Meucci and J. M. Miranda and R. Mirzoyan and H. Miyamoto and J. Mold{\'o}n and M. Moles and A. Moralejo and D. Nieto and K. Nilsson and J. Ninkovic and I. Oya and R. Paoletti and J. M. Paredes and M. Pasanen and D. Pascoli and F. Pauss and R. G. Pegna and M. A. Perez-Torres and M. Persic and L. Peruzzo and F. Prada and E. Prandini and N. Puchades and I. Reichardt and W. Rhode and M. Rib{\'o} and J. Rico and M. Rissi and A. Robert and S. R{\"u}gamer and A. Saggion and T. Y. Saito and M. Salvati and M. Sanchez-Conde and K. Satalecka and V. Scalzotto and V. Scapin and T. Schweizer and M. Shayduk and S. N. Shore and N. Sidro and A. Sierpowska-Bartosik and A. Sillanp{\"a}{\"a} and J. Sitarek and D. Sobczynska and F. Spanier and A. Stamerra and L. S. Stark and L. Takalo and F. Tavecchio and P. Temnikov and D. Tescaro and M. Teshima and D. F. Torres and N. Turini and H. Vankov and R. M. Wagner and V. Zabalza and F. Zandanel and R. Zanin and J. Zapatero},
	doi = {10.1126/science.1175406},
	eprint = {https://www.science.org/doi/pdf/10.1126/science.1175406},
	journal = {Science},
	number = {5939},
	pages = {444-448},
	title = {Radio Imaging of the Very-High-Energy γ-Ray Emission Region in the Central Engine of a Radio Galaxy},
	url = {https://www.science.org/doi/abs/10.1126/science.1175406},
	volume = {325},
	year = {2009}}

@article{Aleksic-2010-ApJ,
	adsurl = {https://ui.adsabs.harvard.edu/abs/2010ApJ...723L.207A},
	archiveprefix = {arXiv},
	author = {{Aleksi{\'c}}, J. and {Antonelli}, L.~A. and {Antoranz}, P. and {Backes}, M. and {Barrio}, J.~A. and {Bastieri}, D. and {Becerra Gonz{\'a}lez}, J. and {Bednarek}, W. and {Berdyugin}, A. and {Berger}, K. and {Bernardini}, E. and {Biland}, A. and {Blanch}, O. and {Bock}, R.~K. and {Boller}, A. and {Bonnoli}, G. and {Bordas}, P. and {Borla Tridon}, D. and {Bosch-Ramon}, V. and {Bose}, D. and {Braun}, I. and {Bretz}, T. and {Camara}, M. and {Ca{\~n}ellas}, A. and {Carmona}, E. and {Carosi}, A. and {Colin}, P. and {Colombo}, E. and {Contreras}, J.~L. and {Cortina}, J. and {Cossio}, L. and {Covino}, S. and {Dazzi}, F. and {De Angelis}, A. and {De Cea del Pozo}, E. and {De Lotto}, B. and {De Maria}, M. and {De Sabata}, F. and {Delgado Mendez}, C. and {Diago Ortega}, A. and {Doert}, M. and {Dom{\'\i}nguez}, A. and {Dominis Prester}, D. and {Dorner}, D. and {Doro}, M. and {Elsaesser}, D. and {Errando}, M. and {Ferenc}, D. and {Fonseca}, M.~V. and {Font}, L. and {Garc{\'\i}a L{\'o}pez}, R.~J. and {Garczarczyk}, M. and {Giavitto}, G. and {Godinovi{\'c}}, N. and {Hadasch}, D. and {Herrero}, A. and {Hildebrand}, D. and {H{\"o}hne-M{\"o}nch}, D. and {Hose}, J. and {Hrupec}, D. and {Jogler}, T. and {Klepser}, S. and {Kr{\"a}henb{\"u}hl}, T. and {Kranich}, D. and {Krause}, J. and {La Barbera}, A. and {Leonardo}, E. and {Lindfors}, E. and {Lombardi}, S. and {Longo}, F. and {L{\'o}pez}, M. and {Lorenz}, E. and {Majumdar}, P. and {Makariev}, M. and {Maneva}, G. and {Mankuzhiyil}, N. and {Mannheim}, K. and {Maraschi}, L. and {Mariotti}, M. and {Mart{\'\i}nez}, M. and {Mazin}, D. and {Meucci}, M. and {Miranda}, J.~M. and {Mirzoyan}, R. and {Miyamoto}, H. and {Mold{\'o}n}, J. and {Moralejo}, A. and {Nieto}, D. and {Nilsson}, K. and {Orito}, R. and {Oya}, I. and {Paoletti}, R. and {Paredes}, J.~M. and {Partini}, S. and {Pasanen}, M. and {Pauss}, F. and {Pegna}, R.~G. and {Perez-Torres}, M.~A. and {Persic}, M. and {Peruzzo}, L. and {Pochon}, J. and {Prada}, F. and {Prada Moroni}, P.~G. and {Prandini}, E. and {Puchades}, N. and {Puljak}, I. and {Reichardt}, I. and {Reinthal}, R. and {Rhode}, W. and {Rib{\'o}}, M. and {Rico}, J. and {R{\"u}gamer}, S. and {Saggion}, A. and {Saito}, K. and {Saito}, T.~Y. and {Salvati}, M. and {S{\'a}nchez-Conde}, M. and {Satalecka}, K. and {Scalzotto}, V. and {Scapin}, V. and {Schultz}, C. and {Schweizer}, T. and {Shayduk}, M. and {Shore}, S.~N. and {Sierpowska-Bartosik}, A. and {Sillanp{\"a}{\"a}}, A. and {Sitarek}, J. and {Sobczynska}, D. and {Spanier}, F. and {Spiro}, S. and {Stamerra}, A. and {Steinke}, B. and {Storz}, J. and {Strah}, N. and {Struebig}, J.~C. and {Suric}, T. and {Takalo}, L. and {Tavecchio}, F. and {Temnikov}, P. and {Terzi{\'c}}, T. and {Tescaro}, D. and {Teshima}, M. and {Torres}, D.~F. and {Vankov}, H. and {Wagner}, R.~M. and {Weitzel}, Q. and {Zabalza}, V. and {Zandanel}, F. and {Zanin}, R. and {Neronov}, A. and {Pfrommer}, C. and {Pinzke}, A. and {Semikoz}, D.~V. and {MAGIC Collaboration}},
	doi = {10.1088/2041-8205/723/2/L207},
	eprint = {1009.2155},
	journal = {\apjl},
	month = nov,
	number = {2},
	pages = {L207-L212},
	primaryclass = {astro-ph.HE},
	title = {{Detection of Very High Energy {\ensuremath{\gamma}}-ray Emission from the Perseus Cluster Head-Tail Galaxy IC 310 by the MAGIC Telescopes}},
	volume = {723},
	year = 2010}

@article{Hervet_2024,
	adsurl = {https://ui.adsabs.harvard.edu/abs/2024ApJ...962..140H},
	archiveprefix = {arXiv},
	author = {{Hervet}, Olivier and {Johnson}, Caitlin A. and {Youngquist}, Adrian},
	doi = {10.3847/1538-4357/ad09c0},
	eid = {140},
	eprint = {2307.08804},
	journal = {\apj},
	month = feb,
	number = {2},
	pages = {140},
	primaryclass = {astro-ph.HE},
	title = {{Bjet\_MCMC: A New Tool to Automatically Fit the Broadband Spectral Energy Distributions of Blazars}},
	volume = {962},
	year = 2024}

@article{Hervet_2015,
	adsurl = {https://ui.adsabs.harvard.edu/abs/2015A&A...578A..69H},
	archiveprefix = {arXiv},
	author = {{Hervet}, O. and {Boisson}, C. and {Sol}, H.},
	doi = {10.1051/0004-6361/201425330},
	eid = {A69},
	eprint = {1503.01377},
	journal = {\aap},
	month = jun,
	pages = {A69},
	primaryclass = {astro-ph.HE},
	title = {{Linking radio and gamma-ray emission in Ap Librae}},
	volume = {578},
	year = 2015}

@article{Franceschini_2017,
	adsurl = {https://ui.adsabs.harvard.edu/abs/2017A&A...603A..34F},
	archiveprefix = {arXiv},
	author = {{Franceschini}, Alberto and {Rodighiero}, Giulia},
	doi = {10.1051/0004-6361/201629684},
	eid = {A34},
	eprint = {1705.10256},
	journal = {\aap},
	month = jul,
	pages = {A34},
	primaryclass = {astro-ph.HE},
	title = {{The extragalactic background light revisited and the cosmic photon-photon opacity}},
	volume = {603},
	year = 2017}

@article{Komissarov-1990-SvAL,
	adsurl = {https://ui.adsabs.harvard.edu/abs/1990SvAL...16..284K},
	author = {{Komissarov}, S.~S.},
	journal = {Soviet Astronomy Letters},
	month = jul,
	pages = {284},
	title = {{Emission by Relativistic Jets with Boundary Layers}},
	volume = {16},
	year = 1990}

@article{Georganopoulos-2003-ApJ,
	adsurl = {https://ui.adsabs.harvard.edu/abs/2003ApJ...594L..27G},
	archiveprefix = {arXiv},
	author = {{Georganopoulos}, Markos and {Kazanas}, Demosthenes},
	doi = {10.1086/378557},
	eprint = {astro-ph/0307404},
	journal = {\apjl},
	month = sep,
	number = {1},
	pages = {L27-L30},
	primaryclass = {astro-ph},
	title = {{Decelerating Flows in TeV Blazars: A Resolution to the BL Lacertae-FR I Unification Problem}},
	volume = {594},
	year = 2003}

@article{Lin-EGRET-1993-ApJ,
	adsurl = {https://ui.adsabs.harvard.edu/abs/1993ApJ...416L..53L},
	author = {{Lin}, Y.~C. and {Bertsch}, D.~L. and {Dingus}, B.~L. and {Fichtel}, C.~E. and {Hartman}, R.~C. and {Hunter}, S.~D. and {Kanbach}, G. and {Kniffen}, D.~A. and {Mattox}, J.~R. and {Mayer-Hasselwander}, H.~A. and {Michelson}, P.~F. and {von Montigny}, C. and {Nolan}, P.~L. and {Schneid}, E. and {Sreekumar}, P. and {Thompson}, D.~J.},
	doi = {10.1086/187069},
	journal = {\apjl},
	month = oct,
	pages = {L53},
	title = {{EGRET Limits on High-Energy Gamma-Ray Emission from X-Ray-- and Low-Energy Gamma-Ray--selected Seyfert Galaxies}},
	volume = {416},
	year = 1993}

@article{Rulten-2022-Galaxies,
	adsurl = {https://ui.adsabs.harvard.edu/abs/2022Galax..10...61R},
	author = {{Rulten}, Cameron},
	doi = {10.3390/galaxies10030061},
	eid = {61},
	journal = {Galaxies},
	month = apr,
	number = {3},
	pages = {61},
	title = {{Radio Galaxies at TeV Energies}},
	volume = {10},
	year = 2022}

@article{Bennett_2014,
	adsurl = {https://ui.adsabs.harvard.edu/abs/2014ApJ...794..135B},
	archiveprefix = {arXiv},
	author = {{Bennett}, C.~L. and {Larson}, D. and {Weiland}, J.~L. and {Hinshaw}, G.},
	doi = {10.1088/0004-637X/794/2/135},
	eid = {135},
	eprint = {1406.1718},
	journal = {\apj},
	month = oct,
	number = {2},
	pages = {135},
	primaryclass = {astro-ph.CO},
	title = {{The 1\% Concordance Hubble Constant}},
	volume = {794},
	year = 2014}

@inproceedings{Cogan:2008ICRC,
	adsurl = {https://ui.adsabs.harvard.edu/abs/2008ICRC....3.1385C},
	archiveprefix = {arXiv},
	author = {{Cogan}, P.},
	booktitle = {International Cosmic Ray Conference},
	doi = {10.48550/arXiv.0709.4233},
	eprint = {0709.4233},
	month = jan,
	pages = {1385-1388},
	primaryclass = {astro-ph},
	series = {International Cosmic Ray Conference},
	title = {{VEGAS, the VERITAS Gamma-ray Analysis Suite}},
	volume = {3},
	year = 2008}

@article{Berge:2007aa,
	adsurl = {https://ui.adsabs.harvard.edu/abs/2007A&A...466.1219B},
	archiveprefix = {arXiv},
	author = {{Berge}, D. and {Funk}, S. and {Hinton}, J.},
	doi = {10.1051/0004-6361:20066674},
	eprint = {astro-ph/0610959},
	journal = {Astronomy and Astrophysics},
	month = may,
	number = {3},
	pages = {1219-1229},
	primaryclass = {astro-ph},
	title = {{Background modelling in very-high-energy {\ensuremath{\gamma}}-ray astronomy}},
	volume = {466},
	year = 2007}

@inproceedings{Hillas:1985-ICRC,
	adsurl = {https://ui.adsabs.harvard.edu/abs/1985ICRC....3..445H},
	author = {{Hillas}, A.~M.},
	booktitle = {19th International Cosmic Ray Conference (ICRC19), Volume 3},
	month = aug,
	pages = {445},
	series = {International Cosmic Ray Conference},
	title = {{Cerenkov Light Images of EAS Produced by Primary Gamma Rays and by Nuclei}},
	volume = {3},
	year = 1985}

@article{VERITAS:2016ATEL9690,
	adsurl = {https://ui.adsabs.harvard.edu/abs/2016ATel.9690....1M},
	author = {{Mukherjee}, Reshmi and {VERITAS Collaboration}},
	journal = {The Astronomer's Telegram},
	month = oct,
	pages = {1},
	title = {{VERITAS detection of the radio galaxy NGC 1275 with elevated very-high-energy gamma-ray emission}},
	volume = {9690},
	year = 2016}

@inproceedings{Benbow:2015ICRC,
	adsurl = {https://ui.adsabs.harvard.edu/abs/2015ICRC...34..821B},
	archiveprefix = {arXiv},
	author = {{Benbow}, W. and {VERITAS Collaboration}},
	booktitle = {34th International Cosmic Ray Conference (ICRC2015)},
	doi = {10.22323/1.236.0821},
	eid = {821},
	eprint = {1508.07251},
	month = jul,
	pages = {821},
	primaryclass = {astro-ph.HE},
	series = {International Cosmic Ray Conference},
	title = {{Highlights from the VERITAS AGN Observation Program}},
	volume = {34},
	year = 2015}

@article{MAGIC:2016ATEL9689,
	adsurl = {https://ui.adsabs.harvard.edu/abs/2016ATel.9689....1M},
	author = {{Mirzoyan}, Razmik},
	journal = {The Astronomer's Telegram},
	month = oct,
	pages = {1},
	title = {{MAGIC detection of an increased activity from NGC 1275 at very-high-energy gamma rays}},
	volume = {9689},
	year = 2016}

@article{Foreman-Mackey-2013-PASP,
	adsurl = {https://ui.adsabs.harvard.edu/abs/2013PASP..125..306F},
	archiveprefix = {arXiv},
	author = {{Foreman-Mackey}, Daniel and {Hogg}, David W. and {Lang}, Dustin and {Goodman}, Jonathan},
	doi = {10.1086/670067},
	eprint = {1202.3665},
	journal = {{Publications of the Astronomical Society of the Pacific}},
	month = mar,
	number = {925},
	pages = {306},
	primaryclass = {astro-ph.IM},
	title = {{emcee: The MCMC Hammer}},
	volume = {125},
	year = 2013}

@inproceedings{Zabalza-2015-ICRC,
	adsurl = {https://ui.adsabs.harvard.edu/abs/2015ICRC...34..922Z},
	archiveprefix = {arXiv},
	author = {{Zabalza}, V.},
	booktitle = {34th International Cosmic Ray Conference (ICRC2015)},
	doi = {10.22323/1.236.0922},
	eid = {922},
	eprint = {1509.03319},
	month = jul,
	pages = {922},
	primaryclass = {astro-ph.HE},
	series = {International Cosmic Ray Conference},
	title = {{Naima: a Python package for inference of particle distribution properties from nonthermal spectra}},
	volume = {34},
	year = 2015}

@article{Ghisellini-2009-MNRAS,
	adsurl = {https://ui.adsabs.harvard.edu/abs/2009MNRAS.397..985G},
	archiveprefix = {arXiv},
	author = {{Ghisellini}, G. and {Tavecchio}, F.},
	doi = {10.1111/j.1365-2966.2009.15007.x},
	eprint = {0902.0793},
	journal = {\mnras},
	month = aug,
	number = {2},
	pages = {985-1002},
	primaryclass = {astro-ph.CO},
	title = {{Canonical high-power blazars}},
	volume = {397},
	year = 2009}

@inproceedings{casa,
	adsurl = {https://ui.adsabs.harvard.edu/abs/2007ASPC..376..127M},
	author = {{McMullin}, J.~P. and {Waters}, B. and {Schiebel}, D. and {Young}, W. and {Golap}, K.},
	booktitle = {Astronomical Data Analysis Software and Systems XVI},
	editor = {{Shaw}, R.~A. and {Hill}, F. and {Bell}, D.~J.},
	month = oct,
	pages = {127},
	series = {Astronomical Society of the Pacific Conference Series},
	title = {{CASA Architecture and Applications}},
	volume = {376},
	year = 2007}

@article{Sikora-2009-ApJ,
	adsurl = {https://ui.adsabs.harvard.edu/abs/2009ApJ...704...38S},
	archiveprefix = {arXiv},
	author = {{Sikora}, Marek and {Stawarz}, {\L}ukasz and {Moderski}, Rafa{\l} and {Nalewajko}, Krzysztof and {Madejski}, Greg M.},
	doi = {10.1088/0004-637X/704/1/38},
	eprint = {0904.1414},
	journal = {\apj},
	month = oct,
	number = {1},
	pages = {38-50},
	primaryclass = {astro-ph.CO},
	title = {{Constraining Emission Models of Luminous Blazar Sources}},
	volume = {704},
	year = 2009}

@article{Giovannini-2018-NatAs,
	adsurl = {https://ui.adsabs.harvard.edu/abs/2018NatAs...2..472G},
	archiveprefix = {arXiv},
	author = {{Giovannini}, G. and {Savolainen}, T. and {Orienti}, M. and {Nakamura}, M. and {Nagai}, H. and {Kino}, M. and {Giroletti}, M. and {Hada}, K. and {Bruni}, G. and {Kovalev}, Y.~Y. and {Anderson}, J.~M. and {D'Ammando}, F. and {Hodgson}, J. and {Honma}, M. and {Krichbaum}, T.~P. and {Lee}, S. -S. and {Lico}, R. and {Lisakov}, M.~M. and {Lobanov}, A.~P. and {Petrov}, L. and {Sohn}, B.~W. and {Sokolovsky}, K.~V. and {Voitsik}, P.~A. and {Zensus}, J.~A. and {Tingay}, S.},
	doi = {10.1038/s41550-018-0431-2},
	eprint = {1804.02198},
	journal = {Nature Astronomy},
	month = apr,
	pages = {472-477},
	primaryclass = {astro-ph.GA},
	title = {{A wide and collimated radio jet in 3C84 on the scale of a few hundred gravitational radii}},
	volume = {2},
	year = 2018}

@article{Tavecchio-2014-MNRAS,
	adsurl = {https://ui.adsabs.harvard.edu/abs/2014MNRAS.443.1224T},
	archiveprefix = {arXiv},
	author = {{Tavecchio}, F. and {Ghisellini}, G.},
	doi = {10.1093/mnras/stu1196},
	eprint = {1404.6894},
	journal = {\mnras},
	month = sep,
	number = {2},
	pages = {1224-1230},
	primaryclass = {astro-ph.HE},
	title = {{On the spine-layer scenario for the very high-energy emission of NGC 1275}},
	volume = {443},
	year = 2014}

@article{Lister-2019-ApJ,
	adsurl = {https://ui.adsabs.harvard.edu/abs/2019ApJ...874...43L},
	archiveprefix = {arXiv},
	author = {{Lister}, M.~L. and {Homan}, D.~C. and {Hovatta}, T. and {Kellermann}, K.~I. and {Kiehlmann}, S. and {Kovalev}, Y.~Y. and {Max-Moerbeck}, W. and {Pushkarev}, A.~B. and {Readhead}, A.~C.~S. and {Ros}, E. and {Savolainen}, T.},
	doi = {10.3847/1538-4357/ab08ee},
	eid = {43},
	eprint = {1902.09591},
	journal = {\apj},
	month = mar,
	number = {1},
	pages = {43},
	primaryclass = {astro-ph.GA},
	title = {{MOJAVE. XVII. Jet Kinematics and Parent Population Properties of Relativistically Beamed Radio-loud Blazars}},
	volume = {874},
	year = 2019}

@article{Nagai-2014-ApJ,
	adsurl = {https://ui.adsabs.harvard.edu/abs/2014ApJ...785...53N},
	archiveprefix = {arXiv},
	author = {{Nagai}, H. and {Haga}, T. and {Giovannini}, G. and {Doi}, A. and {Orienti}, M. and {D'Ammando}, F. and {Kino}, M. and {Nakamura}, M. and {Asada}, K. and {Hada}, K. and {Giroletti}, M.},
	doi = {10.1088/0004-637X/785/1/53},
	eid = {53},
	eprint = {1402.5930},
	journal = {\apj},
	month = apr,
	number = {1},
	pages = {53},
	primaryclass = {astro-ph.HE},
	title = {{Limb-brightened Jet of 3C 84 Revealed by the 43 GHz Very-Long-Baseline-Array Observation}},
	volume = {785},
	year = 2014}

@article{Fermi-4FGL-2020-ApJS,
	adsurl = {https://ui.adsabs.harvard.edu/abs/2020ApJS..247...33A},
	archiveprefix = {arXiv},
	author = {{Abdollahi}, S. and {Acero}, F. and {Ackermann}, M. and {Ajello}, M. and {Atwood}, W.~B. and {Axelsson}, M. and {Baldini}, L. and {Ballet}, J. and {Barbiellini}, G. and {Bastieri}, D. and {Becerra Gonzalez}, J. and {Bellazzini}, R. and {Berretta}, A. and {Bissaldi}, E. and {Bland ford}, R.~D. and {Bloom}, E.~D. and {Bonino}, R. and {Bottacini}, E. and {Brandt}, T.~J. and {Bregeon}, J. and {Bruel}, P. and {Buehler}, R. and {Burnett}, T.~H. and {Buson}, S. and {Cameron}, R.~A. and {Caputo}, R. and {Caraveo}, P.~A. and {Casandjian}, J.~M. and {Castro}, D. and {Cavazzuti}, E. and {Charles}, E. and {Chaty}, S. and {Chen}, S. and {Cheung}, C.~C. and {Chiaro}, G. and {Ciprini}, S. and {Cohen-Tanugi}, J. and {Cominsky}, L.~R. and {Coronado-Bl{\'a}zquez}, J. and {Costantin}, D. and {Cuoco}, A. and {Cutini}, S. and {D'Ammando}, F. and {DeKlotz}, M. and {Torre Luque}, P. de la and {de Palma}, F. and {Desai}, A. and {Digel}, S.~W. and {Lalla}, N. Di and {Mauro}, M. Di and {Venere}, L. Di and {Dom{\'\i}nguez}, A. and {Dumora}, D. and {Dirirsa}, F. Fana and {Fegan}, S.~J. and {Ferrara}, E.~C. and {Franckowiak}, A. and {Fukazawa}, Y. and {Funk}, S. and {Fusco}, P. and {Gargano}, F. and {Gasparrini}, D. and {Giglietto}, N. and {Giommi}, P. and {Giordano}, F. and {Giroletti}, M. and {Glanzman}, T. and {Green}, D. and {Grenier}, I.~A. and {Griffin}, S. and {Grondin}, M. -H. and {Grove}, J.~E. and {Guiriec}, S. and {Harding}, A.~K. and {Hayashi}, K. and {Hays}, E. and {Hewitt}, J.~W. and {Horan}, D. and {J{\'o}hannesson}, G. and {Johnson}, T.~J. and {Kamae}, T. and {Kerr}, M. and {Kocevski}, D. and {Kovac'evic'}, M. and {Kuss}, M. and {Landriu}, D. and {Larsson}, S. and {Latronico}, L. and {Lemoine-Goumard}, M. and {Li}, J. and {Liodakis}, I. and {Longo}, F. and {Loparco}, F. and {Lott}, B. and {Lovellette}, M.~N. and {Lubrano}, P. and {Madejski}, G.~M. and {Maldera}, S. and {Malyshev}, D. and {Manfreda}, A. and {Marchesini}, E.~J. and {Marcotulli}, L. and {Mart{\'\i}-Devesa}, G. and {Martin}, P. and {Massaro}, F. and {Mazziotta}, M.~N. and {McEnery}, J.~E. and {Mereu}, I. and {Meyer}, M. and {Michelson}, P.~F. and {Mirabal}, N. and {Mizuno}, T. and {Monzani}, M.~E. and {Morselli}, A. and {Moskalenko}, I.~V. and {Negro}, M. and {Nuss}, E. and {Ojha}, R. and {Omodei}, N. and {Orienti}, M. and {Orlando}, E. and {Ormes}, J.~F. and {Palatiello}, M. and {Paliya}, V.~S. and {Paneque}, D. and {Pei}, Z. and {Pe{\~n}a-Herazo}, H. and {Perkins}, J.~S. and {Persic}, M. and {Pesce-Rollins}, M. and {Petrosian}, V. and {Petrov}, L. and {Piron}, F. and {Poon}, H. and {Porter}, T.~A. and {Principe}, G. and {Rain{\`o}}, S. and {Rando}, R. and {Razzano}, M. and {Razzaque}, S. and {Reimer}, A. and {Reimer}, O. and {Remy}, Q. and {Reposeur}, T. and {Romani}, R.~W. and {Parkinson}, P.~M. Saz and {Schinzel}, F.~K. and {Serini}, D. and {Sgr{\`o}}, C. and {Siskind}, E.~J. and {Smith}, D.~A. and {Spandre}, G. and {Spinelli}, P. and {Strong}, A.~W. and {Suson}, D.~J. and {Tajima}, H. and {Takahashi}, M.~N. and {Tak}, D. and {Thayer}, J.~B. and {Thompson}, D.~J. and {Tibaldo}, L. and {Torres}, D.~F. and {Torresi}, E. and {Valverde}, J. and {Klaveren}, B. Van and {Zyl}, P. van and {Wood}, K. and {Yassine}, M. and {Zaharijas}, G.},
	doi = {10.3847/1538-4365/ab6bcb},
	eid = {33},
	eprint = {1902.10045},
	journal = {\apjs},
	month = mar,
	number = {1},
	pages = {33},
	primaryclass = {astro-ph.HE},
	title = {{Fermi Large Area Telescope Fourth Source Catalog}},
	volume = {247},
	year = 2020}

@inproceedings{Wood-2017-ICRC,
	adsurl = {https://ui.adsabs.harvard.edu/abs/2017ICRC...35..824W},
	archiveprefix = {arXiv},
	author = {{Wood}, M. and {Caputo}, R. and {Charles}, E. and {Di Mauro}, M. and {Magill}, J. and {Perkins}, J.~S. and {Fermi-LAT Collaboration}},
	booktitle = {35th International Cosmic Ray Conference (ICRC2017)},
	eid = {824},
	eprint = {1707.09551},
	month = jan,
	pages = {824},
	primaryclass = {astro-ph.IM},
	series = {International Cosmic Ray Conference},
	title = {{Fermipy: An open-source Python package for analysis of Fermi-LAT Data}},
	volume = {301},
	year = 2017}

@article{Scargle-2013-ApJ,
	adsurl = {https://ui.adsabs.harvard.edu/abs/2013ApJ...764..167S},
	archiveprefix = {arXiv},
	author = {{Scargle}, Jeffrey D. and {Norris}, Jay P. and {Jackson}, Brad and {Chiang}, James},
	doi = {10.1088/0004-637X/764/2/167},
	eid = {167},
	eprint = {1207.5578},
	journal = {\apj},
	month = feb,
	number = {2},
	pages = {167},
	primaryclass = {astro-ph.IM},
	title = {{Studies in Astronomical Time Series Analysis. VI. Bayesian Block Representations}},
	volume = {764},
	year = 2013}

@article{Lister-2018-ApJS,
	adsurl = {https://ui.adsabs.harvard.edu/abs/2018ApJS..234...12L},
	archiveprefix = {arXiv},
	author = {{Lister}, M.~L. and {Aller}, M.~F. and {Aller}, H.~D. and {Hodge}, M.~A. and {Homan}, D.~C. and {Kovalev}, Y.~Y. and {Pushkarev}, A.~B. and {Savolainen}, T.},
	doi = {10.3847/1538-4365/aa9c44},
	eid = {12},
	eprint = {1711.07802},
	journal = {\apjs},
	month = {Jan},
	number = {1},
	pages = {12},
	primaryclass = {astro-ph.GA},
	title = {{MOJAVE. XV. VLBA 15 GHz Total Intensity and Polarization Maps of 437 Parsec-scale AGN Jets from 1996 to 2017}},
	volume = {234},
	year = {2018}}

@article{MAGIC-2018-AA,
	adsurl = {https://ui.adsabs.harvard.edu/abs/2018A&A...617A..91M},
	archiveprefix = {arXiv},
	author = {{MAGIC Collaboration} and {Ansoldi}, S. and {Antonelli}, L.~A. and {Arcaro}, C. and {Baack}, D. and {Babi{\'c}}, A. and {Banerjee}, B. and {Bangale}, P. and {Barres de Almeida}, U. and {Barrio}, J.~A. and {Becerra Gonz{\'a}lez}, J. and {Bednarek}, W. and {Bernardini}, E. and {Berse}, R. Ch. and {Berti}, A. and {Bhattacharyya}, W. and {Bigongiari}, C. and {Biland}, A. and {Blanch}, O. and {Bonnoli}, G. and {Carosi}, R. and {Ceribella}, G. and {Chatterjee}, A. and {Colak}, S.~M. and {Colin}, P. and {Colombo}, E. and {Contreras}, J.~L. and {Cortina}, J. and {Covino}, S. and {Cumani}, P. and {D'Elia}, V. and {da Vela}, P. and {Dazzi}, F. and {de Angelis}, A. and {de Lotto}, B. and {Delfino}, M. and {Delgado}, J. and {di Pierro}, F. and {Dom{\'\i}nguez}, A. and {Dominis Prester}, D. and {Dorner}, D. and {Doro}, M. and {Einecke}, S. and {Elsaesser}, D. and {Fallah Ramazani}, V. and {Fattorini}, A. and {Fern{\'a}ndez-Barral}, A. and {Ferrara}, G. and {Fidalgo}, D. and {Foffano}, L. and {Fonseca}, M.~V. and {Font}, L. and {Fruck}, C. and {Galindo}, D. and {Gallozzi}, S. and {Garc{\'\i}a L{\'o}pez}, R.~J. and {Garczarczyk}, M. and {Gaug}, M. and {Giammaria}, P. and {Godinovi{\'c}}, N. and {Gora}, D. and {Guberman}, D. and {Hadasch}, D. and {Hahn}, A. and {Hassan}, T. and {Hayashida}, M. and {Herrera}, J. and {Hoang}, J. and {Hose}, J. and {Hrupec}, D. and {Ishio}, K. and {Konno}, Y. and {Kubo}, H. and {Kushida}, J. and {Lamastra}, A. and {Lelas}, D. and {Leone}, F. and {Lindfors}, E. and {Lombardi}, S. and {Longo}, F. and {L{\'o}pez}, M. and {Maggio}, C. and {Majumdar}, P. and {Makariev}, M. and {Maneva}, G. and {Manganaro}, M. and {Mannheim}, K. and {Maraschi}, L. and {Mariotti}, M. and {Mart{\'\i}nez}, M. and {Masuda}, S. and {Mazin}, D. and {Mielke}, K. and {Minev}, M. and {Miranda}, J.~M. and {Mirzoyan}, R. and {Moralejo}, A. and {Moreno}, V. and {Moretti}, E. and {Nagayoshi}, T. and {Neustroev}, V. and {Niedzwiecki}, A. and {Nievas Rosillo}, M. and {Nigro}, C. and {Nilsson}, K. and {Ninci}, D. and {Nishijima}, K. and {Noda}, K. and {Nogu{\'e}s}, L. and {Paiano}, S. and {Palacio}, J. and {Paneque}, D. and {Paoletti}, R. and {Paredes}, J.~M. and {Pedaletti}, G. and {Pe{\~n}il}, P. and {Peresano}, M. and {Persic}, M. and {Pfrang}, K. and {Prada Moroni}, P.~G. and {Prandini}, E. and {Puljak}, I. and {Garcia}, J.~R. and {Reichardt}, I. and {Rhode}, W. and {Rib{\'o}}, M. and {Rico}, J. and {Righi}, C. and {Rugliancich}, A. and {Saha}, L. and {Saito}, T. and {Satalecka}, K. and {Schweizer}, T. and {Sitarek}, J. and {{\v{S}}nidari{\'c}}, I. and {Sobczynska}, D. and {Stamerra}, A. and {Strzys}, M. and {Suri{\'c}}, T. and {Takahashi}, M. and {Tavecchio}, F. and {Temnikov}, P. and {Terzi{\'c}}, T. and {Teshima}, M. and {Torres-Alb{\`a}}, N. and {Tsujimoto}, S. and {Vanzo}, G. and {Vazquez Acosta}, M. and {Vovk}, I. and {Ward}, J.~E. and {Will}, M. and {Zari{\'c}}, D. and {Glawion}, D. and {Takalo}, L.~O. and {Jormanainen}, J.},
	doi = {10.1051/0004-6361/201832895},
	eid = {A91},
	eprint = {1806.01559},
	journal = {\aap},
	month = {Sep},
	pages = {A91},
	primaryclass = {astro-ph.HE},
	title = {{Gamma-ray flaring activity of NGC1275 in 2016-2017 measured by MAGIC}},
	volume = {617},
	year = {2018}}

@article{Fioc_1999,
	adsurl = {https://ui.adsabs.harvard.edu/abs/1999astro.ph.12179F},
	archiveprefix = {arXiv},
	author = {{Fioc}, Michel and {Rocca-Volmerange}, Brigitte},
	eid = {astro-ph/9912179},
	eprint = {astro-ph/9912179},
	journal = {arXiv e-prints},
	month = {Dec},
	pages = {astro-ph/9912179},
	primaryclass = {astro-ph},
	title = {{PEGASE.2, a metallicity-consistent spectral evolution model of galaxies: the documentation and the code}},
	year = {1999}}

@article{Mathews_2006,
	adsurl = {https://ui.adsabs.harvard.edu/abs/2006ApJ...638..659M},
	archiveprefix = {arXiv},
	author = {{Mathews}, William G. and {Faltenbacher}, Andreas and {Brighenti}, Fabrizio},
	doi = {10.1086/499119},
	eprint = {astro-ph/0511151},
	journal = {\apj},
	month = {Feb},
	number = {2},
	pages = {659-667},
	primaryclass = {astro-ph},
	title = {{Heating Cooling Flows with Weak Shock Waves}},
	volume = {638},
	year = {2006}}

@book{deVaucouleurs_1991,
	adsurl = {https://ui.adsabs.harvard.edu/abs/1991rc3..book.....D},
	author = {{de Vaucouleurs}, Gerard and {de Vaucouleurs}, Antoinette and {Corwin}, Herold G., Jr. and {Buta}, Ronald J. and {Paturel}, Georges and {Fouque}, Pascal},
	title = {{Third Reference Catalogue of Bright Galaxies}},
	year = {1991}}

@article{Prestwich_1997,
	adsurl = {https://ui.adsabs.harvard.edu/abs/1997ApJ...477..144P},
	author = {{Prestwich}, Andrea H. and {Joy}, Marshall and {Luginbuhl}, Christian B. and {Sulkanen}, Martin and {Newberry}, Mike},
	doi = {10.1086/303693},
	journal = {\apj},
	month = {Mar},
	number = {1},
	pages = {144-151},
	title = {{A Search for the Cooling Flow Accretion Population: Optical and Near-Infrared Imaging of NGC 1275}},
	volume = {477},
	year = {1997}}

@article{Aleksic-2014-Sci,
	adsurl = {https://ui.adsabs.harvard.edu/#abs/2014Sci...346.1080A},
	archiveprefix = {arXiv},
	author = {{Aleksi{\'c}}, J. and {Ansoldi}, S. and {Antonelli}, L.~A. and {Antoranz}, P. and {Babic}, A. and {Bangale}, P. and {Barrio}, J.~A. and {Gonz{\'a}lez}, J. Becerra and {Bednarek}, W. and {Bernardini}, E. and {Biasuzzi}, B. and {Biland}, A. and {Blanch}, O. and {Bonnefoy}, S. and {Bonnoli}, G. and {Borracci}, F. and {Bretz}, T. and {Carmona}, E. and {Carosi}, A. and {Colin}, P. and {Colombo}, E. and {Contreras}, J.~L. and {Cortina}, J. and {Covino}, S. and {Da Vela}, P. and {Dazzi}, F. and {De Angelis}, A. and {De Caneva}, G. and {De Lotto}, B. and {Wilhelmi}, E. de O{\~n}a and {Mendez}, C. Delgado and {Prester}, D. Dominis and {Dorner}, D. and {Doro}, M. and {Einecke}, S. and {Eisenacher}, D. and {Elsaesser}, D. and {Fonseca}, M.~V. and {Font}, L. and {Frantzen}, K. and {Fruck}, C. and {Galindo}, D. and {L{\'o}pez}, R.~J. Garc{\'\i}a and {Garczarczyk}, M. and {Terrats}, D. Garrido and {Gaug}, M. and {Godinovi{\'c}}, N. and {Mu{\~n}oz}, A. Gonz{\'a}lez and {Gozzini}, S.~R. and {Hadasch}, D. and {Hanabata}, Y. and {Hayashida}, M. and {Herrera}, J. and {Hildebrand}, D. and {Hose}, J. and {Hrupec}, D. and {Idec}, W. and {Kadenius}, V. and {Kellermann}, H. and {Kodani}, K. and {Konno}, Y. and {Krause}, J. and {Kubo}, H. and {Kushida}, J. and {La Barbera}, A. and {Lelas}, D. and {Lewandowska}, N. and {Lindfors}, E. and {Lombardi}, S. and {Longo}, F. and {L{\'o}pez}, M. and {L{\'o }pez-Coto}, R. and {L{\'o}pez-Oramas}, A. and {Lorenz}, E. and {Lozano}, I. and {Makariev}, M. and {Mallot}, K. and {Maneva}, G. and {Mankuzhiyil}, N. and {Mannheim}, K. and {Maraschi}, L. and {Marcote}, B. and {Mariotti}, M. and {Mart{\'\i}nez}, M. and {Mazin}, D. and {Menzel}, U. and {Miranda}, J.~M. and {Mirzoyan}, R. and {Moralejo}, A. and {Munar-Adrover}, P. and {Nakajima}, D. and {Niedzwiecki}, A. and {Nilsson}, K. and {Nishijima}, K. and {Noda}, K. and {Orito}, R. and {Overkemping}, A. and {Paiano}, S. and {Palatiello}, M. and {Paneque}, D. and {Paoletti}, R. and {Paredes}, J.~M. and {Paredes-Fortuny}, X. and {Persic}, M. and {Poutanen}, J. and {Moroni}, P.~G. Prada and {Prandini}, E. and {Puljak}, I. and {Reinthal}, R. and {Rhode}, W. and {Rib{\'o}}, M. and {Rico}, J. and {Garcia}, J. Rodriguez and {R{\"u}gamer}, S. and {Saito}, T. and {Saito}, K. and {Satalecka}, K. and {Scalzotto}, V. and {Scapin}, V. and {Schultz}, C. and {Schweizer}, T. and {Shore}, S.~N. and {Sillanp{\"a}{\"a}}, A. and {Sitarek}, J. and {Snidaric}, I. and {Sobczynska}, D. and {Spanier}, F. and {Stamatescu}, V. and {Stamerra}, A. and {Steinbring}, T. and {Storz}, J. and {Strzys}, M. and {Takalo}, L. and {Takami}, H. and {Tavecchio}, F. and {Temnikov}, P. and {Terzi{\'c}}, T. and {Tescaro}, D. and {Teshima}, M. and {Thaele}, J. and {Tibolla}, O. and {Torres}, D.~F. and {Toyama}, T. and {Treves}, A. and {Uellenbeck}, M. and {Vogler}, P. and {Zanin}, R. and {Kadler}, M. and {Schulz}, R. and {Ros}, E. and {Bach}, U. and {Krau{\ss}}, F. and {Wilms}, J.},
	doi = {10.1126/science.1256183},
	eprint = {1412.4936},
	journal = {Science},
	month = Nov,
	pages = {1080-1084},
	primaryclass = {astro-ph.HE},
	title = {{Black hole lightning due to particle acceleration at subhorizon scales}},
	volume = {346},
	year = 2014}

@inproceedings{Rieger:2017,
	adsurl = {http://adsabs.harvard.edu/abs/2017AIPC.1792b0008R},
	archiveprefix = {arXiv},
	author = {{Rieger}, F.~M.},
	booktitle = {6th International Symposium on High Energy Gamma-Ray Astronomy},
	doi = {10.1063/1.4968893},
	eid = {020008},
	eprint = {1611.02986},
	month = jan,
	pages = {020008},
	primaryclass = {astro-ph.HE},
	series = {American Institute of Physics Conference Series},
	title = {{Gamma-rays from non-blazar AGN}},
	volume = 1792,
	year = 2017}

@article{DermerGiebels:2016,
	adsurl = {http://adsabs.harvard.edu/abs/2016CRPhy..17..594D},
	archiveprefix = {arXiv},
	author = {{Dermer}, C.~D. and {Giebels}, B.},
	doi = {10.1016/j.crhy.2016.04.004},
	eprint = {1602.06592},
	journal = {Comptes Rendus Physique},
	month = jun,
	pages = {594-616},
	primaryclass = {astro-ph.HE},
	title = {{Active galactic nuclei at gamma-ray energies}},
	volume = 17,
	year = 2016}

@article{Wakely:2008,
	adsurl = {http://adsabs.harvard.edu/abs/2008ICRC....3.1341W},
	author = {{Wakely}, S.~P. and {Horan}, D.},
	journal = {International Cosmic Ray Conference},
	pages = {1341-1344},
	title = {{TeVCat: An online catalog for Very High Energy Gamma-Ray Astronomy}},
	volume = 3,
	year = 2008}

@article{Mattox:1996apj,
	adsurl = {http://adsabs.harvard.edu/abs/1996ApJ...461..396M},
	author = {{Mattox}, J.~R. and {Bertsch}, D.~L. and {Chiang}, J. and {Dingus}, B.~L. and {Digel}, S.~W. and {Esposito}, J.~A. and {Fierro}, J.~M. and {Hartman}, R.~C. and {Hunter}, S.~D. and {Kanbach}, G. and {Kniffen}, D.~A. and {Lin}, Y.~C. and {Macomb}, D.~J. and {Mayer-Hasselwander}, H.~A. and {Michelson}, P.~F. and {von Montigny}, C. and {Mukherjee}, R. and {Nolan}, P.~L. and {Ramanamurthy}, P.~V. and {Schneid}, E. and {Sreekumar}, P. and {Thompson}, D.~J. and {Willis}, T.~D.},
	doi = {10.1086/177068},
	journal = {\apj},
	month = apr,
	pages = {396},
	title = {{The Likelihood Analysis of EGRET Data}},
	volume = 461,
	year = 1996}

@article{Atwood:2009apj,
	adsurl = {http://adsabs.harvard.edu/abs/2009ApJ...697.1071A},
	archiveprefix = {arXiv},
	author = {{Atwood}, W.~B. and {Abdo}, A.~A. and {Ackermann}, M. and {Althouse}, W. and {Anderson}, B. and {Axelsson}, M. and {Baldini}, L. and {Ballet}, J. and {Band}, D.~L. and {Barbiellini}, G. and et al.},
	doi = {10.1088/0004-637X/697/2/1071},
	eprint = {0902.1089},
	journal = {\apj},
	month = jun,
	pages = {1071-1102},
	primaryclass = {astro-ph.IM},
	title = {{The Large Area Telescope on the Fermi Gamma-Ray Space Telescope Mission}},
	volume = 697,
	year = 2009}

@article{LiMa:1983apj,
	adsurl = {http://adsabs.harvard.edu/abs/1983ApJ...272..317L},
	author = {{Li}, T.-P. and {Ma}, Y.-Q.},
	doi = {10.1086/161295},
	journal = {\apj},
	month = sep,
	pages = {317-324},
	title = {{Analysis methods for results in gamma-ray astronomy}},
	volume = 272,
	year = 1983}

@article{MAGIC:2017ATEL9929,
	adsurl = {https://ui.adsabs.harvard.edu/abs/2017ATel.9929....1M},
	author = {{Mirzoyan}, Razmik},
	journal = {The Astronomer's Telegram},
	month = jan,
	pages = {1},
	title = {{MAGIC detection of a giant flaring activity from NGC 1275 at very-high-energy gamma rays}},
	volume = {9929},
	year = 2017}

@article{Holder:2011ICRC,
	adsurl = {http://adsabs.harvard.edu/abs/2011ICRC...12..137H},
	archiveprefix = {arXiv},
	author = {{Holder}, J.},
	doi = {10.7529/ICRC2011/V12/H11},
	eprint = {1111.1225},
	journal = {International Cosmic Ray Conference},
	pages = {137},
	primaryclass = {astro-ph.HE},
	title = {{VERITAS: Status and Highlights}},
	volume = 12,
	year = 2011}

@article{Aharonian_2009,
	adsurl = {https://ui.adsabs.harvard.edu/abs/2009ApJ...695L..40A},
	archiveprefix = {arXiv},
	author = {{Aharonian}, F. and {Akhperjanian}, A.~G. and {Anton}, G. and {de Almeida}, U. Barres and {Bazer-Bachi}, A.~R. and {Becherini}, Y. and {Behera}, B. and {Benbow}, W. and {Bernl{\"o}hr}, K. and {Boisson}, C. and {Bochow}, A. and {Borrel}, V. and {Brion}, E. and {Brucker}, J. and {Brun}, P. and {B{\"u}hler}, R. and {Bulik}, T. and {B{\"u}sching}, I. and {Boutelier}, T. and {Chadwick}, P.~M. and {Charbonnier}, A. and {Chaves}, R.~C.~G. and {Cheesebrough}, A. and {Chounet}, L. -M. and {Clapson}, A.~C. and {Coignet}, G. and {Dalton}, M. and {Daniel}, M.~K. and {Davids}, I.~D. and {Degrange}, B. and {Deil}, C. and {Dickinson}, H.~J. and {Djannati-Ata{\"\i}}, A. and {Domainko}, W. and {Drury}, L. O'C. and {Dubois}, F. and {Dubus}, G. and {Dyks}, J. and {Dyrda}, M. and {Egberts}, K. and {Emmanoulopoulos}, D. and {Espigat}, P. and {Farnier}, C. and {Feinstein}, F. and {Fiasson}, A. and {F{\"o}rster}, A. and {Fontaine}, G. and {F{\"u}{\ss}ling}, M. and {Gabici}, S. and {Gallant}, Y.~A. and {G{\'e}rard}, L. and {Giebels}, B. and {Glicenstein}, J. -F. and {Gl{\"u}ck}, B. and {Goret}, P. and {G{\"o}hring}, D. and {Hauser}, D. and {Hauser}, M. and {Heinz}, S. and {Heinzelmann}, G. and {Henri}, G. and {Hermann}, G. and {Hinton}, J.~A. and {Hoffmann}, A. and {Hofmann}, W. and {Holleran}, M. and {Hoppe}, S. and {Horns}, D. and {Jacholkowska}, A. and {de Jager}, O.~C. and {Jahn}, C. and {Jung}, I. and {Katarzy{\'n}ski}, K. and {Katz}, U. and {Kaufmann}, S. and {Kendziorra}, E. and {Kerschhaggl}, M. and {Khangulyan}, D. and {Kh{\'e}lifi}, B. and {Keogh}, D. and {Klu{\'z}niak}, W. and {Kneiske}, T. and {Komin}, Nu. and {Kosack}, K. and {Lamanna}, G. and {Latham}, I.~J. and {Lenain}, J. -P. and {Lohse}, T. and {Marandon}, V. and {Martin}, J.~M. and {Martineau-Huynh}, O. and {Marcowith}, A. and {Maurin}, D. and {McComb}, T.~J.~L. and {Medina}, M.~C. and {Moderski}, R. and {Moulin}, E. and {Naumann-Godo}, M. and {de Naurois}, M. and {Nedbal}, D. and {Nekrassov}, D. and {Niemiec}, J. and {Nolan}, S.~J. and {Ohm}, S. and {Olive}, J. -F. and {de O{\~n}a Wilhelmi}, E. and {Orford}, K.~J. and {Ostrowski}, M. and {Panter}, M. and {Arribas}, M. Paz and {Pedaletti}, G. and {Pelletier}, G. and {Petrucci}, P. -O. and {Pita}, S. and {P{\"u}hlhofer}, G. and {Punch}, M. and {Quirrenbach}, A. and {Raubenheimer}, B.~C. and {Raue}, M. and {Rayner}, S.~M. and {Renaud}, M. and {Rieger}, F. and {Ripken}, J. and {Rob}, L. and {Rosier-Lees}, S. and {Rowell}, G. and {Rudak}, B. and {Rulten}, C.~B. and {Ruppel}, J. and {Sahakian}, V. and {Santangelo}, A. and {Schlickeiser}, R. and {Sch{\"o}ck}, F.~M. and {Schr{\"o}der}, R. and {Schwanke}, U. and {Schwarzburg}, S. and {Schwemmer}, S. and {Shalchi}, A. and {Sikora}, M. and {Skilton}, J.~L. and {Sol}, H. and {Spangler}, D. and {Stawarz}, {\L}. and {Steenkamp}, R. and {Stegmann}, C. and {Superina}, G. and {Szostek}, A. and {Tam}, P.~H. and {Tavernet}, J. -P. and {Terrier}, R. and {Tibolla}, O. and {Tluczykont}, M. and {van Eldik}, C. and {Vasileiadis}, G. and {Venter}, C. and {Venter}, L. and {Vialle}, J.~P. and {Vincent}, P. and {Vink}, J. and {Vivier}, M. and {V{\"o}lk}, H.~J. and {Volpe}, F. and {Wagner}, S.~J. and {Ward}, M. and {Zdziarski}, A.~A. and {Zech}, A.},
	doi = {10.1088/0004-637X/695/1/L40},
	eprint = {0903.1582},
	journal = {\apjl},
	month = apr,
	number = {1},
	pages = {L40-L44},
	primaryclass = {astro-ph.CO},
	title = {{Discovery of Very High Energy {\ensuremath{\gamma}}-Ray Emission from Centaurus a with H.E.S.S.}},
	volume = {695},
	year = 2009}

@article{Abramowski_2012,
	adsurl = {https://ui.adsabs.harvard.edu/abs/2012ApJ...746..151A},
	archiveprefix = {arXiv},
	author = {{Abramowski}, A. and {Acero}, F. and {Aharonian}, F. and {Akhperjanian}, A.~G. and {Anton}, G. and {Balzer}, A. and {Barnacka}, A. and {Barres de Almeida}, U. and {Becherini}, Y. and {Becker}, J. and {Behera}, B. and {Bernl{\"o}hr}, K. and {Birsin}, E. and {Biteau}, J. and {Bochow}, A. and {Boisson}, C. and {Bolmont}, J. and {Bordas}, P. and {Brucker}, J. and {Brun}, F. and {Brun}, P. and {Bulik}, T. and {B{\"u}sching}, I. and {Carrigan}, S. and {Casanova}, S. and {Cerruti}, M. and {Chadwick}, P.~M. and {Charbonnier}, A. and {Chaves}, R.~C.~G. and {Cheesebrough}, A. and {Clapson}, A.~C. and {Coignet}, G. and {Cologna}, G. and {Conrad}, J. and {Dalton}, M. and {Daniel}, M.~K. and {Davids}, I.~D. and {Degrange}, B. and {Deil}, C. and {Dickinson}, H.~J. and {Djannati-Ata{\"\i}}, A. and {Domainko}, W. and {Drury}, L. O'C. and {Dubus}, G. and {Dutson}, K. and {Dyks}, J. and {Dyrda}, M. and {Egberts}, K. and {Eger}, P. and {Espigat}, P. and {Fallon}, L. and {Farnier}, C. and {Fegan}, S. and {Feinstein}, F. and {Fernandes}, M.~V. and {Fiasson}, A. and {Fontaine}, G. and {F{\"o}rster}, A. and {F{\"u}{\ss}ling}, M. and {Gallant}, Y.~A. and {Gast}, H. and {G{\'e}rard}, L. and {Gerbig}, D. and {Giebels}, B. and {Glicenstein}, J.~F. and {Gl{\"u}ck}, B. and {Goret}, P. and {G{\"o}ring}, D. and {H{\"a}ffner}, S. and {Hague}, J.~D. and {Hampf}, D. and {Hauser}, M. and {Heinz}, S. and {Heinzelmann}, G. and {Henri}, G. and {Hermann}, G. and {Hinton}, J.~A. and {Hoffmann}, A. and {Hofmann}, W. and {Hofverberg}, P. and {Holler}, M. and {Horns}, D. and {Jacholkowska}, A. and {de Jager}, O.~C. and {Jahn}, C. and {Jamrozy}, M. and {Jung}, I. and {Kastendieck}, M.~A. and {Katarzy{\'n}ski}, K. and {Katz}, U. and {Kaufmann}, S. and {Keogh}, D. and {Khangulyan}, D. and {Kh{\'e}lifi}, B. and {Klochkov}, D. and {Klu{\'z}niak}, W. and {Kneiske}, T. and {Komin}, Nu. and {Kosack}, K. and {Kossakowski}, R. and {Laffon}, H. and {Lamanna}, G. and {Lennarz}, D. and {Lohse}, T. and {Lopatin}, A. and {Lu}, C. -C. and {Marandon}, V. and {Marcowith}, A. and {Masbou}, J. and {Maurin}, D. and {Maxted}, N. and {Mayer}, M. and {McComb}, T.~J.~L. and {Medina}, M.~C. and {M{\'e}hault}, J. and {Moderski}, R. and {Moulin}, E. and {Naumann}, C.~L. and {Naumann-Godo}, M. and {de Naurois}, M. and {Nedbal}, D. and {Nekrassov}, D. and {Nguyen}, N. and {Nicholas}, B. and {Niemiec}, J. and {Nolan}, S.~J. and {Ohm}, S. and {de O{\~n}a Wilhelmi}, E. and {Opitz}, B. and {Ostrowski}, M. and {Oya}, I. and {Panter}, M. and {Paz Arribas}, M. and {Pedaletti}, G. and {Pelletier}, G. and {Petrucci}, P. -O. and {Pita}, S. and {P{\"u}hlhofer}, G. and {Punch}, M. and {Quirrenbach}, A. and {Raue}, M. and {Rayner}, S.~M. and {Reimer}, A. and {Reimer}, O. and {Renaud}, M. and {de los Reyes}, R. and {Rieger}, F. and {Ripken}, J. and {Rob}, L. and {Rosier-Lees}, S. and {Rowell}, G. and {Rudak}, B. and {Rulten}, C.~B. and {Ruppel}, J. and {Sahakian}, V. and {Sanchez}, D.~A. and {Santangelo}, A. and {Schlickeiser}, R. and {Sch{\"o}ck}, F.~M. and {Schulz}, A. and {Schwanke}, U. and {Schwarzburg}, S. and {Schwemmer}, S. and {Sheidaei}, F. and {Skilton}, J.~L. and {Sol}, H. and {Spengler}, G. and {Stawarz}, {\L}. and {Steenkamp}, R. and {Stegmann}, C. and {Stinzing}, F. and {Stycz}, K. and {Sushch}, I. and {Szostek}, A. and {Tavernet}, J. -P. and {Terrier}, R. and {Tluczykont}, M. and {Valerius}, K. and {van Eldik}, C. and {Vasileiadis}, G. and {Venter}, C. and {Vialle}, J.~P. and {Viana}, A. and {Vincent}, P. and {V{\"o}lk}, H.~J. and {Volpe}, F. and {Vorobiov}, S. and {Vorster}, M. and {Wagner}, S.~J. and {Ward}, M. and {White}, R. and {Wierzcholska}, A. and {Zacharias}, M. and {Zajczyk}, A. and {Zdziarski}, A.~A. and {Zech}, A. and {Zechlin}, H. -S. and {H.~E.~S.~S. Collaboration} and {Aleksi{\'c}}, J. and {Antonelli}, L.~A. and {Antoranz}, P. and {Backes}, M. and {Barrio}, J.~A. and {Bastieri}, D. and {Becerra Gonz{\'a}lez}, J. and {Bednarek}, W. and {Berdyugin}, A. and {Berger}, K. and {Bernardini}, E. and {Biland}, A. and {Blanch}, O. and {Bock}, R.~K. and {Boller}, A. and {Bonnoli}, G. and {Borla Tridon}, D. and {Braun}, I. and {Bretz}, T. and {Ca{\~n}ellas}, A. and {Carmona}, E. and {Carosi}, A. and {Colin}, P. and {Colombo}, E. and {Contreras}, J.~L. and {Cortina}, J. and {Cossio}, L. and {Covino}, S. and {Dazzi}, F. and {De Angelis}, A. and {De Cea del Pozo}, E. and {De Lotto}, B. and {Delgado Mendez}, C. and {Diago Ortega}, A. and {Doert}, M. and {Dom{\'\i}nguez}, A. and {Dominis Prester}, D. and {Dorner}, D. and {Doro}, M. and {Elsaesser}, D. and {Ferenc}, D. and {Fonseca}, M.~V. and {Font}, L. and {Fruck}, C. and {Garc{\'\i}a L{\'o}pez}, R.~J. and {Garczarczyk}, M. and {Garrido}, D. and {Giavitto}, G. and {Godinovi{\'c}}, N. and {Hadasch}, D. and {H{\"a}fner}, D. and {Herrero}, A. and {Hildebrand}, D. and {H{\"o}hne-M{\"o}nch}, D. and {Hose}, J. and {Hrupec}, D. and {Huber}, B. and {Jogler}, T. and {Klepser}, S. and {Kr{\"a}henb{\"u}hl}, T. and {Krause}, J. and {La Barbera}, A. and {Lelas}, D. and {Leonardo}, E. and {Lindfors}, E. and {Lombardi}, S. and {L{\'o}pez}, M. and {Lorenz}, E. and {Makariev}, M. and {Maneva}, G. and {Mankuzhiyil}, N. and {Mannheim}, K. and {Maraschi}, L. and {Mariotti}, M. and {Mart{\'\i}nez}, M. and {Mazin}, D. and {Meucci}, M. and {Miranda}, J.~M. and {Mirzoyan}, R. and {Miyamoto}, H. and {Mold{\'o}n}, J. and {Moralejo}, A. and {Munar}, P. and {Nieto}, D. and {Nilsson}, K. and {Orito}, R. and {Oya}, I. and {Paneque}, D. and {Paoletti}, R. and {Pardo}, S. and {Paredes}, J.~M. and {Partini}, S. and {Pasanen}, M. and {Pauss}, F. and {Perez-Torres}, M.~A. and {Persic}, M. and {Peruzzo}, L. and {Pilia}, M. and {Pochon}, J. and {Prada}, F. and {Prada Moroni}, P.~G. and {Prandini}, E. and {Puljak}, I. and {Reichardt}, I. and {Reinthal}, R. and {Rhode}, W. and {Rib{\'o}}, M. and {Rico}, J. and {R{\"u}gamer}, S. and {Saggion}, A. and {Saito}, K. and {Saito}, T.~Y. and {Salvati}, M. and {Satalecka}, K. and {Scalzotto}, V. and {Scapin}, V. and {Schultz}, C. and {Schweizer}, T. and {Shayduk}, M. and {Shore}, S.~N. and {Sillanp{\"a}{\"a}}, A. and {Sitarek}, J. and {Sobczynska}, D. and {Spanier}, F. and {Spiro}, S. and {Stamerra}, A. and {Steinke}, B. and {Storz}, J. and {Strah}, N. and {Suri{\'c}}, T. and {Takalo}, L. and {Takami}, H. and {Tavecchio}, F. and {Temnikov}, P. and {Terzi{\'c}}, T. and {Tescaro}, D. and {Teshima}, M. and {Thom}, M. and {Tibolla}, O. and {Torres}, D.~F. and {Treves}, A. and {Vankov}, H. and {Vogler}, P. and {Wagner}, R.~M. and {Weitzel}, Q. and {Zabalza}, V. and {Zandanel}, F. and {Zanin}, R. and {MAGIC Collaboration} and {Arlen}, T. and {Aune}, T. and {Beilicke}, M. and {Benbow}, W. and {Bouvier}, A. and {Bradbury}, S.~M. and {Buckley}, J.~H. and {Bugaev}, V. and {Byrum}, K. and {Cannon}, A. and {Cesarini}, A. and {Ciupik}, L. and {Connolly}, M.~P. and {Cui}, W. and {Dickherber}, R. and {Duke}, C. and {Errando}, M. and {Falcone}, A. and {Finley}, J.~P. and {Finnegan}, G. and {Fortson}, L. and {Furniss}, A. and {Galante}, N. and {Gall}, D. and {Godambe}, S. and {Griffin}, S. and {Grube}, J. and {Gyuk}, G. and {Hanna}, D. and {Holder}, J. and {Huan}, H. and {Hui}, C.~M. and {Kaaret}, P. and {Karlsson}, N. and {Kertzman}, M. and {Khassen}, Y. and {Kieda}, D. and {Krawczynski}, H. and {Krennrich}, F. and {Lang}, M.~J. and {LeBohec}, S. and {Maier}, G. and {McArthur}, S. and {McCann}, A. and {Moriarty}, P. and {Mukherjee}, R. and {Nu{\~n}ez}, P.~D. and {Ong}, R.~A. and {Orr}, M. and {Otte}, A.~N. and {Park}, N. and {Perkins}, J.~S. and {Pichel}, A. and {Pohl}, M. and {Prokoph}, H. and {Ragan}, K. and {Reyes}, L.~C. and {Reynolds}, P.~T. and {Roache}, E. and {Rose}, H.~J. and {Ruppel}, J. and {Schroedter}, M. and {Sembroski}, G.~H. and {{\c{S}}ent{\"u}rk}, G.~D. and {Telezhinsky}, I. and {Te{\v{s}}i{\'c}}, G. and {Theiling}, M. and {Thibadeau}, S. and {Varlotta}, A. and {Vassiliev}, V.~V. and {Vivier}, M. and {Wakely}, S.~P. and {Weekes}, T.~C. and {Williams}, D.~A. and {Zitzer}, B. and {VERITAS Collaboration} and {Barres de Almeida}, U. and {Cara}, M. and {Casadio}, C. and {Cheung}, C.~C. and {McConville}, W. and {Davies}, F. and {Doi}, A. and {Giovannini}, G. and {Giroletti}, M. and {Hada}, K. and {Hardee}, P. and {Harris}, D.~E. and {Junor}, W. and {Kino}, M. and {Lee}, N.~P. and {Ly}, C. and {Madrid}, J. and {Massaro}, F. and {Mundell}, C.~G. and {Nagai}, H. and {Perlman}, E.~S. and {Steele}, I.~A. and {Walker}, R.~C. and {Wood}, D.~L.},
	doi = {10.1088/0004-637X/746/2/151},
	eid = {151},
	eprint = {1111.5341},
	journal = {\apj},
	month = feb,
	number = {2},
	pages = {151},
	primaryclass = {astro-ph.CO},
	title = {{The 2010 Very High Energy {\ensuremath{\gamma}}-Ray Flare and 10 Years of Multi-wavelength Observations of M 87}},
	volume = {746},
	year = 2012}

@article{VERITAS:2017ATEL9931,
	author = {Mukherjee, Reshmi and {VERITAS Collaboration}},
	journal = {The Astronomer's Telegram},
	month = jan,
	note = {ADS Bibcode: 2017ATel.9931....1M},
	pages = {1},
	title = {{VERITAS} detection of the radio galaxy {NGC} 1275 with elevated very-high-energy gamma-ray emission},
	url = {https://ui.adsabs.harvard.edu/abs/2017ATel.9931....1M},
	urldate = {2022-11-28},
	volume = {9931},
	year = {2017}}

@article{Acciari-2008-ApJ,
	author = {V. A. Acciari and M. Beilicke and G. Blaylock and S. M. Bradbury and J. H. Buckley and V. Bugaev and Y. Butt and O. Celik and A. Cesarini and L. Ciupik and P. Cogan and P. Colin and W. Cui and M. K. Daniel and C. Duke and T. Ergin and A. D. Falcone and S. J. Fegan and J. P. Finley and G. Finnegan and P. Fortin and L. F. Fortson and K. Gibbs and G. H. Gillanders and J. Grube and R. Guenette and G. Gyuk and D. Hanna and E. Hays and J. Holder and D. Horan and S. B. Hughes and M. C. Hui and T. B. Humensky and A. Imran and P. Kaaret and M. Kertzman and D. B. Kieda and J. Kildea and A. Konopelko and H. Krawczynski and F. Krennrich and M. J. Lang and S. LeBohec and K. Lee and G. Maier and A. McCann and M. McCutcheon and J. Millis and P. Moriarty and R. Mukherjee and T. Nagai and R. A. Ong and D. Pandel and J. S. Perkins and M. Pohl and J. Quinn and K. Ragan and P. T. Reynolds and H. J. Rose and M. Schroedter and G. H. Sembroski and A. W. Smith and D. Steele and S. P. Swordy and A. Syson and J. A. Toner and L. Valcarcel and V. V. Vassiliev and S. P. Wakely and J. E. Ward and T. C. Weekes and A. Weinstein and R. J. White and D. A. Williams and S. A. Wissel and M. D. Wood and B. Zitzer},
	doi = {10.1086/587458},
	journal = {The Astrophysical Journal},
	month = {may},
	number = {1},
	pages = {397},
	title = {Observation of Gamma-Ray Emission from the Galaxy M87 above 250 GeV with VERITAS*},
	url = {https://dx.doi.org/10.1086/587458},
	volume = {679},
	year = {2008}}

@article{Ghisellini:2005aa,
	adsurl = {http://adsabs.harvard.edu/abs/2005A%26A...432..401G},
	arxivurl = {http://arxiv.org/abs/astro-ph/0406093},
	author = {{Ghisellini}, G. and {Tavecchio}, F. and {Chiaberge}, M.},
	doi = {10.1051/0004-6361:20041404},
	eprint = {astro-ph/0406093},
	journal = {\aap},
	month = mar,
	pages = {401-410},
	title = {{Structured jets in TeV BL Lac objects and radiogalaxies. Implications for the observed properties}},
	volume = 432,
	year = 2005}

@article{Ghisellini:1993aa,
	adsurl = {http://adsabs.harvard.edu/abs/1993ApJ...407...65G},
	author = {{Ghisellini}, G. and {Padovani}, P. and {Celotti}, A. and {Maraschi}, L.},
	doi = {10.1086/172493},
	journal = {\apj},
	month = apr,
	pages = {65-82},
	title = {{Relativistic bulk motion in active galactic nuclei}},
	volume = 407,
	year = 1993}

@article{Urry:1995aa,
	adsurl = {http://adsabs.harvard.edu/abs/1995PASP..107..803U},
	arxivurl = {http://arxiv.org/abs/astro-ph/9506063},
	author = {{Urry}, C.~M. and {Padovani}, P.},
	doi = {10.1086/133630},
	eprint = {astro-ph/9506063},
	journal = {\pasp},
	month = sep,
	pages = {803},
	title = {{Unified Schemes for Radio-Loud Active Galactic Nuclei}},
	volume = 107,
	year = 1995}

@article{Aleksic:2012aa,
	adsurl = {http://adsabs.harvard.edu/abs/2012A%26A...539L...2A},
	archiveprefix = {arXiv},
	arxivurl = {http://arxiv.org/abs/1112.3917},
	author = {{Aleksi{\'c}}, J. and {Alvarez}, E.~A. and {Antonelli}, L.~A. and {Antoranz}, P. and {Asensio}, M. and {Backes}, M. and {Barres de Almeida}, U. and {Barrio}, J.~A. and {Bastieri}, D. and {Becerra Gonz{\'a}lez}, J. and {Bednarek}, W. and {Berger}, K. and {Bernardini}, E. and {Biland}, A. and {Blanch}, O. and {Bock}, R.~K. and {Boller}, A. and {Bonnoli}, G. and {Borla Tridon}, D. and {Bretz}, T. and {Ca{\~n}ellas}, A. and {Carmona}, E. and {Carosi}, A. and {Colin}, P. and {Colombo}, E. and {Contreras}, J.~L. and {Cortina}, J. and {Cossio}, L. and {Covino}, S. and {da Vela}, P. and {Dazzi}, F. and {de Angelis}, A. and {de Caneva}, G. and {de Cea Del Pozo}, E. and {de Lotto}, B. and {Delgado Mendez}, C. and {Diago Ortega}, A. and {Doert}, M. and {Dom{\'{\i}}nguez}, A. and {Dominis Prester}, D. and {Dorner}, D. and {Doro}, M. and {Eisenacher}, D. and {Elsaesser}, D. and {Ferenc}, D. and {Fonseca}, M.~V. and {Font}, L. and {Fruck}, C. and {Garc{\'{\i}}a L{\'o}pez}, R.~J. and {Garczarczyk}, M. and {Garrido}, D. and {Giavitto}, G. and {Godinovi{\'c}}, N. and {Gozzini}, S.~R. and {Hadasch}, D. and {H{\"a}fner}, D. and {Herrero}, A. and {Hildebrand}, D. and {H{\"o}hne-M{\"o}nch}, D. and {Hose}, J. and {Hrupec}, D. and {Huber}, B. and {Jogler}, T. and {Kadenius}, V. and {Kellermann}, H. and {Klepser}, S. and {Kr{\"a}henb{\"u}hl}, T. and {Krause}, J. and {La Barbera}, A. and {Lelas}, D. and {Leonardo}, E. and {Lewandowska}, N. and {Lindfors}, E. and {Lombardi}, S. and {L{\'o}pez}, M. and {L{\'o}pez-Coto}, R. and {L{\'o}pez-Oramas}, A. and {Lorenz}, E. and {Makariev}, M. and {Maneva}, G. and {Mankuzhiyil}, N. and {Mannheim}, K. and {Maraschi}, L. and {Mariotti}, M. and {Mart{\'{\i}}nez}, M. and {Mazin}, D. and {Meucci}, M. and {Miranda}, J.~M. and {Mirzoyan}, R. and {Mold{\'o}n}, J. and {Moralejo}, A. and {Munar-Adrover}, P. and {Niedzwiecki}, A. and {Nieto}, D. and {Nilsson}, K. and {Nowak}, N. and {Orito}, R. and {Paiano}, S. and {Paneque}, D. and {Paoletti}, R. and {Pardo}, S. and {Paredes}, J.~M. and {Partini}, S. and {Perez-Torres}, M.~A. and {Persic}, M. and {Peruzzo}, L. and {Pilia}, M. and {Pochon}, J. and {Prada}, F. and {Prada Moroni}, P.~G. and {Prandini}, E. and {Puerto Gimenez}, I. and {Puljak}, I. and {Reichardt}, I. and {Reinthal}, R. and {Rhode}, W. and {Rib{\'o}}, M. and {Rico}, J. and {R{\"u}gamer}, S. and {Saggion}, A. and {Saito}, K. and {Saito}, T.~Y. and {Salvati}, M. and {Satalecka}, K. and {Scalzotto}, V. and {Scapin}, V. and {Schultz}, C. and {Schweizer}, T. and {Shayduk}, M. and {Shore}, S.~N. and {Sillanp{\"a}{\"a}}, A. and {Sitarek}, J. and {Snidaric}, I. and {Sobczynska}, D. and {Spanier}, F. and {Spiro}, S. and {Stamatescu}, V. and {Stamerra}, A. and {Steinke}, B. and {Storz}, J. and {Strah}, N. and {Sun}, S. and {Suri{\'c}}, T. and {Takalo}, L. and {Takami}, H. and {Tavecchio}, F. and {Temnikov}, P. and {Terzi{\'c}}, T. and {Tescaro}, D. and {Teshima}, M. and {Tibolla}, O. and {Torres}, D.~F. and {Treves}, A. and {Uellenbeck}, M. and {Vogler}, P. and {Wagner}, R.~M. and {Weitzel}, Q. and {Zabalza}, V. and {Zandanel}, F. and {Zanin}, R. and {Pfrommer}, C. and {Pinzke}, A.},
	doi = {10.1051/0004-6361/201118668},
	eid = {L2},
	eprint = {1112.3917},
	journal = {\aap},
	month = mar,
	pages = {L2},
	primaryclass = {astro-ph.HE},
	title = {{Detection of very-high energy {$\gamma$}-ray emission from <ASTROBJ>NGC 1275</ASTROBJ> by the MAGIC telescopes}},
	volume = 539,
	year = 2012}

@article{Acciari2010M87,
	adsurl = {https://ui.adsabs.harvard.edu/abs/2010ApJ...716..819A},
	archiveprefix = {arXiv},
	author = {{Acciari}, V.~A. and {Aliu}, E. and {Arlen}, T. and {Aune}, T. and {Beilicke}, M. and {Benbow}, W. and {Boltuch}, D. and {Bradbury}, S.~M. and {Buckley}, J.~H. and {Bugaev}, V. and {Byrum}, K. and {Cannon}, A. and {Cesarini}, A. and {Chow}, Y.~C. and {Ciupik}, L. and {Cogan}, P. and {Cui}, W. and {Dickherber}, R. and {Duke}, C. and {Finley}, J.~P. and {Finnegan}, G. and {Fortin}, P. and {Fortson}, L. and {Furniss}, A. and {Galante}, N. and {Gall}, D. and {Gillanders}, G.~H. and {Godambe}, S. and {Grube}, J. and {Guenette}, R. and {Gyuk}, G. and {Hanna}, D. and {Holder}, J. and {Hui}, C.~M. and {Humensky}, T.~B. and {Imran}, A. and {Kaaret}, P. and {Karlsson}, N. and {Kertzman}, M. and {Kieda}, D. and {Konopelko}, A. and {Krawczynski}, H. and {Krennrich}, F. and {Lang}, M.~J. and {LeBohec}, S. and {Maier}, G. and {McArthur}, S. and {McCann}, A. and {McCutcheon}, M. and {Millis}, J. and {Moriarty}, P. and {Ong}, R.~A. and {Otte}, A.~N. and {Pandel}, D. and {Perkins}, J.~S. and {Pichel}, A. and {Pohl}, M. and {Quinn}, J. and {Ragan}, K. and {Reyes}, L.~C. and {Reynolds}, P.~T. and {Roache}, E. and {Rose}, H.~J. and {Rovero}, A.~C. and {Schroedter}, M. and {Sembroski}, G.~H. and {Senturk}, G. Demet and {Smith}, A.~W. and {Steele}, D. and {Swordy}, S.~P. and {Theiling}, M. and {Thibadeau}, S. and {Varlotta}, A. and {Vincent}, S. and {Wagner}, R.~G. and {Wakely}, S.~P. and {Ward}, J.~E. and {Weekes}, T.~C. and {Weinstein}, A. and {Weisgarber}, T. and {Williams}, D.~A. and {Wissel}, S. and {Wood}, M. and {Zitzer}, B. and {Harris}, D.~E. and {Massaro}, F.},
	doi = {10.1088/0004-637X/716/1/819},
	eprint = {1005.0367},
	journal = {\apj},
	month = jun,
	number = {1},
	pages = {819-824},
	primaryclass = {astro-ph.CO},
	title = {{Veritas 2008-2009 Monitoring of the Variable Gamma-ray Source M 87}},
	volume = {716},
	year = 2010}

@article{Aliu-2012-ApJ,
	adsurl = {https://ui.adsabs.harvard.edu/abs/2012ApJ...746..141A},
	archiveprefix = {arXiv},
	author = {{Aliu}, E. and {Arlen}, T. and {Aune}, T. and {Beilicke}, M. and {Benbow}, W. and {Bouvier}, A. and {Bradbury}, S.~M. and {Buckley}, J.~H. and {Bugaev}, V. and {Byrum}, K. and {Cannon}, A. and {Cesarini}, A. and {Ciupik}, L. and {Collins-Hughes}, E. and {Connolly}, M.~P. and {Cui}, W. and {Dickherber}, R. and {Duke}, C. and {Errando}, M. and {Falcone}, A. and {Finley}, J.~P. and {Finnegan}, G. and {Fortson}, L. and {Furniss}, A. and {Galante}, N. and {Gall}, D. and {Godambe}, S. and {Griffin}, S. and {Grube}, J. and {Guenette}, R. and {Gyuk}, G. and {Hanna}, D. and {Holder}, J. and {Huan}, H. and {Hughes}, G. and {Hui}, C.~M. and {Humensky}, T.~B. and {Imran}, A. and {Kaaret}, P. and {Karlsson}, N. and {Kertzman}, M. and {Kieda}, D. and {Krawczynski}, H. and {Krennrich}, F. and {Lang}, M.~J. and {LeBohec}, S. and {Madhavan}, A.~S. and {Maier}, G. and {Majumdar}, P. and {McArthur}, S. and {McCann}, A. and {Moriarty}, P. and {Mukherjee}, R. and {Nu{\~n}ez}, P.~D. and {Ong}, R.~A. and {Orr}, M. and {Otte}, A.~N. and {Park}, N. and {Perkins}, J.~S. and {Pichel}, A. and {Pohl}, M. and {Prokoph}, H. and {Quinn}, J. and {Ragan}, K. and {Reyes}, L.~C. and {Reynolds}, P.~T. and {Roache}, E. and {Rose}, H.~J. and {Ruppel}, J. and {Saxon}, D.~B. and {Schroedter}, M. and {Sembroski}, G.~H. and {{\c{S}}ent{\"u}rk}, G.~D. and {Skole}, C. and {Staszak}, D. and {Te{\v{s}}i{\'c}}, G. and {Theiling}, M. and {Thibadeau}, S. and {Tsurusaki}, K. and {Tyler}, J. and {Varlotta}, A. and {Vassiliev}, V.~V. and {Vincent}, S. and {Vivier}, M. and {Wakely}, S.~P. and {Ward}, J.~E. and {Weekes}, T.~C. and {Weinstein}, A. and {Weisgarber}, T. and {Williams}, D.~A. and {Zitzer}, B.},
	doi = {10.1088/0004-637X/746/2/141},
	eid = {141},
	eprint = {1112.4518},
	journal = {\apj},
	month = feb,
	number = {2},
	pages = {141},
	primaryclass = {astro-ph.CO},
	title = {{VERITAS Observations of Day-scale Flaring of M 87 in 2010 April}},
	volume = {746},
	year = 2012}

@article{Archer_2020,
	adsurl = {https://ui.adsabs.harvard.edu/abs/2020ApJ...896...41A},
	archiveprefix = {arXiv},
	author = {{Archer}, A. and {Benbow}, W. and {Bird}, R. and {Brill}, A. and {Buchovecky}, M. and {Buckley}, J.~H. and {Carini}, M.~T. and {Christiansen}, J.~L. and {Chromey}, A.~J. and {Daniel}, M.~K. and {Errando}, M. and {Falcone}, A. and {Feng}, Q. and {Fortin}, P. and {Fortson}, L. and {Furniss}, A. and {Gent}, A. and {Georganopoulos}, M. and {Gillanders}, G.~H. and {Giuri}, C. and {Gueta}, O. and {Hanna}, D. and {Hassan}, T. and {Hervet}, O. and {Holder}, J. and {Hughes}, G. and {Humensky}, T.~B. and {Kaaret}, P. and {Kertzman}, M. and {Kieda}, D. and {Krennrich}, F. and {Lang}, M.~J. and {Lin}, T.~T.~Y. and {Lister}, M.~L. and {Lundy}, M. and {Maier}, G. and {Meyer}, E. and {Moriarty}, P. and {Mukherjee}, R. and {Nieto}, D. and {Nievas-Rosillo}, M. and {O'Brien}, S. and {Ong}, R.~A. and {Pfrang}, K. and {Pohl}, M. and {Prado}, R.~R. and {Pueschel}, E. and {Quinn}, J. and {Ragan}, K. and {Ramirez}, K. and {Reynolds}, P.~T. and {Ribeiro}, D. and {Richards}, G.~T. and {Roache}, E. and {Rulten}, C. and {Ryan}, J.~L. and {Sadun}, A. and {Santander}, M. and {Scott}, S.~S. and {Sembroski}, G.~H. and {Shahinyan}, K. and {Shang}, R. and {Stevenson}, B. and {Vassiliev}, V.~V. and {Wakely}, S.~P. and {Weinstein}, A. and {Wilcox}, P. and {Wilhelm}, A. and {Williams}, D.~A. and {Williamson}, T.~J.},
	doi = {10.3847/1538-4357/ab910e},
	eid = {41},
	eprint = {2005.03110},
	journal = {\apj},
	month = jun,
	number = {1},
	pages = {41},
	primaryclass = {astro-ph.HE},
	title = {{VERITAS Discovery of VHE Emission from the Radio Galaxy 3C 264: A Multiwavelength Study}},
	volume = {896},
	year = 2020}

@article{Aleksic:2014ab,
	adsurl = {http://adsabs.harvard.edu/abs/2014A%26A...564A...5A},
	archiveprefix = {arXiv},
	arxivurl = {http://arxiv.org/abs/1310.8500},
	author = {{Aleksi{\'c}}, J. and {Ansoldi}, S. and {Antonelli}, L.~A. and {Antoranz}, P. and {Babic}, A. and {Bangale}, P. and {Barres de Almeida}, U. and {Barrio}, J.~A. and {Becerra Gonz{\'a}lez}, J. and {Bednarek}, W. and {Berger}, K. and {Bernardini}, E. and {Biland}, A. and {Blanch}, O. and {Bock}, R.~K. and {Bonnefoy}, S. and {Bonnoli}, G. and {Borracci}, F. and {Bretz}, T. and {Carmona}, E. and {Carosi}, A. and {Carreto Fidalgo}, D. and {Colin}, P. and {Colombo}, E. and {Contreras}, J.~L. and {Cortina}, J. and {Covino}, S. and {Da Vela}, P. and {Dazzi}, F. and {De Angelis}, A. and {De Caneva}, G. and {De Lotto}, B. and {Delgado Mendez}, C. and {Doert}, M. and {Dom{\'{\i}}nguez}, A. and {Dominis Prester}, D. and {Dorner}, D. and {Doro}, M. and {Einecke}, S. and {Eisenacher}, D. and {Elsaesser}, D. and {Farina}, E. and {Ferenc}, D. and {Fonseca}, M.~V. and {Font}, L. and {Frantzen}, K. and {Fruck}, C. and {Garc{\'{\i}}a L{\'o}pez}, R.~J. and {Garczarczyk}, M. and {Garrido Terrats}, D. and {Gaug}, M. and {Giavitto}, G. and {Godinovi{\'c}}, N. and {Gonz{\'a}lez Mu{\~n}oz}, A. and {Gozzini}, S.~R. and {Hadamek}, A. and {Hadasch}, D. and {Herrero}, A. and {Hildebrand}, D. and {Hose}, J. and {Hrupec}, D. and {Idec}, W. and {Kadenius}, V. and {Kellermann}, H. and {Knoetig}, M.~L. and {Krause}, J. and {Kushida}, J. and {La Barbera}, A. and {Lelas}, D. and {Lewandowska}, N. and {Lindfors}, E. and {Lombardi}, S. and {L{\'o}pez}, M. and {L{\'o}pez-Coto}, R. and {L{\'o}pez-Oramas}, A. and {Lorenz}, E. and {Lozano}, I. and {Makariev}, M. and {Mallot}, K. and {Maneva}, G. and {Mankuzhiyil}, N. and {Mannheim}, K. and {Maraschi}, L. and {Marcote}, B. and {Mariotti}, M. and {Mart{\'{\i}}nez}, M. and {Mazin}, D. and {Menzel}, U. and {Meucci}, M. and {Miranda}, J.~M. and {Mirzoyan}, R. and {Moralejo}, A. and {Munar-Adrover}, P. and {Nakajima}, D. and {Niedzwiecki}, A. and {Nilsson}, K. and {Nowak}, N. and {Orito}, R. and {Overkemping}, A. and {Paiano}, S. and {Palatiello}, M. and {Paneque}, D. and {Paoletti}, R. and {Paredes}, J.~M. and {Paredes-Fortuny}, X. and {Partini}, S. and {Persic}, M. and {Prada}, F. and {Prada Moroni}, P.~G. and {Prandini}, E. and {Preziuso}, S. and {Puljak}, I. and {Reinthal}, R. and {Rhode}, W. and {Rib{\'o}}, M. and {Rico}, J. and {Rodriguez Garcia}, J. and {R{\"u}gamer}, S. and {Saggion}, A. and {Saito}, T. and {Saito}, K. and {Salvati}, M. and {Satalecka}, K. and {Scalzotto}, V. and {Scapin}, V. and {Schultz}, C. and {Schweizer}, T. and {Shore}, S.~N. and {Sillanp{\"a}{\"a}}, A. and {Sitarek}, J. and {Snidaric}, I. and {Sobczynska}, D. and {Spanier}, F. and {Stamatescu}, V. and {Stamerra}, A. and {Steinbring}, T. and {Storz}, J. and {Sun}, S. and {Suri{\'c}}, T. and {Takalo}, L. and {Tavecchio}, F. and {Terzi{\'c}}, T. and {Tescaro}, D. and {Teshima}, M. and {Thaele}, J. and {Tibolla}, O. and {Torres}, D.~F. and {Toyama}, T. and {Treves}, A. and {Uellenbeck}, M. and {Vogler}, P. and {Wagner}, R.~M. and {Zandanel}, F. and {Zanin}, R. and {MAGIC Collaboration} and {Balmaverde}, B. and {Kataoka}, J. and {Rekola}, R. and {Takahashi}, Y.},
	doi = {10.1051/0004-6361/201322951},
	eid = {A5},
	eprint = {1310.8500},
	journal = {\aap},
	month = apr,
	pages = {A5},
	primaryclass = {astro-ph.HE},
	title = {{Contemporaneous observations of the radio galaxy NGC 1275 from radio to very high energy {$\gamma$}-rays}},
	volume = 564,
	year = 2014}

@article{Acciari:2008aa,
	adsurl = {http://adsabs.harvard.edu/abs/2008ApJ...679.1427A},
	archiveprefix = {arXiv},
	arxivurl = {http://arxiv.org/abs/0802.2363},
	author = {{Acciari}, V.~A. and {Beilicke}, M. and {Blaylock}, G. and {Bradbury}, S.~M. and {Buckley}, J.~H. and {Bugaev}, V. and {Butt}, Y. and {Byrum}, K.~L. and {Celik}, O. and {Cesarini}, A. and {Ciupik}, L. and {Chow}, Y.~C.~K. and {Cogan}, P. and {Colin}, P. and {Cui}, W. and {Daniel}, M.~K. and {Duke}, C. and {Ergin}, T. and {Falcone}, A.~D. and {Fegan}, S.~J. and {Finley}, J.~P. and {Fortin}, P. and {Fortson}, L.~F. and {Gall}, D. and {Gibbs}, K. and {Gillanders}, G.~H. and {Grube}, J. and {Guenette}, R. and {Hanna}, D. and {Hays}, E. and {Holder}, J. and {Horan}, D. and {Hughes}, S.~B. and {Hui}, C.~M. and {Humensky}, T.~B. and {Kaaret}, P. and {Kieda}, D.~B. and {Kildea}, J. and {Konopelko}, A. and {Krawczynski}, H. and {Krennrich}, F. and {Lang}, M.~J. and {LeBohec}, S. and {Lee}, K. and {Maier}, G. and {McCann}, A. and {McCutcheon}, M. and {Millis}, J. and {Moriarty}, P. and {Mukherjee}, R. and {Nagai}, T. and {Ong}, R.~A. and {Pandel}, D. and {Perkins}, J.~S. and {Pizlo}, F. and {Pohl}, M. and {Quinn}, J. and {Ragan}, K. and {Reynolds}, P.~T. and {Rose}, H.~J. and {Schroedter}, M. and {Sembroski}, G.~H. and {Smith}, A.~W. and {Steele}, D. and {Swordy}, S.~P. and {Toner}, J.~A. and {Valcarcel}, L. and {Vassiliev}, V.~V. and {Wagner}, R. and {Wakely}, S.~P. and {Ward}, J.~E. and {Weekes}, T.~C. and {Weinstein}, A. and {White}, R.~J. and {Williams}, D.~A. and {Wissel}, S.~A. and {Wood}, M. and {Zitzer}, B.},
	doi = {10.1086/587736},
	eid = {1427-1432},
	eprint = {0802.2363},
	journal = {\apj},
	month = jun,
	pages = {1427-1432},
	title = {{VERITAS Observations of the {$\gamma$}-Ray Binary LS I +61 303}},
	volume = 679,
	year = 2008}

@article{Schlafly:2011apj,
	adsurl = {http://adsabs.harvard.edu/abs/2011ApJ...737..103S},
	archiveprefix = {arXiv},
	author = {{Schlafly}, E.~F. and {Finkbeiner}, D.~P.},
	doi = {10.1088/0004-637X/737/2/103},
	eid = {103},
	eprint = {1012.4804},
	journal = {\apj},
	month = aug,
	pages = {103},
	primaryclass = {astro-ph.GA},
	title = {{Measuring Reddening with Sloan Digital Sky Survey Stellar Spectra and Recalibrating SFD}},
	volume = 737,
	year = 2011}

@inproceedings{Breeveld:2011aipc,
	adsurl = {http://adsabs.harvard.edu/abs/2011AIPC.1358..373B},
	archiveprefix = {arXiv},
	author = {{Breeveld}, A.~A. and {Landsman}, W. and {Holland}, S.~T. and {Roming}, P. and {Kuin}, N.~P.~M. and {Page}, M.~J.},
	booktitle = {American Institute of Physics Conference Series},
	doi = {10.1063/1.3621807},
	editor = {{J.~E.~McEnery, J.~L.~Racusin, \& N.~Gehrels}},
	eprint = {1102.4717},
	month = aug,
	pages = {373-376},
	primaryclass = {astro-ph.IM},
	series = {American Institute of Physics Conference Series},
	title = {{An Updated Ultraviolet Calibration for the Swift/UVOT}},
	volume = 1358,
	year = 2011}

@article{Fitzpatrick:1999PASP,
	adsurl = {https://ui.adsabs.harvard.edu/abs/1999PASP..111...63F},
	author = {{Fitzpatrick}, E.~L.},
	doi = {10.1086/316293},
	eprint = {astro-ph/9809387},
	journal = {\pasp},
	month = jan,
	pages = {63-75},
	title = {{Correcting for the Effects of Interstellar Extinction}},
	volume = 111,
	year = 1999}

@article{Walker:1994ApJ,
	adsurl = {https://ui.adsabs.harvard.edu/abs/1994ApJ...430L..45W},
	author = {{Walker}, R.~C. and {Romney}, J.~D. and {Benson}, J.~M.},
	doi = {10.1086/187434},
	journal = {\apjl},
	month = jul,
	pages = {L45},
	title = {{Detection of a VLBI Counterjet in NGC 1275: A Possible Probe of the Parsec-Scale Accretion Region}},
	volume = {430},
	year = 1994}

@article{Fujita:2017mnras,
	author = {Yutaka Fujita and Hiroshi Nagai},
	doi = {10.1093/mnrasl/slw217},
	journal = {Monthly Notices of the Royal Astronomical Society: Letters},
	month = {nov},
	number = {1},
	pages = {L94--L98},
	publisher = {Oxford University Press ({OUP})},
	title = {Discovery of a new subparsec counterjet in {NGC} 1275: the inclination angle and the environment},
	url = {https://doi.org/10.1093%2Fmnrasl%2Fslw217},
	volume = {465},
	year = 2016}

@article{Jorstad:2017ApJ,
	adsurl = {https://ui.adsabs.harvard.edu/abs/2017ApJ...846...98J},
	archiveprefix = {arXiv},
	author = {{Jorstad}, Svetlana G. and {Marscher}, Alan P. and {Morozova}, Daria A. and {Troitsky}, Ivan S. and {Agudo}, Iv{\'a}n and {Casadio}, Carolina and {Foord}, Adi and {G{\'o}mez}, Jos{\'e} L. and {MacDonald}, Nicholas R. and {Molina}, Sol N. and {L{\"a}hteenm{\"a}ki}, Anne and {Tammi}, Joni and {Tornikoski}, Merja},
	doi = {10.3847/1538-4357/aa8407},
	eid = {98},
	eprint = {1711.03983},
	journal = {\apj},
	month = sep,
	number = {2},
	pages = {98},
	primaryclass = {astro-ph.GA},
	title = {{Kinematics of Parsec-scale Jets of Gamma-Ray Blazars at 43 GHz within the VLBA-BU-BLAZAR Program}},
	volume = {846},
	year = 2017}

@article{Nagai:2010PASJ,
	adsurl = {https://ui.adsabs.harvard.edu/abs/2010PASJ...62L..11N},
	archiveprefix = {arXiv},
	author = {{Nagai}, Hiroshi and {Suzuki}, Kenta and {Asada}, Keiichi and {Kino}, Motoki and {Kameno}, Seiji and {Doi}, Akihiro and {Inoue}, Makoto and {Kataoka}, Jun and {Bach}, Uwe and {Hirota}, Tomoya and {Matsumoto}, Naoko and {Honma}, Mareki and {Kobayashi}, Hideyuki and {Fujisawa}, Kenta},
	doi = {10.1093/pasj/62.2.L11},
	eprint = {1001.3852},
	journal = {\pasj},
	month = apr,
	pages = {L11},
	primaryclass = {astro-ph.HE},
	title = {{VLBI Monitoring of 3C 84 (NGC 1275) in Early Phase of the 2005 Outburst}},
	volume = {62},
	year = 2010}

@article{Kino2021ApJ,
	adsurl = {https://ui.adsabs.harvard.edu/abs/2021ApJ...920L..24K},
	author = {{Kino}, Motoki and {Niinuma}, Kotaro and {Kawakatu}, Nozomu and {Nagai}, Hiroshi and {Giovannini}, Gabriele and {Orienti}, Monica and {Wajima}, Kiyoaki and {D'Ammando}, Filippo and {Hada}, Kazuhiro and {Giroletti}, Marcello and {Gurwell}, Mark},
	doi = {10.3847/2041-8213/ac24fa},
	eid = {L24},
	journal = {\apjl},
	month = oct,
	number = {1},
	pages = {L24},
	title = {{Morphological Transition of the Compact Radio Lobe in 3C 84 via the Strong Jet-Cloud Collision}},
	volume = {920},
	year = 2021}

@article{Nagai:2017ApJ,
	adsurl = {https://ui.adsabs.harvard.edu/abs/2017ApJ...849...52N},
	archiveprefix = {arXiv},
	author = {{Nagai}, H. and {Fujita}, Y. and {Nakamura}, M. and {Orienti}, M. and {Kino}, M. and {Asada}, K. and {Giovannini}, G.},
	doi = {10.3847/1538-4357/aa8e43},
	eid = {52},
	eprint = {1709.06708},
	journal = {\apj},
	month = nov,
	number = {1},
	pages = {52},
	primaryclass = {astro-ph.HE},
	title = {{Enhanced Polarized Emission from the One-parsec-scale Hotspot of 3C 84 as a Result of the Interaction with the Clumpy Ambient Medium}},
	volume = {849},
	year = 2017}

@article{Kino:2018ApJ,
	adsurl = {https://ui.adsabs.harvard.edu/abs/2018ApJ...864..118K},
	archiveprefix = {arXiv},
	author = {{Kino}, M. and {Wajima}, K. and {Kawakatu}, N. and {Nagai}, H. and {Orienti}, M. and {Giovannini}, G. and {Hada}, K. and {Niinuma}, K. and {Giroletti}, M.},
	doi = {10.3847/1538-4357/aad6e3},
	eid = {118},
	eprint = {1808.08855},
	journal = {\apj},
	month = sep,
	number = {2},
	pages = {118},
	primaryclass = {astro-ph.HE},
	title = {{Evidence of Jet-Clump Interaction: A Flip of the Radio Jet Head of 3C 84}},
	volume = {864},
	year = 2018}

@article{Paraschos2023,
	adsurl = {https://ui.adsabs.harvard.edu/abs/2023A&A...669A..32P},
	archiveprefix = {arXiv},
	author = {{Paraschos}, G.~F. and {Mpisketzis}, V. and {Kim}, J. -Y. and {Witzel}, G. and {Krichbaum}, T.~P. and {Zensus}, J.~A. and {Gurwell}, M.~A. and {L{\"a}hteenm{\"a}ki}, A. and {Tornikoski}, M. and {Kiehlmann}, S. and {Readhead}, A.~C.~S.},
	doi = {10.1051/0004-6361/202244814},
	eid = {A32},
	eprint = {2210.09795},
	journal = {\aap},
	month = jan,
	pages = {A32},
	primaryclass = {astro-ph.HE},
	title = {{A multi-band study and exploration of the radio wave-{\ensuremath{\gamma}}-ray connection in 3C 84}},
	volume = {669},
	year = 2023}

@article{Ahnen2016-lowstate,
	adsurl = {https://ui.adsabs.harvard.edu/abs/2016A&A...589A..33A},
	archiveprefix = {arXiv},
	author = {{Ahnen}, M.~L. and {Ansoldi}, S. and {Antonelli}, L.~A. and {Antoranz}, P. and {Babic}, A. and {Banerjee}, B. and {Bangale}, P. and {Barres de Almeida}, U. and {Barrio}, J.~A. and {Becerra Gonz{\'a}lez}, J. and {Bednarek}, W. and {Bernardini}, E. and {Biasuzzi}, B. and {Biland}, A. and {Blanch}, O. and {Bonnefoy}, S. and {Bonnoli}, G. and {Borracci}, F. and {Bretz}, T. and {Buson}, S. and {Carmona}, E. and {Carosi}, A. and {Chatterjee}, A. and {Clavero}, R. and {Colin}, P. and {Colombo}, E. and {Contreras}, J.~L. and {Cortina}, J. and {Covino}, S. and {Da Vela}, P. and {Dazzi}, F. and {De Angelis}, A. and {De Lotto}, B. and {de O{\~n}a Wilhelmi}, E. and {Delgado Mendez}, C. and {Di Pierro}, F. and {Dom{\'\i}nguez}, A. and {Dominis Prester}, D. and {Dorner}, D. and {Doro}, M. and {Einecke}, S. and {Eisenacher Glawion}, D. and {Elsaesser}, D. and {Fern{\'a}ndez-Barral}, A. and {Fidalgo}, D. and {Fonseca}, M.~V. and {Font}, L. and {Frantzen}, K. and {Fruck}, C. and {Galindo}, D. and {Garc{\'\i}a L{\'o}pez}, R.~J. and {Garczarczyk}, M. and {Garrido Terrats}, D. and {Gaug}, M. and {Giammaria}, P. and {Godinovi{\'c}}, N. and {Gonz{\'a}lez Mu{\~n}oz}, A. and {Gora}, D. and {Guberman}, D. and {Hadasch}, D. and {Hahn}, A. and {Hanabata}, Y. and {Hayashida}, M. and {Herrera}, J. and {Hose}, J. and {Hrupec}, D. and {Hughes}, G. and {Idec}, W. and {Kodani}, K. and {Konno}, Y. and {Kubo}, H. and {Kushida}, J. and {La Barbera}, A. and {Lelas}, D. and {Lindfors}, E. and {Lombardi}, S. and {Longo}, F. and {L{\'o}pez}, M. and {L{\'o}pez-Coto}, R. and {Lorenz}, E. and {Majumdar}, P. and {Makariev}, M. and {Mallot}, K. and {Maneva}, G. and {Manganaro}, M. and {Mannheim}, K. and {Maraschi}, L. and {Marcote}, B. and {Mariotti}, M. and {Mart{\'\i}nez}, M. and {Mazin}, D. and {Menzel}, U. and {Miranda}, J.~M. and {Mirzoyan}, R. and {Moralejo}, A. and {Moretti}, E. and {Nakajima}, D. and {Neustroev}, V. and {Niedzwiecki}, A. and {Nievas Rosillo}, M. and {Nilsson}, K. and {Nishijima}, K. and {Noda}, K. and {Orito}, R. and {Overkemping}, A. and {Paiano}, S. and {Palacio}, J. and {Palatiello}, M. and {Paneque}, D. and {Paoletti}, R. and {Paredes}, J.~M. and {Paredes-Fortuny}, X. and {Pedaletti}, G. and {Persic}, M. and {Poutanen}, J. and {Prada Moroni}, P.~G. and {Prandini}, E. and {Puljak}, I. and {Rhode}, W. and {Rib{\'o}}, M. and {Rico}, J. and {Rodriguez Garcia}, J. and {Saito}, T. and {Satalecka}, K. and {Schultz}, C. and {Schweizer}, T. and {Sillanp{\"a}{\"a}}, A. and {Sitarek}, J. and {Snidaric}, I. and {Sobczynska}, D. and {Stamerra}, A. and {Steinbring}, T. and {Strzys}, M. and {Takalo}, L. and {Takami}, H. and {Tavecchio}, F. and {Temnikov}, P. and {Terzi{\'c}}, T. and {Tescaro}, D. and {Teshima}, M. and {Thaele}, J. and {Torres}, D.~F. and {Toyama}, T. and {Treves}, A. and {Vazquez Acosta}, M. and {Verguilov}, V. and {Vovk}, I. and {Ward}, J.~E. and {Will}, M. and {Wu}, M.~H. and {Zanin}, R. and {Pfrommer}, C. and {Pinzke}, A. and {Zandanel}, F.},
	doi = {10.1051/0004-6361/201527846},
	eid = {A33},
	eprint = {1602.03099},
	journal = {\aap},
	month = may,
	pages = {A33},
	primaryclass = {astro-ph.HE},
	title = {{Deep observation of the NGC 1275 region with MAGIC: search of diffuse {\ensuremath{\gamma}}-ray emission from cosmic rays in the Perseus cluster}},
	volume = {589},
	year = 2016}

@article{Maier2017ICRC,
	author = {Maier, Gernot and Holder, Jamie},
	doi = {10.22323/1.301.0747},
	journal = {Proceedings of 35th International Cosmic Ray Conference --- {PoS}({ICRC}2017)},
	pages = {747},
	title = {Eventdisplay: An Analysis and Reconstruction Package for Ground-based Gamma-ray Astronomy},
	year = {2017}}

@ARTICLE{Zensus1997,
       author = {{Zensus}, J. Anton},
        title = "{Parsec-Scale Jets in Extragalactic Radio Sources}",
      journal = {\araa},
         year = 1997,
        month = jan,
       volume = {35},
        pages = {607-636},
          doi = {10.1146/annurev.astro.35.1.607},
       adsurl = {https://ui.adsabs.harvard.edu/abs/1997ARA&A..35..607Z}
}

@ARTICLE{SPOL1992,
       author = {{Schmidt}, Gary D. and {Stockman}, H.~S. and {Smith}, Paul S.},
        title = "{Discovery of a Sub-Megagauss Magnetic White Dwarf through Spectropolarimetry}",
      journal = {\apjl},
         year = 1992,
        month = oct,
       volume = {398},
        pages = {L57},
          doi = {10.1086/186576},
       adsurl = {https://ui.adsabs.harvard.edu/abs/1992ApJ...398L..57S}
}

\appendix

\section{Observations and spectral parameters for power-law fit}
\label{Appendix::spec}

\begin{deluxetable}{ccccccc}[H]
\caption{Summary of \swiftXRT{} observations and spectral parameters for power-law fit.}
\label{table:xrt_obervations}
\centering
\tabletypesize{\scriptsize}
    \tablehead{
        \colhead{Observation} & \colhead{Mode} & \colhead{Date} & \colhead{Exposure [s]} & \colhead{PL Index} & \colhead{PL norm} & \colhead{Flux [ergs cm$^{-2}$s$^{-1}$]}
    }
    \startdata
  00031770011 & WT   & 2017-01-01  & 980  & 1.81 $\pm$ 0.06 & 0.025 $\pm$ 0.001   & (1.59 $\pm$ 0.10)$\times 10^{-10}$ \\
00031770012 & WT   & 2017-01-03  & 1000  & 1.96 $\pm$ 0.06  & 0.024 $\pm$ 0.001  & (1.40 $\pm$ 0.09)$\times 10^{-10}$ \\
00031770013 & WT   & 2017-01-05  & 300  & 2.01 $\pm$ 0.16 & 0.017  $\pm$ 0.002   & (9.70 $\pm$ 1.30)$ \times 10^{-11}$ \\
00031770014 & WT   & 2017-01-07  & 200  & 1.88 $\pm$ 0.16 & 0.021 $\pm$ 0.003 & (1.33 $\pm$ 0.20)$ \times 10^{-10}$ \\
00031770015 & WT   & 2017-01-12 & 240  & 2.04  $\pm$ 0.18  & 0.019  $\pm$ 0.003 & (1.08 $\pm$ 0.20)$ \times 10^{-10}$ \\
00031770016 & WT   & 2017-01-17 &  575 & 1.85 $\pm$ 0.14  & 0.012  $\pm$ 0.002 & (7.67 $\pm$ 1.10)$\times 10^{-11}$ \\
00031770017 & WT   & 2017-01-20 & 775   & 1.84 $\pm$ 0.10 & 0.018  $\pm$ 0.001 & (1.11 $\pm$ 0.10)$ \times 10^{-10}$ \\
00031770018 & WT   & 2017-01-22 & 1000  & 1.97$\pm$ 0.08 & 0.020  $\pm$ 0.001  & (1.13  $\pm$ 0.08)$ \times 10^{-10}$ \\
00031770019 & WT   & 2017-01-24 & 670  & 2.02 $\pm$ 0.12  & 0.016 $\pm$ 0.001  & (9.17 $\pm$ 0.90)$ \times 10^{-11}$ \\
00031770020 & WT   & 2017-01-26 & 910  & 1.87 $\pm$ 0.08 & 0.017 $\pm$ 0.001  & (1.07 $\pm$ 0.09)$ \times 10^{-10}$ \\
00031770021 & WT   & 2017-01-29 & 1020 & 1.96$\pm$ 0.07 & 0.020 $\pm$ 0.001  & (1.18 $\pm$ 0.08)$ \times 10^{-10}$ \\
00031770022 & WT   & 2017-02-07 & 1140   & 1.89$\pm$ 0.08 & 0.017  $\pm$ 0.001  & (1.04 $\pm$ 0.08)$ \times 10^{-10}$ \\
00031770023 & WT   & 2017-02-10 & 1460  & 1.95 $\pm$ 0.06  & 0.021  $\pm$ 0.001 & (1.22  $\pm$ 0.07)$ \times 10^{-10}$ \\
00031770024 & WT   & 2017-02-13 & 1170  & 2.14 $\pm$ 0.07 & 0.024  $\pm$ 0.001  & (1.25 $\pm$ 0.07)$ \times 10^{-10}$ \\
00031770025 & WT   & 2017-02-16 & 1400 & 2.03 $\pm$ 0.07 & 0.020 $\pm$ 0.001 & (1.14 $\pm$ 0.07)$ \times 10^{-10}$  \\
00031770026 & WT   & 2017-02-24 & 1430 & 1.90 $\pm$ 0.06 & 0.020  $\pm$ 0.001 & (1.19 $\pm$ 0.07)$ \times 10^{-10}$ \\
00031770027 & WT   & 2017-02-27 & 900 & 1.91 $\pm$ 0.08 & 0.019 $\pm$ 0.001  & (1.11 $\pm$ 0.10)$ \times 10^{-10}$ \\
00031770029 & WT   & 2017-03-05 & 300 & 1.70 $\pm$ 0.20 & 0.013  $\pm$ 0.002  & (9.11 $\pm$ 1.80)$\times 10^{-11}$ \\
00031770030 & WT   & 2017-03-08 & 330  & 1.90 $\pm$ 0.10  & 0.018  $\pm$ 0.002  & (1.09 $\pm$ 0.20)$ \times 10^{-10}$ \\
00031770031 & WT   & 2017-03-09 & 600 & 1.96 $\pm$ 0.12 & 0.016 $\pm$ 0.002  & (9.50 $\pm$ 1.10)$\times 10^{-11}$ \\
00031770032 & WT   & 2017-03-11 & 1020 & 1.90 $\pm$ 0.08 & 0.019  $\pm$ 0.001  & (1.12 $\pm$ 0.09)$ \times 10^{-10}$ \\
00031770034 & WT   & 2017-03-20 & 960 & 1.72 $\pm$ 0.10 & 0.014  $\pm$ 0.001  & (9.63 $\pm$ 1.10)$ \times 10^{-11}$ \\
00031770035 & WT   & 2017-03-23 & 980 & 1.75 $\pm$ 0.08 & 0.016 $\pm$ 0.001  & (1.09 $\pm$ 0.10)$ \times 10^{-10}$ \\
00031770036 & WT   & 2017-03-26 & 1130 & 1.76 $\pm$ 0.07 & 0.017 $\pm$ 0.001 & (1.13 $\pm$ 0.09)$ \times 10^{-10}$ \\
00031770037 & WT   & 2017-03-29 & 1080 & 2.02 $\pm$ 0.08 & 0.019 $\pm$ 0.001  & (1.07 $\pm$ 0.08)$ \times 10^{-10}$ \\
00031770038 & WT   & 2017-04-01 & 970 & 1.85 $\pm$ 0.08 & 0.018  $\pm$ 0.001   & (1.11 $\pm$ 0.10)$ \times 10^{-10}$\\
00031770039 & WT   & 2017-04-04 & 780 & 1.74 $\pm$ 0.09 & 0.017 $\pm$ 0.001  & (1.14 $\pm$ 0.10)$ \times 10^{-10}$ \\
00088027001 & WT   & 2017-02-01 & 1660 & 2.02    $\pm$ 0.06 & 0.022   $\pm$ 0.001 & (1.21 $\pm$ 0.06)$ \times 10^{-10}$\\
00088027002 & WT   & 2017-02-04 & 1600 & 1.95  $\pm$ 0.06 & 0.020   $\pm$ 0.001 & (1.17 $\pm$ 0.07)$ \times 10^{-10}$ \\
00034765001 & PC   & 2016-10-30 & 1980 & 1.80  $\pm$ 0.09  & 0.016   $\pm$ 0.001  & (1.04 $\pm$ 0.10)$ \times 10^{-10}$ \\
00034765002 & PC   & 2016-10-31 & 1970 & 1.92  $\pm$ 0.09 & 0.016   $\pm$ 0.001  & (9.24 $\pm$ 0.80)$ \times 10^{-11}$ \\
00034765003 & PC   & 2016-11-01 & 1870 & 1.81  $\pm$ 0.10 & 0.014   $\pm$ 0.001 & (9.03 $\pm$ 1.00)$ \times 10^{-11}$ \\
00034765004 & PC   & 2016-11-02 & 1630 & 1.70  $\pm$ 0.10 & 0.013   $\pm$ 0.001  & (9.30 $\pm$ 1.30)$ \times 10^{-11}$ \\
00034765006 & PC   & 2016-11-04 & 1680  & 1.89  $\pm$ 0.11  & 0.014   $\pm$ 0.001  & (8.70 $\pm$ 1.00)$ \times 10^{-11}$ \\
00034765007 & PC   & 2016-11-05 & 1800 & 1.79  $\pm$ 0.09 & 0.014  $\pm$ 0.001  & (9.01 $\pm$ 1.00)$ \times 10^{-11}$ \\
00034765008 & PC   & 2016-11-06 & 2000 & 1.81  $\pm$ 0.11   & 0.014    $\pm$ 0.001  & (9.23 $\pm$ 1.00)$ \times 10^{-11}$ \\
00034765009 & PC   & 2016-11-07 & 1950 & 1.94   $\pm$ 0.09 & 0.019  $\pm$ 0.001  & (1.10 $\pm$ 0.09)$ \times 10^{-10}$ \\
00034765010 & PC   & 2016-11-08 & 1600 & 1.95  $\pm$ 0.09 & 0.020    $\pm$ 0.001  & (1.15 $\pm$ 0.10)$ \times 10^{-10}$ \\
00034765011 & PC   & 2016-11-09 & 1550 & 1.99  $\pm$ 0.10  & 0.018   $\pm$ 0.001  & (1.01 $\pm$ 0.09)$ \times 10^{-10}$ \\
00034765012 & PC   & 2016-11-10 & 1920 & 2.02  $\pm$ 0.09 & 0.018   $\pm$ 0.001  & (9.84 $\pm$ 0.70)$ \times 10^{-11}$ \\
00087311001 & PC   & 2017-01-01 & 630 & 1.69  $\pm$ 0.13 & 0.021   $\pm$ 0.002  & (1.45  $\pm$ 0.20)$ \times 10^{-10}$ \\
00087311002 & PC   & 2017-01-02 & 800 & 1.73  $\pm$ 0.11  & 0.021   $\pm$ 0.002  & (1.45   $\pm$ 0.20)$ \times 10^{-10}$ \\
00087311003 & PC   & 2017-03-15 & 1070 & 1.88  $\pm$ 0.13  & 0.017   $\pm$ 0.002 & (1.04 $\pm$ 0.10)$ \times 10^{-10}$ \\
00087311004 & PC   & 2017-03-21 & 860 & 1.80  $\pm$ 0.18   & 0.011  $\pm$ 0.001  & (7.12 $\pm$ 1.40)$ \times 10^{-11}$ \\
00087311005 & PC   & 2017-03-24 & 2400 & 1.78  $\pm$ 0.09 & 0.012   $\pm$ 0.001  & (8.00 $\pm$ 0.90)$ \times 10^{-11}$ \\
00087312001 & PC   & 2016-12-30 & 940 & 1.86  $\pm$ 0.12  & 0.018  $\pm$ 0.002  & (1.16  $\pm$ 0.10)$ \times 10^{-10}$ \\
00087312002 & PC   & 2017-03-21 & 1470 & 1.74  $\pm$ 0.12  & 0.012   $\pm$ 0.001  & (7.98  $\pm$ 1.10)$ \times 10^{-11}$ \\
00087312003 & PC   & 2017-03-25 & 800 & 1.97  $\pm$ 0.22 & 0.012   $\pm$ 0.002  & (7.23 $\pm$ 1.50)$ \times 10^{-11}$ \\
00087312005 & PC   & 2017-03-31 & 2130 & 1.89 $\pm$ 0.10 & 0.014  $\pm$ 0.001   & (8.41 $\pm$ 0.90)$\times 10^{-11}$\\
\hline
\enddata
\tablenotetext{}{\emph{Observation} lists the \swiftXRT{ }observation ID; \emph{Mode} indicates whether the observation was taken in the Photon Counting (PC) or Window Timing (WT) mode; \emph{Date} is the date of observation (UTC); \emph{Exp} is the effective on-source time (in seconds); \emph{PL Index} is the best-fitting power-law photon index; \emph{PL norm} is the best-fitting power-law normalization constant at 1 keV (ph cm$^{-2}$s$^{-1}$keV$^{-1}$); and \emph{Flux} is the deabsorbed 0.3-10 keV energy flux in units of ergs cm$^{-2}$s$^{-1}.$}
\end{deluxetable}

\section{MCMC corner plots}
\label{Appendix::corner_plots}

\begin{figure*}[h]
\centering
\includegraphics[width=12cm]{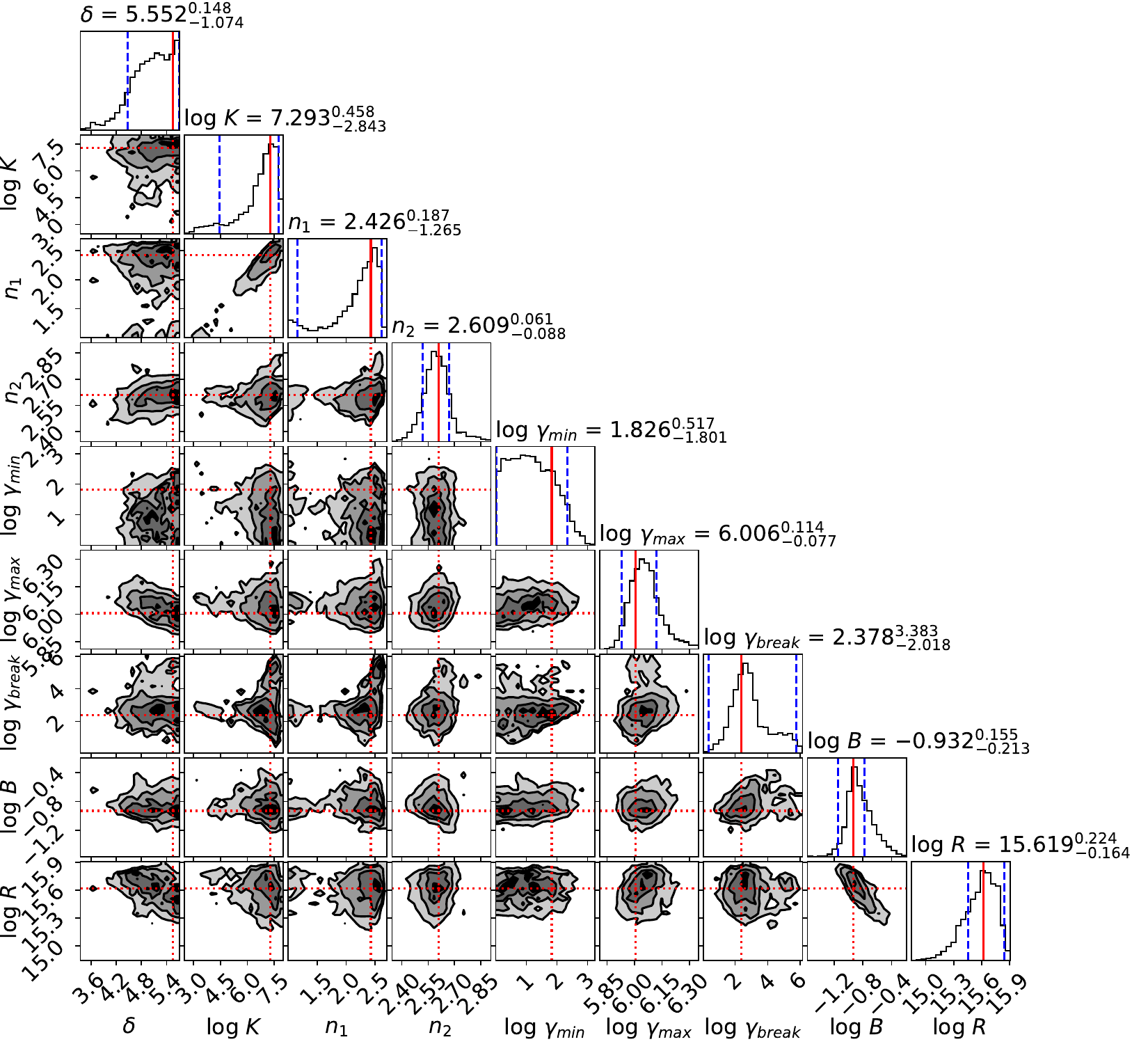}
\includegraphics[width=12cm]{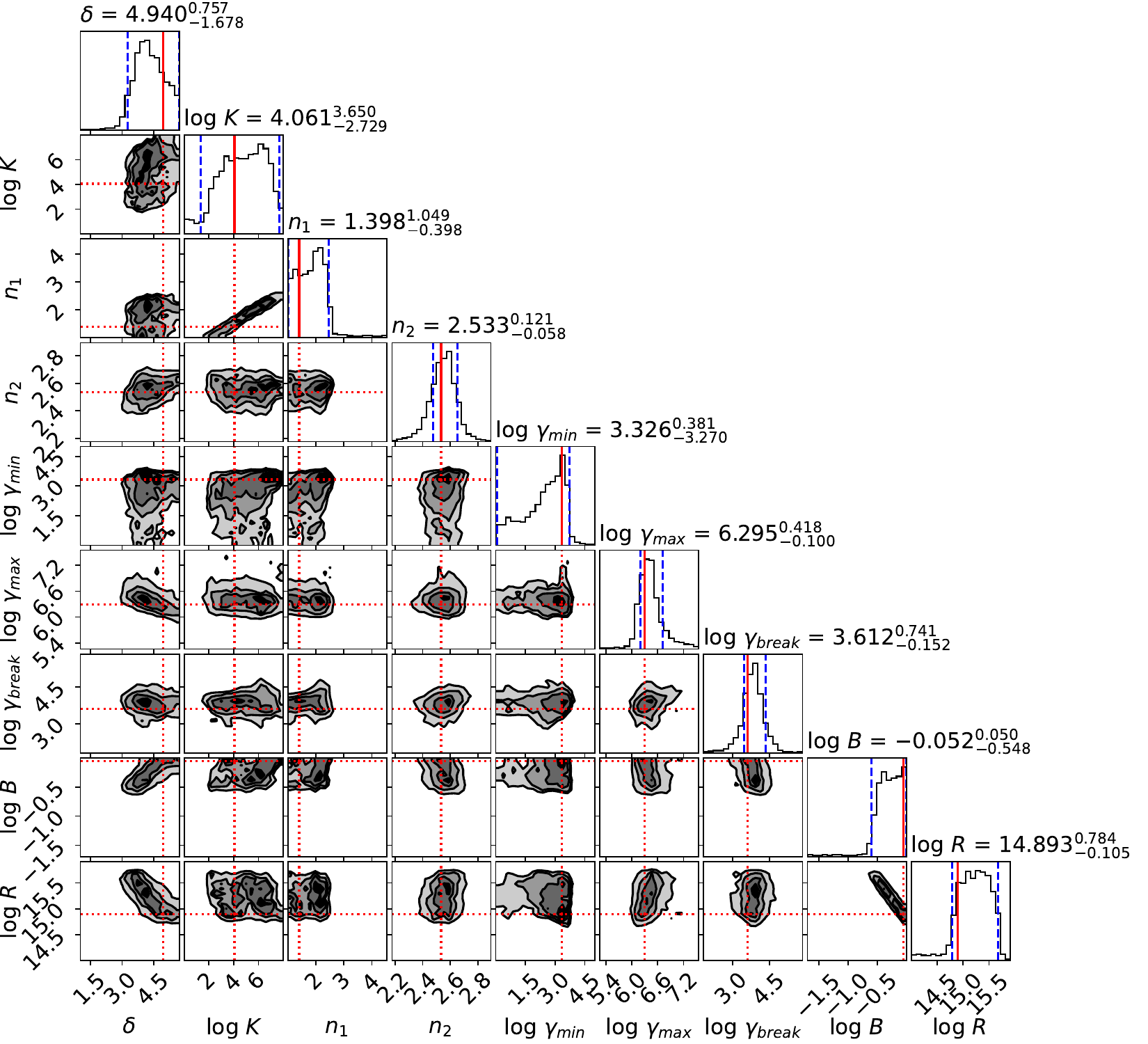}
\caption{Corner plots of the posterior distribution of the free parameters in the SED model fit of \decthirtyfirst{}/\janfirst{} (top panel) and \jansecond{} (bottom panel).}
\vspace{-20ex}
\label{figure:sed_modelling_corner}
\end{figure*}
\end{document}